\documentclass[prx, onecolumn, superscriptaddress, notitlepage]{revtex4-2}

\usepackage{amsmath,amsfonts, amssymb, amsthm, dsfont}
\usepackage{yfonts}
\usepackage{bm,bbm}
\usepackage{mathrsfs}
\usepackage{graphicx}
\usepackage{verbatim}
\usepackage{hyperref}
\usepackage{multirow}
\usepackage[normalem]{ulem}
\usepackage{sseq}
\usepackage{array}
\usepackage{shuffle}
\usepackage{todonotes}
\usepackage[justification=Justified]{caption}
\usepackage{subcaption}
\usepackage{amssymb}
\usepackage{tikz}
	\usetikzlibrary{calc}
	\usetikzlibrary{shapes.multipart}
\usepackage{tikz-cd}

\theoremstyle{theorem}
\newtheorem{theorem}{Theorem}

\newtheorem{proposition}[theorem]{Proposition}

\theoremstyle{definition}
\newtheorem{definition}[theorem]{Definition}
\newtheorem{remark}[theorem]{Remark}

\numberwithin{theorem}{section}

\newcommand{\dd}{\mathrm{d}}
\newcommand{\ii}{\mathrm{i}}

\newcommand{\comments}[1]{}

\newcommand{\eq}[1]{Eq.~(\ref{#1})}

\renewcommand{\Vec}{\mathrm{Vec}}
\newcommand{\Rep}{\mathrm{Rep}}
\newcommand{\C}{\mathcal C}
\newcommand{\ACA}{{}_A\mathcal C_A}

\def\U{\mathrm{U}(1)}

\def\M{{{\mathcal{M}}}}
\def\T{{{\mathcal{T}}}}

\def\Z{\mathbb{Z}}

\usetikzlibrary{decorations.pathreplacing,decorations.markings}
\tikzset{middlearrow/.style={
        decoration={markings,
            mark= at position 0.55 with {\arrow{#1}} ,
        },
        postaction={decorate}
    }
}

\begin{document}

\title{$E_\infty^{1,2}$-type Lieb-Schultz-Mattis anomalies, deconfined quantum critical points, and non-invertible symmetry breaking}
\date{\today}

\author{Hao-Ran Zhang}
\affiliation{State Key Laboratory of Low-Dimensional Quantum Physics, Department of Physics, Tsinghua University, Beijing 100084, China}

\author{Hanlin Lin}
\affiliation{Qiuzhen College, Tsinghua University, Beijing 100084, China}

\author{Shuo Yang}
\affiliation{State Key Laboratory of Low-Dimensional Quantum Physics, Department of Physics, Tsinghua University, Beijing 100084, China}
\affiliation{Frontier Science Center for Quantum Information, Beijing, China}
\affiliation{Hefei National Laboratory, Hefei 230088, China}

\author{Qing-Rui Wang}
\email{wangqr@mail.tsinghua.edu.cn}
\affiliation{Yau Mathematical Sciences Center, Tsinghua University, Beijing 100084, China}

\begin{abstract}
We study deconfined quantum critical points (DQCP) associated with Lieb-Schultz-Mattis (LSM) anomalies in one-dimensional spin chains. Our starting point is a structural characterization of the LSM anomaly in the Lyndon-Hochschild-Serre spectral sequence:
$
\omega_{\mathrm{LSM}}\in E_\infty^{1,2}= H^1(\mathbb Z_{\mathrm{trans}},H^2(G_{\mathrm{int}},\U))\subseteq H^3(G_{\mathrm{int}}\rtimes_{\rho}\mathbb Z_{\mathrm{trans}},\U)
$.
Physically, this class decorates a translation defect with a projective representation of the internal symmetry $G_\mathrm{int}$. We show that gauging the internal symmetry in the presence of an $E_\infty^{1,2}$-type anomaly necessarily produces a non-invertible dual symmetry. This gives a general mechanism for type-II DQCP: in contrast to type-I examples with $E_\infty^{2,1}$-type anomalies which are dual to ordinary group-like symmetry breaking, type-II transitions are dual to spontaneous breaking of a non-invertible symmetry. We illustrate the mechanism using a spin-$1/2$ chain with an anomalous $D_8$ LSM symmetry. We construct a dimer-to-ferromagnet DQCP candidate, provide numerical evidence for a critical theory with central charge $c\approx 1$, and show, using both category theory and explicit lattice constructions, that gauging the internal symmetry yields the non-invertible $\Rep(H_8)$ dual symmetry.
\end{abstract}

\maketitle
\tableofcontents

\section{Introduction}

The Landau-Ginzburg-Wilson paradigm describes a continuous phase transition in terms of fluctuations of a local order parameter. Deconfined quantum critical points (DQCPs) provide a different possibility: the two adjacent gapped phases may break different, non-nested symmetries, while the critical point is more naturally described by fractionalized degrees of freedom, emergent gauge fields, or an anomaly-enforced critical theory \cite{SenthilScience2004,SenthilPRB2004,MotrunichVishwanath2004,WangNahumMetlitskiXuSenthil2017}. The original and most familiar examples involve the N\'eel-valence-bond-solid transition in two spatial dimensions. This proposal has been extensively explored in lattice realizations and numerical studies of $J$-$Q$-type models, emergent $U(1)$ symmetry, and scaling behavior near the putative DQCP \cite{Sandvik2007,LouSandvikBalents2007,MelkoKaul2008,NahumChalkerSernaOrtunoSomoza2015}. More recently, one-dimensional realizations and exactly solvable boundary constructions have made it possible to study the same logic in a sharper algebraic setting, where duality, anomaly, and spontaneous symmetry breaking (SSB) can be tracked explicitly \cite{JiangMotrunich2019,PhysRevB.99.165143,ZhangLevin2023,Chatterjee_2023}.

The anomaly viewpoint is especially useful in one dimension because it connects DQCPs with Lieb-Schultz-Mattis (LSM) constraints. The original LSM theorem shows that a half-odd-integer spin chain cannot have a unique gapped translation-invariant ground state \cite{LiebSchultzMattis1961}; related rigorous spin-chain extensions include the Affleck-Lieb argument for half-odd-integer spin chains \cite{AffleckLieb1986}. Oshikawa's flux insertion argument and Hastings' higher-dimensional extension reveal the topological nature of this obstruction \cite{Oshikawa2000,Hastings2004}. For general internal symmetries, the obstruction is controlled by the projective representation carried by each primitive unit cell, or equivalently by a mixed anomaly between internal symmetry and lattice translation \cite{OgataTachikawaTasaki2021,YaoOshikawa2021,ChoHsiehRyu2017,FuruyaOshikawa2017,YaoHsiehOshikawa2019,ElseThorngren2020,ChengSeiberg2023,AksoyMudryFurusakiTiwari2024}. From the viewpoint of symmetry-protected topological (SPT) phases, this LSM anomaly can be realized as the boundary anomaly of a weak crystalline SPT phase, whose translation defect is decorated by a one-dimensional internal-symmetry SPT state \cite{ChenGuWen2010LocalUnitary,ChenGuWen2011,ChenGuWen2011Complete,SchuchPerezGarciaCirac2011,PollmannBergTurnerOshikawa2012,LevinGu2012,ChenGuLiuWen2013,LanKongWen2017Classification,ChenLuVishwanath2014,HuangSongHuangHermele2017,ThorngrenElse2018,ElseThorngren2019}.

Mathematically, the natural bookkeeping device for this kind of anomaly is the Lyndon-Hochschild-Serre (LHS) spectral sequence associated with a symmetry extension \cite{Lyndon1948,HochschildSerre1953,Brown1982,WangNingCheng2021}
\begin{align}
1\longrightarrow A\longrightarrow G\longrightarrow Q\longrightarrow 1 .
\end{align}
The LSM anomaly considered in this work, denoted by $\omega_{\mathrm{LSM}}$, is an $E_\infty^{1,2}$-type class: it is represented, on the $E_2$ page, by a class in $H^1(Q,H^2(A,\U))$ and survives to the $E_\infty$ page. For a spin chain with internal symmetry $G_{\mathrm{int}}$ and translation symmetry $\mathbb Z_{\mathrm{trans}}$, the central point is
\begin{align}
\omega_{\mathrm{LSM}}
\in E_\infty^{1,2}
= E_2^{1,2}
=H^1(\mathbb Z_{\mathrm{trans}},H^2(G_{\mathrm{int}},\U))
\subseteq H^3(G_{\mathrm{int}}\rtimes_{\rho}\mathbb Z_{\mathrm{trans}},\U)
\end{align}
of the LHS spectral sequence, where the displayed inclusion denotes the surviving part of the $E_2$ page. Here $\mathbb Z_{\mathrm{trans}}$ denotes the one-dimensional lattice translation symmetry, and $\mathbb Z_{\mathrm{trans}}$ acts on $H^2(G_{\mathrm{int}},\U)$ through its action on $G_{\mathrm{int}}$. Physically, the $H^2(G_{\mathrm{int}},\U)$ factor records the projective representation, or one-dimensional $G_{\mathrm{int}}$-SPT decoration, while the $H^1(\mathbb Z_{\mathrm{trans}},\cdot)$ factor records the translation defect on which this decoration is placed. This is distinct from $E_\infty^{2,1}$-type anomalies, where the decoration is a zero-dimensional charge attached to an intersection or endpoint of domain walls. The distinction will be important because it controls what kind of symmetry appears after gauging.

A useful point of comparison is the exactly solvable DQCP model of Zhang and Levin, which involves only internal symmetries \cite{ZhangLevin2023}. With respect to the subgroup being gauged, the anomaly in that model is of $E_\infty^{2,1}$ type rather than the $E_\infty^{1,2}$ LSM type studied here, and this is why gauging produces an invertible dual symmetry. There the boundary theory has an anomalous $\Z_2\times\Z_2$ symmetry: a domain wall of one $\Z_2$ carries charge under the other. Gauging one $\Z_2$ maps the anomalous boundary problem to an ordinary $\Z_4$ spin chain. Thus a conventional $\Z_4$ symmetry-breaking transition in the dual variables becomes, before gauging, a DQCP between two gapped edges that break different $\Z_2$ subgroups. From the bulk point of view, the same operation relates the twisted Dijkgraaf-Witten theory associated with the $\Z_2\times\Z_2$ SPT edge to an untwisted $\Z_4$ gauge-theory description \cite{DijkgraafWitten1990,kapustin2014anomalies,kapustin2014anomalous}. This example is closely related to the broader idea that Kramers-Wannier duality, order-disorder duality, and order/disorder parameters in spin chains can be understood through gauging and duality operations \cite{KramersWannier1941,FradkinSusskind1978,Levin2020OrderDisorder,bhardwaj2018finite}. It shows that a DQCP may become an ordinary symmetry-breaking transition after a suitable gauging, provided the dual symmetry remains an ordinary invertible group symmetry.

The goal of this paper is to generalize this picture to LSM anomalies and to identify what changes when the relevant anomaly is of $E_\infty^{1,2}$ type. We use the language of generalized and fusion-category symmetries, in which topological defect lines need not be invertible \cite{LevinWen2005StringNet,GaiottoKapustinSeibergWillett2015,KongWen2014BraidedFusion,CordovaDumitrescuIntriligatorShao2022Snowmass,BhardwajBottiniFraserTalienteGladdenGouldPlatschorreTillim2024Lectures,McGreevy2023GeneralizedSymmetries,BrennanHong2023Introduction,SchaferNameki2024ICTP,Shao2023TASI,thorngren1912fusion,AasenFendleyMong2020,luo2024lecture,FrohlichFuchsRunkelSchweigert2004,FuchsRunkelSchweigert2009DefectLines,ChangLinShaoWangYin2019TopologicalDefectLines}. An anomalous finite group symmetry with anomaly $\omega\in H^3(G,\U)$ is described by the pointed (meaning every simple object has an tensor inverse) fusion category $\Vec_G^{\omega}$ \cite{DijkgraafWitten1990,etingof2016tensor,BhardwajBottiniSchaferNamekiTiwari2023HigherCategorical}. Gapped phases are classified by module categories over the symmetry category. Condensable algebras give algebra representatives of these module categories and, when chosen as concrete gaugeable algebras, define generalized gauging operations \cite{kitaev2012models,ostrik2002module,ostrik2003module,bruguieres2011exact,natale2020notion,uribe2017classification,kong2014anyon,kong2022invitation,davydov2010modular,davydov2017lagrangian,bhardwaj2018finite,tachikawa2020gauging,RoumpedakisSeifnashriShao2023HigherGauging,ChoiCordovaHsinLamShao2022DualityDefects,ChoiCordovaHsinLamShao2023CondensationDefects,PerezLonaRobbinsSharpeVandermeulenYu2024GaugingPartI,ChoiLuSun2024SelfDuality,diatlyk2024gauging,diatlyk2026gaugingnoninvertiblesymmetriestopological,SeifnashriShaoYang2025LatticeGauging}. Gauging a chosen condensable algebra $A$ changes the symmetry category from $\C$ to the bimodule category ${}_A\C_A$. When ${}_A\C_A$ is pointed, the dual symmetry is again an ordinary group symmetry. When ${}_A\C_A$ is not pointed, the dual symmetry is non-invertible. This criterion lets us separate several distinct mechanisms by which gauging produces non-invertible symmetry:
\begin{itemize}
\item Gauging a non-Abelian subgroup can leave non-invertible representation-sector defects.
\item Gauging a non-normal subgroup can leave non-invertible double-coset defects.
\item In this work, even gauging an Abelian normal subgroup gives a non-invertible dual symmetry, if the anomaly has a nontrivial $E_\infty^{1,2}$ component.
\end{itemize}
By contrast, in the usual subgroup-gauging setting, gauging a normal Abelian subgroup with anomaly outside the $E_\infty^{1,2}$ position always gives a pointed, hence invertible, dual symmetry \cite{tachikawa2020gauging,Naidu2007CategoricalMorita,uribe2017classification,Mu_oz_2018}.
Such non-invertible symmetries have recently become an important organizing principle for one-dimensional spin chains, generalized gauging, and symmetry-protected or symmetry-breaking critical phenomena \cite{tambara1998tensor,tambara2000representations,SeibergShao2024MajoranaIsing,Seifnashri2024LSMObstructions,SeibergSeifnashriShao2024LSMNonInvertible,BhardwajBottiniPajerSchaferNameki2024CategoricalLandau,bhardwaj2403hasse,10.21468/SciPostPhys.17.4.115,PhysRevLett.133.116601}.

Our main general observation is thus that $E_\infty^{1,2}$-type LSM anomalies naturally lead, after gauging the internal symmetry, to non-invertible dual symmetries. We refer to DQCPs of this kind as ``type-II'': they are dual to symmetry breaking for a non-invertible fusion-category symmetry. By contrast, previously studied DQCPs that become ordinary symmetry-breaking transitions of an invertible dual symmetry after gauging will be called ``type-I''. This type-I situation is represented by $E_\infty^{2,1}$-type mixed anomalies, where gauging can produce a non-anomalous invertible dual symmetry such as the $\Z_4$ symmetry in the Zhang-Levin construction \cite{ZhangLevin2023}. The difference is visible already at the level of the decorated-domain-wall picture of the bulk SPT: an $E_\infty^{2,1}$ anomaly decorates defects with charges, while an $E_\infty^{1,2}$ LSM anomaly decorates defects with projective representations. Proliferating such decorated defects therefore forces the dual symmetry data to remember a higher-dimensional local degeneracy, which is naturally encoded by non-invertible simple objects.

We illustrate the mechanism in a concrete spin-$1/2$ chain. The microscopic symmetry is generated by two internal $\Z_2$ operations $U_{XY}$ and $U_{YX}$, together with one-site translation $T$, where $T$ exchanges the two internal $\Z_2$ factors. In the sector relevant to phases preserving $T^2$, the effective finite symmetry group is
\begin{align}
D_8
=
(\Z_2^{XY}\times \Z_2^{YX})\rtimes \Z_2^{T}.
\end{align}
Because the local representatives of $U_{XY}$ and $U_{YX}$ are Pauli $X$ and $Y$ operators, each primitive unit cell carries the nontrivial projective representation of $\Z_2^{XY}\times\Z_2^{YX}$. In this concrete $D_8$ realization, the LSM anomaly $\omega_{\mathrm{LSM}}$ is the class $2\zeta$ in the $\mathbb Z_4$ summand of $H^3(D_8,\U)$ \cite{tomoda2008remarks,handel1993products}. Thus the anomalous symmetry of the spin chain is the fusion category $\Vec_{D_8}^{\omega_{\mathrm{LSM}}}$.

We then classify the gapped phases with $\Vec_{D_8}^{\omega_{\mathrm{LSM}}}$ symmetry in terms of anomaly-free subgroups of $D_8$ and construct explicit representative ground states with their parent Hamiltonians. Among these phases we focus on a dimer phase and a $z$-ferromagnetic phase. A simple interpolation between their parent Hamiltonians gives a one-dimensional DQCP candidate; after a unitary transformation, it becomes a nearest-neighbor-$z$-anisotropic Majumdar-Ghosh-type spin chain \cite{MajumdarGhosh1969I,MajumdarGhosh1969II}. Using VUMPS and finite-entanglement scaling \cite{TagliacozzoOliveiraIblisdirLatorre2008,PhysRevLett.102.255701,PhysRevB.97.045145}, we find a growing correlation length and an entanglement scaling consistent with a $c\approx 1$ conformal field theory. This provides numerical evidence that the transition is continuous.

Finally, we analyze generalized gauging in this example. Gauging the central subgroup $\Z_2^{a^2}$ gives a self-dual, still pointed, $\Vec_{D_8}^{\omega_{\mathrm{LSM}}}$ symmetry, closely analogous to an ordinary Kramers-Wannier duality. Gauging the single subgroup $\Z_2^{XY}$ realizes the familiar non-normal-subgroup route to non-invertibility. Most importantly, gauging the full normal Abelian internal symmetry $\Z_2^{XY}\times\Z_2^{YX}$ also produces the non-pointed fusion category $\Rep(H_8)$, where $H_8$ is the eight-dimensional Kac-Paljutkin Hopf algebra \cite{tambara2000representations,etingof2021tensor}. The same $\Rep(H_8)$ fusion category has also appeared in recent discussions of non-invertible symmetry and its gauging, including generalized orbifold constructions and self-dualities of $c=1$ orbifold CFTs \cite{diatlyk2024gauging,ChoiLuSun2024SelfDuality}. This last gauging is the concrete LSM mechanism: the gauged subgroup is as ordinary as possible, but the $E_\infty^{1,2}$ anomaly forces the dual symmetry to be non-invertible. Under this gauging, the dimer and ferromagnetic phases map to different condensable algebras in $\Rep(H_8)$, and the DQCP is reinterpreted as a transition controlled by non-invertible symmetry breaking. This gives a lattice realization of the general principle that $E_\infty^{1,2}$ LSM anomalies are naturally dual to non-invertible symmetry-breaking criticality.

The rest of the paper is organized as follows. In Sec.~II we review lattice symmetry actions and formulate LSM anomalies as $E_\infty^{1,2}$ classes in the LHS spectral sequence, including the decorated-domain-wall picture. In Sec.~III we discuss how gauging an $E_\infty^{1,2}$-type anomaly produces non-invertible dual symmetry. In Sec.~IV we study the concrete $\Vec_{D_8}^{\omega_{\mathrm{LSM}}}$ spin chain, classify its gapped phases, construct the dimer-to-ferromagnet DQCP Hamiltonian, and analyze its dual $\Rep(H_8)$ description. The appendices collect the category-theoretic background, the subgroup-gauging computations, and the explicit group-cohomology data used in the main text.

\section{Lattice symmetry actions and $E_\infty^{1,2}$-type LSM anomalies}
\label{sec:symm}

In this section, we formulate a general framework for the lattice symmetries underlying LSM anomalies in one-dimensional systems. We start from a general internal symmetry $G_{\mathrm{int}}$ together with either the full translation group $\mathbb Z_{\mathrm{trans}}$ or a finite translation quotient $\mathbb Z_N$, allowing translations to act nontrivially on $G_{\mathrm{int}}$. After spelling out the corresponding symmetry group structure, we turn to the anomaly data and show that the LSM anomaly is naturally described by an $E_\infty^{1,2}$ class in the Lyndon-Hochschild-Serre spectral sequence. We then give the equivalent decorated-domain-wall interpretation from the viewpoint of a bulk weak crystalline SPT phase. This formulation provides the starting point for our general gauging construction and the emergence of non-invertible dual symmetries.

\subsection{Lattice symmetry actions}

\subsubsection{For $\Z$ translational symmetry}

We begin with a one-dimensional quantum spin system on a translation-invariant lattice. The microscopic Hilbert space is
$\mathcal H=\bigotimes_{i\in\mathbb Z}V_i$, where the finite-dimensional vector spaces on different sites are isomorphic, $V_i\simeq V$.
Let $G_{\mathrm{int}}$ be a unitary internal symmetry group. We allow the one-site translation $T$ to act nontrivially on $G_{\mathrm{int}}$ by an automorphism $\rho\in \operatorname{Aut}(G_{\mathrm{int}})$. Thus the global symmetry operators obey
\begin{align}
T g T^{-1}=\rho(g),
\qquad
g\in G_{\mathrm{int}} .
\end{align}
If $\rho$ is trivial, then the internal symmetry commutes with translation. In general, the full symmetry group fits into the short exact sequence
\begin{align}\label{sesGtot}
1\longrightarrow
G_{\mathrm{int}}
\longrightarrow
G_{\mathrm{tot}}
\longrightarrow
\mathbb Z_{\mathrm{trans}}
\longrightarrow 1 .
\end{align}
After choosing a lift of the generator of $\mathbb Z_{\mathrm{trans}}$ to the physical translation $T$, the group can be written as the semidirect product
\begin{align}
G_{\mathrm{tot}}
=
G_{\mathrm{int}}\rtimes_{\rho}\mathbb Z_{\mathrm{trans}} .
\end{align}
The important point is that the group structure of $G_{\mathrm{tot}}$ alone does not yet specify the microscopic lattice realization. The local Hilbert space $V_i$ in each primitive unit cell may transform projectively under $G_{\mathrm{int}}$ \cite{ChenGuWen2011,SchuchPerezGarciaCirac2011}.

Let $U_i(g)$ be the local action of $g\in G_{\mathrm{int}}$ on $V_i$. A projective representation is characterized by a $2$-cocycle $\omega_2^{(i)}\in Z^2(G_{\mathrm{int}},\mathrm U(1))$:
\begin{align}
U_i(g)U_i(h)
=
\omega_2^{(i)}(g,h)U_i(gh),
\qquad
\dd \omega_2^{(i)}=1 .
\end{align}
Changing the phase convention $U_i(g)\mapsto \lambda_i(g)U_i(g)$ shifts $\omega_2^{(i)}$ by a coboundary. Therefore the invariant datum is the cohomology class $[\omega_2^{(i)}]\in H^2(G_{\mathrm{int}},\mathrm U(1))$ \cite{ChenGuWen2011,ChenGuWen2011Complete,PollmannBergTurnerOshikawa2012}. Translation covariance implies that the projective class at neighboring sites is related by the induced translation action on $H^2(G_{\mathrm{int}},\mathrm U(1))$. We denote this action by $T\cdot$, so in additive notation $[\omega_2^{(i+1)}]=T\cdot[\omega_2^{(i)}]$.
For a strictly uniform chain with trivial $\rho$, all primitive unit cells carry the same projective class $[\omega_2^{(i)}]=[\omega_2]\in H^2(G_{\mathrm{int}},\mathrm U(1))$.

For a finite periodic chain, the total action of $G_{\mathrm{int}}$ on the full Hilbert space must be an honest linear representation. This usually requires the total projective class over all sites to vanish. This is not in conflict with the Lieb-Schultz-Mattis mechanism: the obstruction is controlled by the projective class per primitive unit cell, not by the total projective class on a finite ring \cite{LiebSchultzMattis1961,OgataTachikawaTasaki2021,YaoOshikawa2021}. For example, a spin-$1/2$ chain has a projective representation of $SO(3)$ per site, while an even-length periodic chain has a linear global $SO(3)$ action.

\subsubsection{For $\Z_N$ translational symmetry}

It is often useful to focus on phases in which translation is spontaneously broken from $\mathbb Z_{\mathrm{trans}}$ to $N\mathbb Z_{\mathrm{trans}}$. In such a phase, a pure ground state is still invariant under $T^N$, while $T$ acts on the $N$ symmetry-related ground states. Equivalently, on the ground-state multiplet one keeps the finite quotient
\begin{align}
\mathbb Z_{\mathrm{trans}}/\langle T^N\rangle
\cong
\mathbb Z_N .
\end{align}
The quotient by $T^N$ can be interpreted as the effective symmetry acting on the low-energy symmetry-breaking multiplet. Equivalently, for the purposes of classifying such phases, one may restrict attention to the sector in which $T^N$ acts trivially. This is not meant to describe the entire microscopic Hilbert space, but rather the subsector relevant to phases with period dividing $N$. In the case $N=2$, which will be our main example below, we keep states invariant under the two-site translation $T^2$, while allowing the one-site translation $T$ to act nontrivially. In particular, dimerized states are included, since they preserve $T^2$ but break $T$. By contrast, states with, for example, a tripled unit cell are outside this reduced description and will not be considered here.

To replace $G_{\mathrm{tot}}$ by a finite symmetry group while keeping the same internal subgroup $G_{\mathrm{int}}$, we assume that the translation action $\rho$ has period dividing $N$, namely $\rho^N=\mathrm{id}_{G_{\mathrm{int}}}$.
Under this assumption, the $\mathbb Z_{\mathrm{trans}}$-action on $G_{\mathrm{int}}$ descends to a well-defined $\mathbb Z_N$-action, and the effective finite symmetry group is
\begin{align}
G
=
G_{\mathrm{int}}\rtimes_{\rho}\mathbb Z_N .
\end{align}
It fits into the short exact sequence
\begin{align}
1\longrightarrow
G_{\mathrm{int}}
\longrightarrow
G
\longrightarrow
\mathbb Z_N
\longrightarrow 1 .
\label{sesG}
\end{align}
This sequence should be viewed as the finite quotient of the infinite translation extension in Eq.~\eqref{sesGtot}, restricted to the sector where $T^N$ acts trivially.

In the following, we will often use the finite extension in Eq.~\eqref{sesG} as a convenient replacement for the full extension in Eq.~\eqref{sesGtot}, whenever we restrict to phases whose translation period divides $N$. If $G_{\mathrm{int}}$ is finite, then $G$ is finite, so the relevant group-cohomology and Lyndon-Hochschild-Serre spectral sequence computations \cite{Lyndon1948,HochschildSerre1953,Brown1982} reduce to finite-group computations. This finite reduction is also useful categorically: the category of $G$-graded vector spaces $\mathrm{Vec}_{G}^{\omega}$ with associator $\omega$ is a fusion category \cite{DijkgraafWitten1990,etingof2016tensor}, whereas $\mathrm{Vec}_{G_{\mathrm{tot}}}^{\omega}$ is not a fusion category because $G_{\mathrm{tot}}$ contains the infinite translation group and hence has infinitely many simple objects.

\subsection{$E_\infty^{1,2}$-type LSM anomaly in the LHS spectral sequence}

Although the group structure of $G_{\mathrm{tot}}$ or of its finite quotient $G$ is fixed by the symmetry extension, the way in which the symmetry is realized on a microscopic lattice may still carry a LSM anomaly. The relevant datum is the projective representation of the internal symmetry $G_{\mathrm{int}}$ carried by each primitive unit cell. If this projective representation is nontrivial, a local Hamiltonian preserving both $G_{\mathrm{int}}$ and primitive translation symmetry cannot have a unique, gapped, symmetry-preserving, short-range-entangled ground state \cite{OgataTachikawaTasaki2021}.

In one spatial dimension, a nontrivial LSM index implies that a symmetric local Hamiltonian must either be gapless or have degenerate ground states. The degeneracy may arise from spontaneous breaking of the internal symmetry, from translation symmetry breaking, or from a combination of the two. In higher spatial dimensions, there is an additional possibility: the system may have a symmetric gapped phase with intrinsic topological order and fractionalized excitations.
The original LSM theorem was formulated for half-odd-integer spin chains \cite{LiebSchultzMattis1961}. Oshikawa's flux-insertion argument and Hastings' higher-dimensional generalization showed that this obstruction is not special to exactly solvable one-dimensional models \cite{Oshikawa2000,Hastings2004}. For discrete internal symmetries, the same mechanism can be formulated in terms of projective representations per unit cell and symmetry-twisted boundary conditions \cite{OgataTachikawaTasaki2021,YaoOshikawa2021}. The modern anomaly viewpoint identifies this obstruction as a mixed anomaly between internal symmetry and lattice translation \cite{ChoHsiehRyu2017,FuruyaOshikawa2017,YaoHsiehOshikawa2019,ElseThorngren2020,ChengSeiberg2023,AksoyMudryFurusakiTiwari2024}.

Mathematically, the LSM anomaly is naturally organized by the LHS spectral sequence associated with the extension in Eq.~\eqref{sesGtot} or \eqref{sesG} \cite{Lyndon1948,HochschildSerre1953,Brown1982}. Since it is located at the $(1,2)$-position of the LHS spectral sequence, we refer to this LSM anomaly as being of $E_\infty^{1,2}$-type. In what follows, we discuss the LHS spectral sequence for infinite $\Z$ and finite $\Z_N$ translational symmetries separately, as they are slightly different from a technical standpoint.

\subsubsection{For $\Z$ translational symmetry}

For $G_{\mathrm{tot}}=G_{\mathrm{int}}\rtimes_{\rho}\mathbb Z_{\mathrm{trans}}$, the second page of the LHS spectral sequence associated with the extension in Eq.~\eqref{sesGtot} is
\begin{align}
E_2^{p,q}
=
H^p\!\left(
\mathbb Z_{\mathrm{trans}},
H^q(G_{\mathrm{int}},\mathrm U(1))
\right)
\Longrightarrow
H^{p+q}(G_{\mathrm{tot}},\mathrm U(1)) .
\end{align}
Here $\mathbb Z_{\mathrm{trans}}$ acts on $H^q(G_{\mathrm{int}},\mathrm U(1))$ through the automorphism $\rho$. The LSM data lives on the $(p,q)=(1,2)$ entry:
\begin{align}\label{E212}
\omega_{\mathrm{LSM}}
\in
E_2^{1,2}
&=
H^1\left(
\mathbb Z_{\mathrm{trans}},
H^2(G_{\mathrm{int}},\mathrm U(1))
\right)
=
H^2(G_{\mathrm{int}},\mathrm U(1))_{\mathbb Z_{\mathrm{trans}}}
\nonumber\\
&=
\frac{H^2(G_{\mathrm{int}},\mathrm U(1))}{(T-1)\cdot H^2(G_{\mathrm{int}},\mathrm U(1))}.
\end{align}
Here $M_{\mathbb Z_{\mathrm{trans}}}$ denotes the coinvariants of the coefficient module $M$ under translation. More explicitly, the relation $TgT^{-1}=\rho(g)$ for $g\in G_{\mathrm{int}}$ induces an action of translation on $H^2(G_{\mathrm{int}},\mathrm U(1))$. With the convention that cohomology classes are written additively, we denote this action by $T\cdot[\omega_2]$. Thus $E_2^{1,2}$ is obtained from $H^2(G_{\mathrm{int}},\mathrm U(1))$ by imposing the equivalence relation $T\cdot[\omega_2]\sim[\omega_2]$ for all $[\omega_2]$.

Physically, a representative $[\omega_2]\in H^2(G_{\mathrm{int}},\mathrm U(1))$ labels the projective representation of the internal symmetry carried by one translation unit cell, or equivalently the one-dimensional $G_{\mathrm{int}}$-SPT decoration attached to a fundamental translation defect in a one-dimension-higher weak SPT bulk \cite{ChenGuWen2011,ChenGuWen2011Complete,SchuchPerezGarciaCirac2011,PollmannBergTurnerOshikawa2012,ChenGuLiuWen2013,HuangSongHuangHermele2017}. The quotient by $(T-1)\cdot H^2(G_{\mathrm{int}},\mathrm U(1))$ expresses the freedom to change the microscopic choice of unit-cell cut, or to attach a locally trivial pattern of lower-dimensional SPT layers. If two projective representations are related by a lattice translation, then they represent the same physical LSM anomaly. Therefore, the difference $T\cdot[\omega_2]-[\omega_2]$ should be understood as a trivial anomaly and should be quotiented out.

If the action of $T$ on $H^2(G_{\mathrm{int}},\mathrm U(1))$ is trivial, even though the automorphism $\rho$ of $G_{\mathrm{int}}$ itself may still be nontrivial, the above expression reduces to
\begin{align}
\omega_{\mathrm{LSM}}
\in
E_2^{1,2}
=
H^2(G_{\mathrm{int}},\mathrm U(1)),
\quad
\text{for trivial $\mathbb Z_{\mathrm{trans}}$-action on $H^2(G_{\mathrm{int}},\mathrm U(1))$}.
\end{align}
In this common case, the LSM invariant is simply the projective representation class per unit cell, with no further identification. This is the situation most often assumed in standard discussions of LSM constraints.

Although $\omega_{\mathrm{LSM}}$ is first identified as an element of the $(1,2)$ entry on the $E_2$-page, the actual anomaly should be viewed as a class in $H^3(G_{\mathrm{tot}},\mathrm U(1))$. Therefore one has to check whether this $E_2^{1,2}$ class survives to the $E_\infty$ page of the LHS spectral sequence. In the present case the base group is $\mathbb Z_{\mathrm{trans}}$. For any $\mathbb Z_{\mathrm{trans}}$-module $A$, including modules with nontrivial translation action, one has $H^p(\mathbb Z_{\mathrm{trans}},A)=0$ for all $p>1$ \cite{Brown1982}. Therefore the spectral sequence has no nonzero columns with $p>1$. Hence no higher differential can either hit or leave $E_r^{1,2}$. It follows that $E_2^{1,2}=E_\infty^{1,2}$. Thus every LSM class in \eq{E212} can be equivalently viewed as
\begin{align}
\omega_\mathrm{LSM}
\in
E_\infty^{1,2} = E_2^{1,2}
\subseteq H^3(G_\mathrm{tot},\U),
\end{align}
which survives to the $E_\infty$-page of the LHS spectral sequence. This is why we call the LSM anomaly $E_\infty^{1,2}$-type.

\subsubsection{For $\Z_N$ translational symmetry}

For the finite translation quotient $G=G_{\mathrm{int}}\rtimes_{\rho}\mathbb Z_N$ with $\rho^N=\mathrm{id}$, the corresponding LHS spectral sequence associated with the extension in Eq.~\eqref{sesG} is \cite{Lyndon1948,HochschildSerre1953,Brown1982}
\begin{align}
E_2^{p,q}
=
H^p\!\left(
\mathbb Z_N,
H^q(G_{\mathrm{int}},\mathrm U(1))
\right)
\Longrightarrow
H^{p+q}(G,\mathrm U(1)) .
\end{align}
Using the standard cyclic-group cohomology formula \cite{Brown1982}, the analogue of the LSM component is
\begin{align}
\omega_{\mathrm{LSM}}
\in
E_2^{1,2}
=
H^1(\mathbb Z_N,M) 
=
\frac{\ker N_T}{(T-1)\cdot M},
\qquad
N_T:=1+T+T^2+\cdots+T^{N-1},
\end{align}
where $M=H^2(G_{\mathrm{int}},\mathrm U(1))$. Thus a representative $[\omega_2]\in M$ defines a $\mathbb Z_N$-translation-invariant decoration only if $N_T[\omega_2] = [\omega_2]+T\cdot[\omega_2]+\cdots+T^{N-1}\cdot[\omega_2] = 0$. The physical statement is that the total projective class accumulated around one finite translation orbit must be trivial. When the action of $T$ on $M=H^2(G_{\mathrm{int}},\mathrm U(1))$ is trivial, this reduces to
\begin{align}
\omega_{\mathrm{LSM}}
\in
E_2^{1,2}
=
\mathrm{ker}\ N_T
:=
\{\, [\omega_2]\in H^2(G_{\mathrm{int}},\mathrm U(1)) \mid N[\omega_2]=0\,\}.
\end{align}

There is a small but important distinction between finite and infinite translation symmetries. For a finite translation group $\mathbb Z_N$, a class on the $E_2$ page of the LHS spectral sequence does not automatically define a class in the final cohomology group. It must survive all relevant differentials and hence determine a class on the $E_\infty$ page. Therefore, one should in principle check that the candidate anomaly is not killed by any differential.
We use the convention $d_r:E_r^{p,q}\to E_r^{p+r,q-r+1}$. For the bidegree $(1,2)$, the first possible outgoing differential is
\begin{align}
d_2:E_2^{1,2}\longrightarrow E_2^{3,1}.
\end{align}
The next outgoing differential is $d_3:E_3^{1,2}\to E_3^{4,0}$. Since $E_2^{4,0}=H^4(\mathbb Z_N,\mathrm U(1))=0$, its target also vanishes on the $E_3$ page, so $d_3$ is trivial. For $r\geq 4$, the outgoing differential from $E_r^{1,2}$ has target $E_r^{1+r,3-r}$, whose second degree is negative. Hence all such higher outgoing differentials vanish for degree reasons. Similarly, any incoming differential to $E_r^{1,2}$ with $r\geq 2$ would have source $E_r^{1-r,r+1}$, whose first degree is negative, and therefore no nontrivial incoming differential exists.
Consequently, the only potentially nontrivial differential involving the $(1,2)$ term is the above $d_2$, and we have
\begin{align}
\omega_{\mathrm{LSM}} \in
E_\infty^{1,2}
=
E_3^{1,2}
=
\ker\!\left(d_2:E_2^{1,2}\to E_2^{3,1}\right)
\subseteq E_2^{1,2}.
\end{align}
In the $D_8$ applications below, the relevant LSM anomaly class lies in this kernel and therefore survives to the $E_\infty$ page. This is the precise sense in which the LSM anomaly is of $E_\infty^{1,2}$ type.

\subsection{Decorated-domain-wall interpretation from the bulk weak crystalline SPT}

From the modern viewpoint, the LSM obstruction is a crystalline 't Hooft anomaly \cite{ChoHsiehRyu2017,ThorngrenElse2018,ElseThorngren2019}. The anomalous (1+1)-dimensional lattice theory can be realized as the boundary of a (2+1)-dimensional weak crystalline SPT phase protected by $G_{\mathrm{int}}$ and translation along the $x$-direction \cite{HuangSongHuangHermele2017,ThorngrenElse2018,ElseThorngren2019}. In the decorated-domain-wall construction of SPT phases \cite{ChenLuVishwanath2014,HuangSongHuangHermele2017,WangNingCheng2021}, a spatial translation defect line in the two-dimensional bulk is decorated by a one-dimensional $G_{\mathrm{int}}$-SPT phase labeled by $[\omega_2]\in H^2(G_{\mathrm{int}},\mathrm U(1))$. Equivalently, the decoration data is encoded by
\begin{align}
n_1
\in
H^1\!\left(
\mathbb Z_{\mathrm{trans}},
H^2(G_{\mathrm{int}},\mathrm U(1))
\right)
\quad \text{or} \quad
n_1
\in
H^1\!\left(
\mathbb Z_N,
H^2(G_{\mathrm{int}},\mathrm U(1))
\right) .
\end{align}
In principle, the decoration data $n_1$ must satisfy obstruction and trivialization conditions. In the LHS spectral sequence language, these conditions are precisely the requirement that the $E_2^{1,2}$ class survive to the $E_\infty$-page, modulo equivalence relations that become trivial in the full group cohomology $H^3(G_{\mathrm{tot}},\mathrm U(1))$ (or $H^3(G,\mathrm U(1))$ for the finite translation quotient) \cite{WangNingCheng2021}.

When a decorated translation defect ends on the boundary, its endpoint carries the projective representation class $[\omega_2]$ \cite{ChenGuWen2011,PollmannBergTurnerOshikawa2012}. Therefore a boundary Hilbert space with this projective class per unit cell cannot admit a completely trivial symmetric short-range-entangled ground state.

Schematically, when the translation action on $G_{\mathrm{int}}$ is trivial, the corresponding bulk response may be written as
\begin{align}
S_{\mathrm{bulk}}
=
\int_{M_3}
A_{\mathrm{trans}}
\smile
\omega_2(A_{\mathrm{int}}) .
\end{align}
Here $A_{\mathrm{trans}}$ is a background one-cocycle for translation and $\omega_2(A_{\mathrm{int}})$ is the pullback of a representative group cocycle $\omega_2$ along the background $G_{\mathrm{int}}$ gauge field \cite{DijkgraafWitten1990,ChenGuLiuWen2013,kapustin2014anomalies}. The factor $A_{\mathrm{trans}}$ detects the translation defect, while $\omega_2(A_{\mathrm{int}})$ records the one-dimensional internal-symmetry SPT that decorates it. For nontrivial $\rho$, the same expression should be understood with the corresponding twisted local coefficient system: across a translation branch cut, the $G_{\mathrm{int}}$ background is glued by $\rho$, and the cocycle $\omega_2$ is transported by the induced $T$-action.

This picture is analogous to the decorated-domain-wall construction of $(2+1)$-dimensional fermionic SPT phases \cite{TarantinoFidkowski2016,WareSonChengMishmashAliceaBauer2016,WangGu2018,WangGu2020}. In that setting, one may decorate $G_b$-symmetry domain walls with Kitaev chains \cite{Kitaev2001}. Up to equivalence, this Majorana-chain decoration is specified by a class
\begin{align}
    n_1\in H^1(G_b,\mathbb Z_2).
\end{align}
Here the coefficient $\mathbb Z_2$ is the group of invertible $(1+1)$D fermionic phases generated by the Kitaev chain. The analogy with the present LSM construction is that the endpoint of a $(1+1)$D bosonic $G_{\mathrm{int}}$-SPT carries a projective $G_{\mathrm{int}}$ representation, while the endpoint of a Kitaev chain carries a Majorana zero mode, or equivalently a fermion-parity anomaly. The crucial difference is geometric. For an ordinary internal symmetry, domain walls can be summed over and become gauge-flux defects after gauging the symmetry. By contrast, a translation defect is a crystalline defect tied to the spatial lattice structure; in the present geometry, it is a defect line running parallel to the $y$-direction. This rigidity is precisely why the LSM obstruction is naturally interpreted as a crystalline, or weak-SPT, anomaly rather than as an ordinary onsite anomaly \cite{ThorngrenElse2018,ElseThorngren2019,ChoHsiehRyu2017,ElseThorngren2020,ChengSeiberg2023,AksoyMudryFurusakiTiwari2024}.

\section{$E_\infty^{1,2}$-type anomaly and the non-invertible dual symmetry}
\label{sec:E12}

\begin{figure}[h!]
\centering
\includegraphics[width=0.5\textwidth]{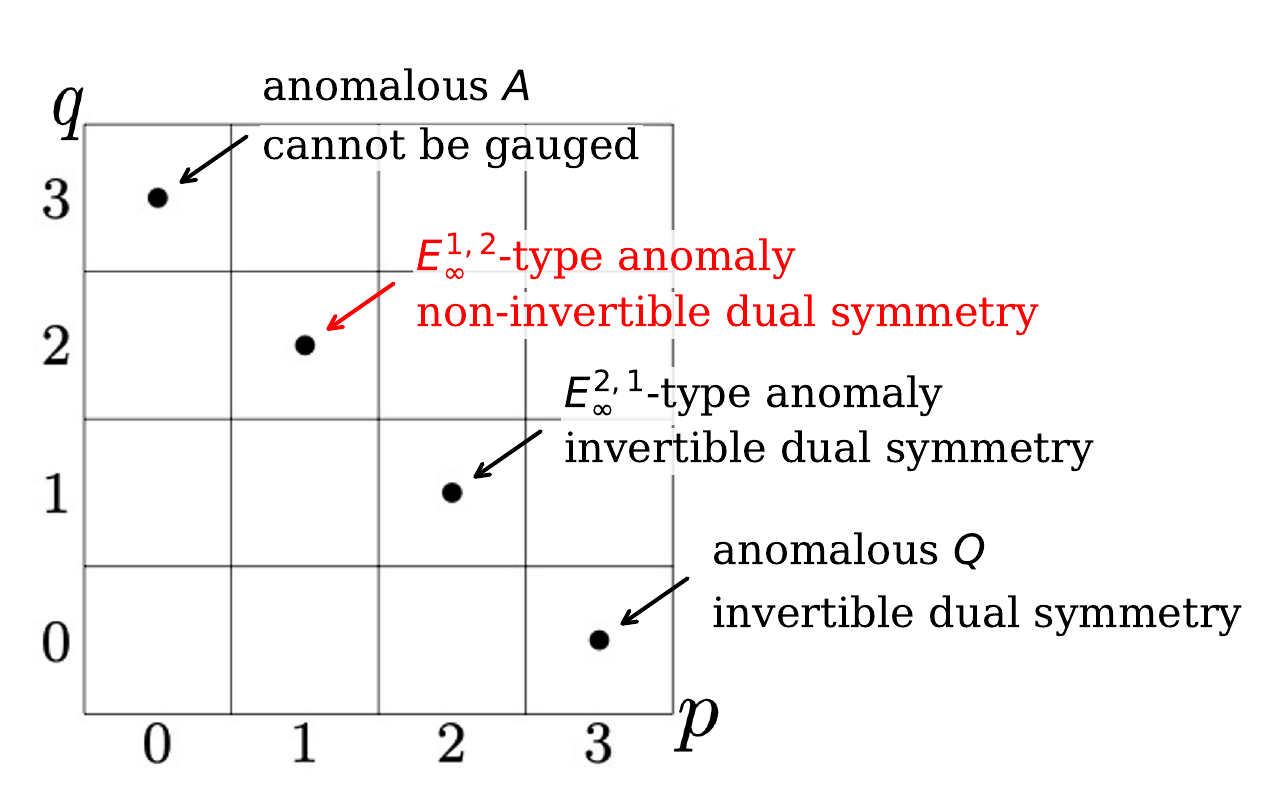}
\caption{Effect of LHS anomaly components under gauging an Abelian normal subgroup $A$ of $G$. The key result is that a nontrivial $E_\infty^{1,2}$ component produces projective $A$-sectors and hence a non-invertible dual symmetry, whereas $E_\infty^{3,0}$ and $E_\infty^{2,1}$ data keep the dual symmetry invertible.
}
\label{fig:LHS-anomaly-classes}
\end{figure}

The previous section identified the LSM obstruction as an $E_\infty^{1,2}$ component of the LHS spectral sequence. We now explain the categorical consequence of this statement. The full translation group $G_{\mathrm{tot}}$ is usually infinite, so $\Vec_{G_{\mathrm{tot}}}^{\omega_{\mathrm{LSM}}}$ has infinitely many simple objects and is not a fusion category in the usual finite sense. In the applications below, and more generally whenever we restrict to a finite translation quotient, we work instead with a finite group $G$ and an anomalous pointed fusion category $\C=\Vec_G^\omega$. This is the setting in which generalized gauging is described by condensable algebras and Morita dual fusion categories. The categorical tools needed in this section are summarized in Appendix~\ref{App:A}, where we also spell out the dictionary between the mathematical objects--fusion categories, module categories, condensable algebras, and Morita duals--and their physical meanings as symmetries, gapped phases, gauging interfaces, and dual symmetries.

Throughout this section we use a short exact sequence
\begin{align}\label{ses-general-E12}
1\longrightarrow A\longrightarrow G\longrightarrow Q\longrightarrow 1 ,
\end{align}
whose associated LHS spectral sequence is
\begin{align}\label{LHS-general-E12}
E_2^{p,q}=H^p\!\left(Q,H^q(A,\U)\right)\Longrightarrow H^{p+q}(G,\U).
\end{align}
Here $A$ is a finite normal subgroup. We assume $A$ to be Abelian for simplicity. The quotient $Q$ may be non-Abelian, and $G$ need not be a direct product. We assume that the restriction of the anomaly to $A$ is trivial, so that $A$ is gaugeable. For notational simplicity we choose a trivialization of $\omega|_A$ and denote the corresponding condensable algebra in $\C$ by the same letter $A$; explicitly, it is the algebra supported on the objects $a\in A$. Gauging $A$ changes the symmetry category from $\C$ to the Morita dual category ${}_A\C_A$ \cite{ostrik2002module,ostrik2003module,kong2014anyon,bhardwaj2018finite,tachikawa2020gauging}. If ${}_A\C_A$ is pointed, the dual symmetry is an ordinary invertible symmetry. If it is not pointed, the dual symmetry is non-invertible.

Fig.~\ref{fig:LHS-anomaly-classes} summarizes the main point of this section: for Abelian normal subgroup gauging, the $E_\infty^{1,2}$ component is precisely the source of non-invertibility. An $E_\infty^{2,1}$ component can change the group law of the dual invertible symmetry, and an $E_\infty^{3,0}$ component can leave an anomaly for the quotient symmetry, but neither of them forces higher-dimensional simple objects. By contrast, a nontrivial $E_\infty^{1,2}$ component makes some dual defect sector carry a nontrivial projective representation of $A$. Such a sector has quantum dimension greater than one, and hence the dual symmetry is non-invertible.

In this section, we first explain this statement from the decorated-domain-wall construction of the bulk SPT. We then review the group-like cases covered by subgroup gauging: the $E_\infty^{0,3}$ entry is the obstruction to gauging $A$, while the $E_\infty^{3,0}$ and $E_\infty^{2,1}$ entries leave a pointed dual category with a dual group and anomaly. The remaining subsections treat the genuinely new $E_\infty^{1,2}$ case. We describe the simple objects of the dual category, derive their fusion rules, and finally state an explicit criterion for when this non-invertible dual symmetry is anomaly-free, or equivalently admits a fully symmetric gapped phase.

\subsection{Decorated-domain-wall picture}

The LHS spectral sequence has a useful physical realization in terms of the decorated-domain-wall construction of the $(2+1)$D bulk SPT whose boundary carries the anomaly \cite{ChenLuVishwanath2014,WangNingCheng2021}. The entry $E_2^{p,q}=H^p(Q,H^q(A,\U))$ should be read as follows: the $Q$ gauge-field configuration supplies the domain-wall network, while the coefficient $H^q(A,\U)$ supplies a lower-dimensional $A$-SPT used to decorate this network. Survival to $E_\infty$ means that these local decorations can be assembled into a consistent, obstruction-free bulk SPT and remain nontrivial, rather than being trivialized by the boundary anomalous SPT \cite{WangQiGu2019ASPT}.

\begin{figure}[h!]
    \centering
    \begin{subfigure}[t]{0.42\textwidth}
        \centering
        \includegraphics[width=\linewidth]{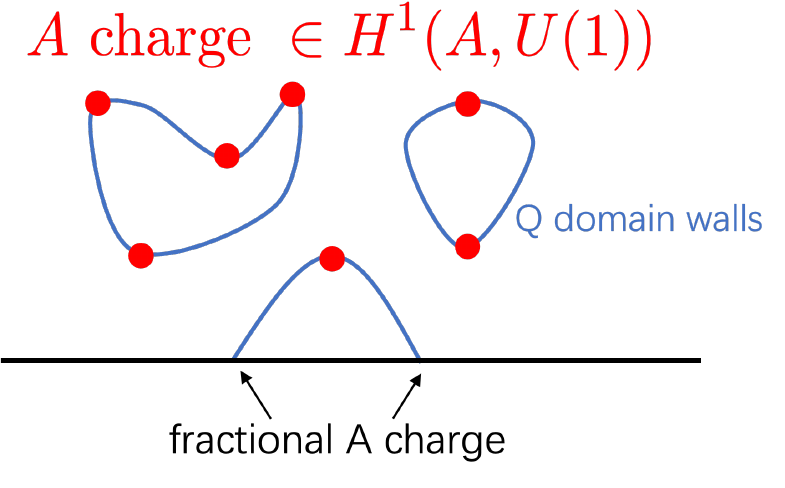}
        \caption{$E_\infty^{2,1}$}
    \end{subfigure}
    \hspace{0.04\textwidth}
    \begin{subfigure}[t]{0.42\textwidth}
        \centering
        \includegraphics[width=\linewidth]{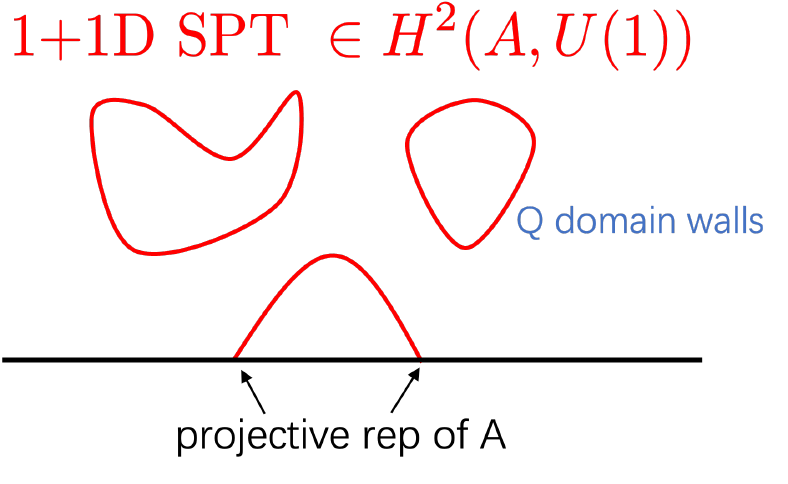}
        \caption{$E_\infty^{1,2}$}
    \end{subfigure}
    \caption{
Decorated-domain-wall representatives for a 2+1D bulk SPT with
$1\to A\to G\to Q\to 1$. Left: an $E_\infty^{2,1}$ decoration puts
$A$ charges on $Q$-wall junctions, so gauging $A$ keeps the dual symmetry
invertible. Right: an $E_\infty^{1,2}$ decoration puts a 1+1D $A$-SPT on
each $Q$ wall; its boundary endpoint carries a projective $A$ representation,
which becomes a non-invertible dual defect after gauging $A$.
    }
    \label{fig:ddw-E21-E12}
\end{figure}

Fig.~\ref{fig:ddw-E21-E12} shows the decorated-domain-wall construction of the $(2+1)$D bulk SPT wavefunction on a two-dimensional space with a one-dimensional boundary: the blue loops represent $Q$ domain walls, while the red dots or lines denote lower-dimensional $A$-SPT decorations.

The left panel illustrates the simplest case $Q=\mathbb Z_2$ for an $E_\infty^{2,1}$ representative $n_2\in H^2(Q,\widehat A)$. In general, this class decorates codimension-two junctions of the $Q$-domain-wall network by a $0+1$D $A$-SPT, equivalently an ordinary $A$ charge or character. For $Q=\mathbb Z_2$, the same decoration is represented in the wavefunction picture by placing red dots at the maxima and minima of a blue $Q$-domain-wall loop. When the blue domain wall meets the boundary, a bulk minimum can be opened into two boundary $Q$-domain-wall defects. Thus the two boundary defects together carry one unit of $A$ charge; in the usual $\mathbb Z_2$ picture, each boundary $Q$ domain wall carries half of this $A$ charge. After gauging $A$, these are still ordinary character sectors of the dual gauge symmetry. The dual defects are therefore invertible: the $E_\infty^{2,1}$ data can modify the dual group extension, but it does not create higher-dimensional local sectors.

The right panel shows the parallel construction for an $E_\infty^{1,2}$ representative $n_1\in H^1(Q,H^2(A,\U))$. Instead of putting a $0+1$D $A$ charge at a junction, this class decorates the blue $Q$ domain wall itself by a one-dimensional $A$-SPT phase, drawn as a red line. When the decorated $Q$ domain wall ends on the boundary, the red $A$-SPT line also ends there, so the boundary $Q$-domain-wall defect carries the edge mode of a one-dimensional $A$-SPT. This edge mode transforms projectively under $A$, rather than as an ordinary $A$ charge. After gauging $A$, the corresponding dual defect is labeled by an irreducible projective representation of $A$; if the projective class is nontrivial, this representation has dimension larger than one, and the dual defect is non-invertible.

This logic is closely parallel to decorated-domain-wall constructions of fermionic SPT phases \cite{TarantinoFidkowski2016,WareSonChengMishmashAliceaBauer2016,WangGu2018,WangGu2020}. A Kitaev-chain \cite{Kitaev2001} decoration of a bosonic domain wall leaves a Majorana zero mode at the boundary, whereas a charge decoration leaves an ordinary fractional charge. In the present bosonic setting, the role of the Majorana edge mode is played by the projective $A$ representation at the endpoint of an $A$-SPT decoration. The important point is not the fermionic nature of the analogy, but the distinction between decorating a defect by a charge and decorating it by a one-dimensional invertible phase with a nontrivial edge.

\subsection{Invertible dual symmetry}

For the LHS spectral sequence in Eq.~\eqref{LHS-general-E12}, the first issue is whether the subgroup $A$ can be gauged at all. The $E_\infty^{0,3}$ component is the restriction of the anomaly to $A$. If it is nonzero, there is no algebra supported on $A$ whose multiplication cancels $\omega|_A$, so ordinary subgroup gauging is obstructed. In this subsection we assume that this obstruction vanishes and fix a trivialization $\gamma\in C^2(A,\U)$ with $\dd\gamma=\omega|_A$. This choice is the local counterterm entering the gauging algebra supported on $A$.

We now ask when the dual symmetry after gauging is still an ordinary group symmetry. The answer is the $1+1$D version of Tachikawa's subgroup-gauging result \cite{tachikawa2020gauging}: once $A$ is gaugeable, the dual symmetry remains invertible as long as no $Q$ defect carries a projective representation of $A$. In the LHS language this means that the $E_\infty^{1,2}$ component of $\omega$ vanishes. Equivalently, the Morita dual ${}_A(\Vec_G^\omega)_A$ is pointed; the categorical statement is reviewed in Appendices~\ref{App:A1pointed} and~\ref{App:method} \cite{ostrik2002module,Naidu2007CategoricalMorita,gelaki2009some,natale2017equivalence,uribe2017classification,Mu_oz_2018}. Thus the gauged theory has an ordinary finite dual group $\widehat G$:
\begin{align}
{}_A(\Vec_G^\omega)_A\simeq \Vec_{\widehat G}^{\widehat\omega},
\end{align}
with
\begin{align}
1\longrightarrow \widehat A\longrightarrow \widehat G\longrightarrow Q\longrightarrow 1 ,
\qquad
\widehat A=H^1(A,\U).
\end{align}
Physically, $\widehat A$ is the usual quantum symmetry created by gauging the Abelian group $A$.

The simple dual defects can then be described without invoking the full categorical machinery. A defect is labeled by two pieces of data: a quotient domain wall $q\in Q$ and a character $\chi\in\widehat A$. The character records the dual gauge charge attached to the defect. Because the $E_\infty^{1,2}$ component is absent, the $q$ defect does not carry a projective $A$ edge mode; all such defects are one-dimensional and therefore invertible.

Fusion of these defects gives the dual group law. After choosing representatives, one may write
\begin{align}
(\chi,q)\cdot(\chi',q')
=
\left(\chi+q\cdot\chi'+n_2(q,q'),qq'\right),
\qquad
\chi,\chi'\in\widehat A .
\end{align}
Here the $Q$ action on $\widehat A$ is induced from conjugation in $G$, and $n_2\in Z^2(Q,\widehat A)$ represents the $E_\infty^{2,1}$ component. This component remains completely group-like after gauging: it changes the extension class of the dual group $\widehat G$, but it does not make any defect non-invertible. Equivalently, gauging converts the original $E_\infty^{2,1}$ anomaly data represented by $n_2$ into the symmetry-extension data of the dual symmetry.

The anomaly of the dual ordinary group symmetry is encoded in the associator $\widehat\omega\in Z^3(\widehat G,\U)$. A pure $E_\infty^{3,0}$ component is simply the quotient anomaly pulled back from $Q$. There is also the standard Tachikawa mixed anomaly when the original group extension
\eqref{ses-general-E12} is non-split. The original short exact sequence is specified by a group-extension class $[e]\in H^2(Q,A)$; choosing a cocycle representative $e\in Z^2(Q,A)$, one representative of this mixed anomaly is
\begin{align}
\widehat\omega_{\mathrm{mix}}\bigl((\chi_1,q_1),(\chi_2,q_2),(\chi_3,q_3)\bigr)
=
\left\langle \chi_1,e(q_2,q_3)\right\rangle ,
\end{align}
up to the usual changes by coboundaries and the choice of cocycle representative. These terms affect the $F$-symbol of the ordinary group symmetry, not the invertibility of its defects. For example, in a split extension with only a quotient anomaly $\nu\in H^3(Q,\U)$, one obtains the pointed category $\Vec_{\widehat A\rtimes Q}^{p_Q^*\nu}$, where $p_Q:\widehat A\rtimes Q\to Q$ is the projection; with $n_2\neq0$ but no anomaly, one obtains a nontrivial ordinary dual group extension $\widehat G$.

The lesson is that $E_\infty^{0,3}$ controls whether gauging is allowed, $E_\infty^{2,1}$ controls the dual group law, and $E_\infty^{3,0}$ controls an ordinary quotient anomaly. The original extension class of $G$ contributes Tachikawa's mixed anomaly. None of these ingredients forces a single $Q$ defect to carry a projective $A$ representation. That is the new ingredient in the $E_\infty^{1,2}$ case, to which we now turn.

\subsection{Non-invertible dual symmetry}

Assume now that the anomaly has a nontrivial $E_\infty^{1,2}$ component. On the $E_2$ page, this component is represented by a class
\[
\eta\in H^1\!\left(Q,H^2(A,\U)\right).
\]
This means that each quotient defect $q\in Q$ is assigned a projective $A$ representation class, which we denote by $[\eta_q]\in H^2(A,\U)$. Physically, the $q$ defect is decorated by the edge state of a one-dimensional $A$-SPT phase, and $[\eta_q]$ is the projective class carried by that edge state. In the LSM application, the fundamental quotient defect is translation, and $[\eta_q]$ is exactly the projective representation carried by one primitive unit cell.

Let us now gauge $A$. The quotient symmetry $Q$ is not gauged, so its domain-wall sectors remain as sectors of the dual symmetry. Meanwhile, gauging the Abelian group $A$ normally produces the dual quantum symmetry $\widehat A$: in the untwisted sector, its simple charges are ordinary characters of $A$. The $E_\infty^{1,2}$ anomaly changes this statement in a very concrete way. In a sector with quotient defect $q$, the endpoint of the decorated domain wall carries the projective class $[\eta_q]$. Therefore the ``dual $A$ charge'' attached to that defect is not an ordinary character of $A$, but an irreducible projective representation of $A$ in the class $[\eta_q]$.

It is useful to see how this follows from the general group-theoretical description of the Morita dual. Simple objects of an $A$-$A$ bimodule category in $\Vec_G^\omega$ are generally labeled by an $A$-$A$ double coset in $G$, together with a projective representation of the corresponding stabilizer subgroup; see Appendix~\ref{App:method}. In the case considered here, $A$ is normal, so the double cosets reduce to ordinary quotient elements, $A\backslash G/A = G/A=Q$. Moreover, because $A$ is Abelian and normal, the stabilizer associated with a representative $g\in G$ of each quotient sector is $A\cap gAg^{-1}=A$. Thus the general double-coset-plus-stabilizer classification collapses to the simple physical labeling described above: a quotient wall $q$, together with a projective $A$ representation living on that wall.

Thus an elementary dual defect is labeled by
\[
(q,\rho),
\]
where $q$ tells us which quotient wall sector we are in, and $\rho$ is the projective $A$ multiplet living on that wall. More precisely,
\begin{align}\label{dual-simple-objects}
\mathrm{Irr}\!\left({}_A(\Vec_G^\omega)_A\right)
=
\left\{
X_{q,\rho}\ \middle|\ 
q\in Q,\ 
\rho\in \mathrm{Irr}_{\eta_q}(A)
\right\},
\qquad
d(X_{q,\rho})=\dim(\rho).
\end{align}
Here $\mathrm{Irr}_{\eta_q}(A)$ denotes irreducible projective representations of $A$ in the projective class $[\eta_q]$. Equation~\eqref{dual-simple-objects} is the normal-Abelian specialization of the standard group-theoretical description of the Morita dual; the general statement is reviewed in Appendices~\ref{App:A1gauging} and~\ref{App:method}.

This description directly matches the decorated-domain-wall picture. Before gauging, the $q$ wall carries the endpoint of an $A$-SPT decoration. After gauging, this endpoint cannot disappear; it becomes an internal Hilbert space attached to the dual defect line. If the decoration is trivial, this internal space is just an ordinary one-dimensional $A$ charge, equivalently a character in $\widehat A$. The defect is then group-like, as in the invertible dual symmetry discussed above. If the decoration is nontrivial, the endpoint transforms projectively under $A$. For finite Abelian $A$, a nontrivial projective multiplet is higher-dimensional: the $A$ symmetry operators on the edge cannot all be represented as ordinary phases. This higher-dimensional edge multiplet is precisely what the quantum dimension in Eq.~\eqref{dual-simple-objects} counts.

We can therefore phrase the mechanism as a pointedness criterion for the dual category. In this normal-Abelian subgroup-gauging setting, the only way for the dual defects to remain all group-like is for every quotient wall to carry an ordinary, rather than projective, $A$ edge mode. Equivalently, the $E_\infty^{1,2}$ component must be absent.

\begin{proposition}\label{prop:E12-nonpointed}
Let $A$ be a finite Abelian normal subgroup of $G$, and suppose that $A$ is gaugeable in $\Vec_G^\omega$. Then the Morita dual ${}_A(\Vec_G^\omega)_A$ is pointed if and only if the $E_\infty^{1,2}$ component of $\omega$ vanishes. In particular, a nonzero $E_\infty^{1,2}$ component forces the dual symmetry after gauging $A$ to be non-invertible.
\end{proposition}

The statement follows directly from Eq.~\eqref{dual-simple-objects}. If the $E_\infty^{1,2}$ component is nonzero, some quotient defect carries a nontrivial projective $A$ edge mode. The corresponding dual defect has quantum dimension equal to the ordinary dimension of that projective representation. This dimension can be larger than one, but it is still an integer; this integrality is a general feature of group-theoretical fusion categories. A simple object with $d>1$ is a non-Abelian defect and cannot be invertible, so the dual category is not pointed. Conversely, if no quotient defect carries such a projective edge mode, all projective representations in Eq.~\eqref{dual-simple-objects} reduce to ordinary characters of $A$, and every simple object has dimension one. Thus the non-invertible object is not produced by gauging alone; it is produced by gauging a symmetry whose defects already carry protected projective edge degrees of freedom. This is the categorical version of the decorated-domain-wall statement that proliferating $A$ gauge configurations leaves behind projective edge modes on the remaining $Q$ defects.

\subsection{Fusion rules}

We now describe the fusion rules in the same normal-Abelian setting. The mathematical result is the following. Up to associator phases, which do not affect fusion multiplicities, two simple defects in Eq.~\eqref{dual-simple-objects} fuse as
\begin{align}\label{fusion-general-E12}
X_{q,\rho}\otimes X_{q',\rho'}
\simeq
\bigoplus_{\sigma\in \mathrm{Irr}_{\eta_{qq'}}(A)}
N^{\sigma}_{\rho,\,q\cdot\rho'\otimes \lambda_{q,q'}}\,
X_{qq',\sigma}.
\end{align}
Here the right-hand side should be read as follows. The quotient-wall labels fuse first, giving the sector $qq'$. The projective edge mode on the second wall must be viewed in the $A$ frame transported across the first wall; this gives $q\cdot\rho'$. The factor $\lambda_{q,q'}$ is only an ordinary one-dimensional $A$ charge, coming from possible $E_\infty^{2,1}$ data and from the choice of section. If this component is absent, one may set $\lambda_{q,q'}=1$. The coefficient $N^{\sigma}_{\rho,\,q\cdot\rho'\otimes \lambda_{q,q'}}$ is then the number of times the projective multiplet $\sigma$ appears after the edge Hilbert spaces are combined.

In physical terms, Eq.~\eqref{fusion-general-E12} is the rule for stacking two decorated quotient domain walls. The topological wall labels multiply, but the endpoint Hilbert spaces also have to be fused. The fused endpoint Hilbert space is
\[
\rho\otimes(q\cdot\rho')\otimes \lambda_{q,q'} .
\]
The $E_\infty^{1,2}$ cocycle condition says that this Hilbert space has exactly the projective class appropriate to the fused wall, namely $[\eta_{qq'}]$. The last step is therefore a familiar physical operation: resolve the local edge Hilbert space into irreducible $A$ multiplets. Each irreducible channel $\sigma$ becomes a possible topological line $X_{qq',\sigma}$ in the fusion outcome.

This explains why the absence or presence of the $E_\infty^{1,2}$ component changes the nature of the dual symmetry. If all $[\eta_q]$ vanish, every endpoint carries an ordinary one-dimensional charge. Stacking two defects just multiplies two wall labels and two characters, producing a single new defect. This is an ordinary group law. If some $[\eta_q]$ is nontrivial, the endpoint can be a higher-dimensional projective multiplet. When such multiplets are fused, the local Hilbert space can split into several irreducible channels. The direct sum in Eq.~\eqref{fusion-general-E12} is precisely this physical splitting of the endpoint Hilbert space.

The $D_8$ example below gives the smallest concrete realization. Gauge the normal Abelian subgroup $A\cong\Z_2\times\Z_2$ with quotient $Q\cong\Z_2$. There are only two quotient-wall sectors. In the identity sector there is no nontrivial quotient wall, so the dual defects are just the ordinary $A$ gauge charges. These are the four characters of $A$, denoted by $X_1,X_2,X_3,X_4$, with $X_1$ the tensor unit and with fusion given by the group law of $\widehat A\cong\Z_2\times\Z_2$. In the nontrivial quotient sector, the $E_\infty^{1,2}$ decoration attaches the nontrivial $A$-SPT edge mode. For $A=\Z_2\times\Z_2$, that edge mode is a single two-dimensional irreducible projective representation, giving one non-invertible defect $X_0$.

The fusion rules can now be read off without further category theory. Fusing a character line $X_i$ with $X_0$ only attaches an ordinary $A$ charge to the same projective edge multiplet. Since there is only one irreducible projective multiplet in this projective class, the result is still $X_0$. Fusing two $X_0$ lines is more interesting. The two nontrivial $Q$ walls combine to the identity sector, while the two projective edge multiplets combine into an ordinary four-dimensional $A$ representation. That ordinary representation decomposes into the four one-dimensional characters of $A$. Therefore
\begin{align}
X_i\otimes X_j &= X_{i\cdot j},
\qquad i,j=1,2,3,4,\\
X_i\otimes X_0&=X_0\otimes X_i=X_0,
\qquad i=1,2,3,4,\\
X_0\otimes X_0
&=
X_1\oplus X_2\oplus X_3\oplus X_4 .
\end{align}
Here $i\cdot j$ denotes the product in the group of the four invertible objects. These are the Tambara-Yamagami fusion rules for $\Z_2\times\Z_2$. With the associator and higher Frobenius-Schur indicator data computed for the present group-theoretical category, this is the $\Rep(H_8)$ fusion category, where $H_8$ is the eight-dimensional Kac-Paljutkin Hopf algebra \cite{tambara1998tensor,tambara2000representations,etingof2021tensor}; see also Appendix~\ref{App:D8ACA}.

\subsection{Conditions for anomaly-free dual symmetry}

We now ask a question that is distinct from invertibility. After gauging $A$, the dual symmetry category may be non-pointed, hence non-invertible. It may nevertheless be anomaly-free in the sense relevant for a strictly one-dimensional system. Physically, anomaly-free means that the symmetry admits a fully symmetric short-range-entangled gapped phase. In such a phase there is a unique ground-state sector, and all symmetry defects can consistently end on it without producing protected degeneracy.

The categorical language is a compact way to express this physical statement. A gapped phase with fusion-category symmetry $\mathcal D$ is described by a $\mathcal D$-module category; for a physics reader, this module category records the possible ground-state sectors and the ways in which symmetry defects act on or end on them. A fully symmetric short-range-entangled phase has only one such sector, so the corresponding module category has a single simple object \cite{thorngren1912fusion}. Categorically, this one-object module category is $\Vec$. Giving an action of $\mathcal D$ on $\Vec$ is the same as giving a tensor functor $F:\mathcal D\to \Vec$, namely a fiber functor \cite{etingof2016tensor}. Thus the dual category $\mathcal D={}_A(\Vec_G^\omega)_A$ is anomaly-free precisely when it admits a fiber functor. This is different from asking whether $\mathcal D$ is invertible: invertibility asks whether every defect line is group-like, while anomaly-freeness asks whether the whole symmetry can act on a unique symmetric gapped ground state.

We now turn this condition into subgroup data. Let the gauging algebra be represented by the pair $(A,\psi)$, with $\omega|_A=\dd\psi$. Since $\mathcal D={}_A(\Vec_G^\omega)_A$ is Morita dual to $\Vec_G^\omega$, a candidate gapped phase after gauging can be obtained by starting from a gapped phase of the original pointed theory and then passing it through the gauging interface. Such a phase of $\Vec_G^\omega$ is represented by a pair $(A',\psi')$, where $A'\leq G$ and $\omega|_{A'}=\dd\psi'$. Physically, $A'$ is the unbroken subgroup of the candidate phase before gauging, and $\psi'$ is the local SPT counterterm that cancels the restricted anomaly. The corresponding $\mathcal D$-module category after gauging is the bimodule category ${}_A(\Vec_G^\omega)_{A'}$ \cite{muger2003subfactors,ostrik2002module}. Therefore a fiber functor is obtained exactly when this bimodule category has a single simple object.

The standard group-theoretical description of bimodules over pointed categories now gives a useful physical test \cite{ostrik2002module,Naidu2007CategoricalMorita,natale2017equivalence}. A simple object of ${}_A(\Vec_G^\omega)_{A'}$ has two labels. The first label is a double coset in $A\backslash G/A'$. It tells us which defect sector remains after we are allowed to slide gauged $A$ lines on one side and use the boundary condition $A'$ on the other side. The second label is an irreducible projective representation of the stabilizer $K_g=A\cap gA'g^{-1}$ for a representative $g$ of the double coset. This is the local quantum number that can live at the endpoint of that defect. Its factor set is the relative two-cocycle determined by $\psi$, $\psi'$, and $\omega$. Hence ${}_A(\Vec_G^\omega)_{A'}$ has a single simple object if and only if both labels are reduced to a unique choice \cite{ostrik2002module}:
\begin{itemize}
\item There is only one double coset, so $A\backslash G/A'$ has one element, equivalently $G=AA'$. Physically, no defect-sector label is left to distinguish degenerate ground states.
\item For that double coset, the relevant projective representation category of the stabilizer has only one irreducible object. Equivalently, the corresponding relative two-cocycle on the stabilizer is nondegenerate. Physically, no additional local projective multiplet label is left.
\end{itemize}
Together with the gaugeability of $A'$, this is the fiber-functor criterion.

In the normal Abelian case considered in this section, this criterion becomes explicit. Let $p:G\to Q=G/A$ be the quotient map. Because $A$ is a normal subgroup of $G$, the one-double-coset condition $G=AA'$ is equivalent to $p(A')=Q$. Thus $A'$ must contain representatives for all quotient defects in $Q$. After this condition is imposed, we can use the representative $g=1$, and the stabilizer becomes $K=A\cap A'$. This is the subgroup that is seen both by the gauged symmetry $A$ and by the candidate boundary condition $A'$.

The only remaining possible degeneracy is local to this overlap subgroup $K$. Since $K\leq A$ and $A$ is Abelian, $K$ is Abelian. On $K$, the two trivializations $\psi$ and $\psi'$ have the same coboundary $\omega|_K$, so their ratio $\beta=(\psi\cdot{\psi'}^{-1})|_K$ is a genuine two-cocycle on $K$. In the standard group-theoretical formula, the stabilizer factor set is cohomologous to this relative cocycle. Equivalently, after a harmless coboundary change, the endpoint quantum numbers are precisely the $\beta$-projective representations of $K$. The endpoint label is unique exactly when the twisted group algebra $\mathbb C_\beta[K]$ is a single irreducible block, equivalently a full matrix algebra; this is what it means for $\beta$ to be nondegenerate.

For Abelian $K$, this gives a simple definition: a two-cocycle $\beta$ on $K$ is nondegenerate if the alternating bicharacter $\alpha_\beta(k,k')=\beta(k,k')/\beta(k',k)$ defines an isomorphism $K\to \widehat K$, $k\mapsto \alpha_\beta(k,-)$. Equivalently, every nontrivial element of $K$ has a nontrivial projective commutator with some other element of $K$. Physically, this means that no nontrivial $K$ quantum number remains as an independent endpoint label. If $K=1$, the condition is automatic. If $K$ is nontrivial, it must be a finite symplectic Abelian group, $K\cong\bigoplus_i(\Z_{N_i}\times\Z_{N_i})$. After choosing paired generators and changing $\beta$ within its cohomology class, a standard representative is $\beta(k,k')=\prod_i\exp(2\pi\ii\, m_i k_i^a {k'_i}^b/N_i)$, with $m_i\in\Z_{N_i}^{\times}$, whose commutator pairing is nondegenerate \cite{etingof2016tensor,davydov2010modular}. We have therefore arrived at a concrete criterion: in the normal Abelian case, the dual symmetry is anomaly-free if and only if there exists a subgroup $A'\leq G$ and a cochain $\psi'$ such that
\begin{align}
p(A')=Q,\qquad
\omega|_{A'}=\dd\psi',\qquad
(\psi\cdot{\psi'}^{-1})|_{A\cap A'}\ \text{is nondegenerate}.
\end{align}

Let us finally translate the three conditions back into physics. The subgroup $A'$ supplies a candidate fully symmetric gapped boundary condition after gauging. The condition $p(A')=Q$ says that every quotient defect sector can end on this boundary condition, so there is no residual degeneracy from distinct $Q$ sectors. The condition $\omega|_{A'}=\dd\psi'$ says that this candidate boundary condition is itself anomaly-free before gauging. The nondegenerate relative cocycle on $A\cap A'$ says that the remaining endpoint projective degeneracy on the overlap is also removed: there is a unique irreducible projective representation in that sector, although it may be higher-dimensional. This last point is important. A unique projective multiplet can still have dimension larger than one, so anomaly-free does not imply invertible.

This also clarifies the $D_8$ example. Non-invertibility means that the dual category is not pointed; in $\Rep(H_8)$ this is visible in the two-dimensional simple object $X_0$. Anomaly-free means something different: the category admits a fiber functor, or equivalently a fully symmetric short-range-entangled phase. Since $\Rep(H_8)$ is the representation category of the Kac-Paljutkin Hopf algebra, it has the usual forgetful tensor functor to $\Vec$ \cite{tambara2000representations,etingof2021tensor}. Thus the gauged type-II DQCP is controlled by symmetry breaking for a genuinely non-invertible, but anomaly-free, dual symmetry.

\section{A $\Vec_{D_8}^{\omega_{\mathrm{LSM}}}$ spin-chain DQCP and its $\Rep(H_8)$ dual symmetry}
\label{sec:D8}

In this section we make the general mechanism concrete in a spin-$1/2$ chain. We first identify the microscopic lattice symmetry and explain why, for the period-two phases relevant to the DQCP, it is natural to work with an effective $D_8$ quotient. We then determine the corresponding LSM anomaly and describe the resulting anomalous symmetry as $\Vec_{D_8}^{\omega_{\mathrm{LSM}}}$. With this symmetry data fixed, we classify all of the gapped phases by anomaly-free subgroups. This identifies the dimer and ferromagnetic phases as two distinct symmetry-breaking descendants of the same anomalous $D_8$ symmetry. We then construct an explicit interpolation between them and use it as a lattice realization of a dimer-to-ferromagnet DQCP. The last part of the section studies what happens when different subgroups are gauged. This gauging changes the categorical symmetry acting on the model, and in particular produces the non-invertible dual symmetry $\Rep(H_8)$. In this way the critical point admits both the original $\Vec_{D_8}^{\omega_{\mathrm{LSM}}}$ DQCP description and a dual description in terms of $\Rep(H_8)$ symmetry breaking.

\subsection{Lattice symmetry actions and the $D_8$ quotient}

Consider a spin-$1/2$ chain with an even number of lattice sites. We take the thermodynamic limit $L\rightarrow \infty$. The microscopic Hamiltonian is assumed to be invariant under the following symmetry generators:
\begin{equation}\label{symm}
\begin{gathered}
    U_{XY} = \prod_i X_{2i}Y_{2i+1},\qquad
    U_{YX} = \prod_i Y_{2i}X_{2i+1},\\
    T\sigma_i^\alpha T^{-1} = \sigma_{i+1}^\alpha \quad(\alpha=x,y,z).
\end{gathered}
\end{equation}
Here $X_i,Y_i,Z_i$ are Pauli operators on site $i$.
Up to a relabeling of Pauli axes, this symmetry action also appears as the effective symmetry action on the $\alpha'\gamma$ interface of the weak non-invertible SPT construction in Ref.~\onlinecite{Furukawa2025SubsystemWeakNISPT}.
The operators $U_{XY}$ and $U_{YX}$ generate an internal symmetry $G_\mathrm{int}=\mathbb Z_2^{XY}\times \mathbb Z_2^{YX}$, while $T$ denotes one-site lattice translation. They satisfy $U_{XY}^2=U_{YX}^2=1$, $T U_{XY} T^{-1}=U_{YX}$, and $T U_{YX} T^{-1}=U_{XY}$.
Thus one-site translation acts nontrivially on the internal symmetry by exchanging the two $\mathbb Z_2$ factors. The full microscopic symmetry group fits into the short exact sequence
\begin{align}
1\longrightarrow
\mathbb Z_2^{XY}\times \mathbb Z_2^{YX}
\longrightarrow
G_\mathrm{tot}
\longrightarrow
\mathbb Z_{\mathrm{trans}}
\longrightarrow 1 ,
\end{align}
and is explicitly given by the semidirect product
$G_\mathrm{tot}=
\left(\mathbb Z_2^{XY}\times \mathbb Z_2^{YX}\right)
\rtimes
\mathbb Z_{\mathrm{trans}}$,
where the generator of $\mathbb Z_{\mathrm{trans}}$ acts on $\mathbb Z_2^{XY}\times \mathbb Z_2^{YX}$ by interchanging the two factors.

The phases of interest below, including the dimer phase, preserve two-site translation while allowing one-site translation to be spontaneously broken. Following the finite-translation reduction discussed in Sec.~\ref{sec:symm}, we therefore keep the effective translation quotient $\mathbb Z_{\mathrm{trans}}/\langle T^2\rangle\cong \mathbb Z_2$.
This does not mean that the microscopic translation group is absent; rather, it is the finite symmetry acting faithfully on the period-two low-energy multiplet. In this reduced description the symmetry group is
\begin{align}
G=G_\mathrm{tot}/\langle T^2\rangle
=
(\mathbb Z_2^{XY}\times\mathbb Z_2^{YX})\rtimes \mathbb Z_2
\cong D_8 .
\end{align}
We use the presentation $D_8=\langle a,x\mid a^4=x^2=1,\ xax=a^{-1}\rangle$ and identify $T\mapsto x$, $U_{XY}\mapsto ax$, and $U_{YX}\mapsto a^3x$.
Indeed, the subgroup generated by $ax$ and $a^3x$ is $\langle ax,a^3x\rangle=\{1,a^2,ax,a^3x\}\cong \mathbb Z_2\times\mathbb Z_2$, and conjugation by $x$ exchanges the two generators: $x(ax)x^{-1}=a^3x$ and $x(a^3x)x^{-1}=ax$.
Thus the reduced symmetry fits into the short exact sequence
\begin{align}\label{sesD8}
1\rightarrow \Z_2^{ax}\times\Z_2^{a^3x} \rightarrow D_8\rightarrow \Z_2^x\rightarrow 1.
\end{align}
In the following, we will use the microscopic generators of $G_\mathrm{tot}$ and their images in the quotient $D_8$ interchangeably when no confusion can arise. This is the effective finite symmetry on which the explicit anomaly and gauging computations are based.

\subsection{Identifying $E_\infty^{1,2}$-type LSM anomaly}

The purpose of this subsection is to identify the LSM anomaly associated with the symmetry action in Eq.~\eqref{symm}. We do this in two complementary ways. First, we read off the projective representation carried by one primitive unit cell and place the resulting class in the LHS spectral sequence for $G=D_8$. Second, we check that the same class is singled out by the pattern of gaugeable subgroups and their Morita dual symmetry categories.

\subsubsection{From LHS spectral sequence}

Although the period-two reduction gives an ordinary finite group $G=D_8$, the action on the spin chain still carries an LSM obstruction. This obstruction is already visible in the on-site representation. The internal subgroup is $G_\mathrm{int}=\Z_2^{XY}\times\Z_2^{YX}$, whose image in $G$ is generated by $ax$ and $a^3x$. On an even site, the local representatives of $ax$ and $a^3x$ are $X$ and $Y$, while on an odd site they are $Y$ and $X$. In both cases the two generators anticommute. Thus each primitive unit cell carries the nontrivial projective representation class $[\omega_2]\in H^2(G_\mathrm{int},\U)=\Z_2$, which is the microscopic LSM index.

This local projective class becomes a mixed anomaly between $G_\mathrm{int}$ and the translation quotient $\Z_2^x$. For the extension in Eq.~\eqref{sesD8}, the relevant LHS entry is
\begin{align}\label{D8-E12}
E_2^{1,2}
=
H^1\!\left(\Z_2^x,H^2(G_\mathrm{int},\U)\right)
=
H^1(\Z_2^x,\Z_2)
=
\Z_2 .
\end{align}
Here $\Z_2^x$ acts on $G_\mathrm{int}$ by conjugation, exchanging $ax$ and $a^3x$ in $G$. This exchange fixes the nontrivial class $[\omega_2]\in H^2(G_\mathrm{int},\U)$, since $H^2(G_\mathrm{int},\U)$ has only one nonzero element. Hence the induced action of translation on $H^2(G_\mathrm{int},\U)$ is trivial, and $[\omega_2]$ gives the nontrivial element of Eq.~\eqref{D8-E12}.

As shown in Appendix~\ref{App:LHS}, this $E_2^{1,2}$ class is not killed by later differentials and therefore survives to the $E_\infty$ page. The surviving class embeds into $H^3(G,\U)=H^3(D_8,\U)=\Z_2^2\times\Z_4$. In the convention of Appendix~\ref{App:LHS}, it is represented by
\begin{align}\label{omegaLSM2zeta}
\omega_{\mathrm{LSM}}
=
2\zeta\in H^3(G,\U).
\end{align}
Here $\zeta$ generates the relevant $\Z_4$ summand, and an explicit cocycle representative is given in Appendix~\ref{App:nu3D8}. For the original microscopic group $G_\mathrm{tot}$, the same anomaly is the pullback of $2\zeta$ along the quotient map $G_\mathrm{tot}\to G=G_\mathrm{tot}/\langle T^2\rangle=D_8$. We use the same notation $\omega_{\mathrm{LSM}}$ for both the $G=D_8$ cocycle and its pullback to $G_\mathrm{tot}$ when no confusion can arise.

\subsubsection{From subgroup gauging}

The same anomaly can be identified from subgroup gauging. A subgroup $H\subseteq G$ is gaugeable in $\Vec_G^{\omega}$ precisely when the restricted cocycle $\omega|_H$ is a coboundary, i.e., when $H$ is anomaly-free. For the present lattice symmetry action, the gaugeable subgroups and their Morita duals have a characteristic pattern.

The corresponding lattice gauging operations are discussed in Sec.~\ref{sec:latticegauging}. We need the following three cases:
\begin{itemize}
\item The diagonal internal element $U_{XY}U_{YX}=\prod_i Z_i$ is represented in $G=D_8$ by the central element $a^2$. The subgroup $\Z_2^{a^2}=\langle a^2\rangle$ is anomaly-free, and gauging it gives a self-dual invertible symmetry, namely another copy of $\Vec_G^{\omega_{\mathrm{LSM}}}$.
\item Each single internal $\Z_2$ generated by $U_{XY}$ or $U_{YX}$ is anomaly-free. In $G=D_8$ these are the subgroups $\langle ax\rangle$ and $\langle a^3x\rangle$. Gauging either one produces a non-invertible dual category.
\item The full internal subgroup $G_\mathrm{int}$, whose image in $G$ is $\langle ax,a^3x\rangle=\langle ax,a^2\rangle$, is also gaugeable. This gauging can be viewed as first gauging $\langle a^2\rangle$ and then gauging the remaining single internal $\Z_2$, so it combines the two operations above. The resulting dual symmetry is again non-invertible.
\end{itemize}

Appendix~\ref{App:D8ACA} tabulates the gaugeable subgroups and Morita duals for all anomaly classes $\omega\in H^3(G,\U)$ with $G=D_8$. The pattern above singles out the same class as the LHS computation. Requiring $\langle a^2\rangle$ to be gaugeable with self-dual invertible dual symmetry leaves the three possibilities $2\zeta$, $2\zeta+\alpha$, and $2\zeta+\beta$ in the notation of Appendix~\ref{App:D8ACA}. Imposing in addition that the internal subgroups $\langle ax\rangle$ and $\langle a^3x\rangle$ are gaugeable selects only $2\zeta$. The gaugeability of the full internal subgroup $G_\mathrm{int}$ and its non-invertible Morita dual are then consistent with this same choice. Thus the subgroup-gauging criterion independently identifies the LSM anomaly as $\omega_{\mathrm{LSM}}=2\zeta$, in agreement with Eq.~\eqref{omegaLSM2zeta}.

\subsection{Classification of gapped phases}

In this subsection, we classify the gapped phases allowed by the anomalous symmetry $\Vec_{D_8}^{\omega_{\mathrm{LSM}}}$ and then give representative ground states together with parent Hamiltonians. The special dimer and $z$-ferromagnetic ($z$FM) representatives identified here will be used in the next subsection to construct the dimer-to-ferromagnet DQCP.

\subsubsection{Module-category classification}

Let $\mathcal C=\Vec_{D_8}^{\omega_{\mathrm{LSM}}}$ with $\omega_{\mathrm{LSM}}=2\zeta$. As reviewed in Appendix~\ref{App:A}, a 1+1D gapped phase with fusion-category symmetry $\mathcal C$ is described by an indecomposable semisimple module category over $\mathcal C$. After choosing algebra representatives, such phases can be presented by Morita classes of condensable algebras in $\mathcal C$. For the pointed category $\Vec_{D_8}^{\omega_{\mathrm{LSM}}}$, an algebra representative is specified by a pair $(H,\gamma)$, where $H\subseteq D_8$ and $\gamma\in C^2(H,U(1))$ satisfy $d\gamma=\omega_{\mathrm{LSM}}|_H$. Equivalently, the restricted anomaly on $H$ must be canceled by the local counterterm $\gamma$. Physically, $H$ is the unbroken subgroup in the corresponding phase, while changing the cohomology class of $\gamma$ stacks a one-dimensional $H$-SPT phase on top of the same symmetry-breaking pattern.

We now apply this criterion to the present anomaly. Among the conjugacy classes of subgroups of $D_8$, the two represented by $\langle a\rangle$ and $D_8$ itself are anomalous, so neither can occur as the unbroken symmetry of a gapped phase. The remaining anomaly-free subgroups fall into six conjugacy classes. Two of these representatives are isomorphic to $\mathbb Z_2\times\mathbb Z_2$, so one might expect an additional label from $H^2(H,U(1))=\mathbb Z_2$, whose nontrivial class is the usual Haldane-chain SPT protected by $\mathbb Z_2\times\mathbb Z_2$ symmetry \cite{ChenGuWen2011Complete,PollmannBergTurnerOshikawa2012}. However, the Morita-class computation in Appendix~\ref{App:D8ACA}, or equivalently the identification discussed in \cite{diatlyk2026gaugingnoninvertiblesymmetriestopological}, shows that the possible nontrivial choices of $\gamma$ are Morita equivalent to the trivial choice and therefore do not give additional gapped phases for this anomaly. We may therefore choose $\gamma$ in the trivial cohomology class for all six cases. The phase labels then reduce to the six anomaly-free unbroken subgroups $H\subseteq D_8$, modulo conjugacy.

Table~\ref{Table:phases} lists the corresponding conjugacy classes of phases. For each representative subgroup $H$, the ground-state subspace is generated by the $D_8$ orbit of a representative state $|\Omega_H\rangle$, giving $\mathrm{GSD}=|D_8/H|$. When a row lists conjugate subgroups, the displayed state preserves the representative subgroup shown in the table, while the other states in its $D_8$ orbit preserve the conjugate subgroups. In what follows, we give the unbroken symmetry, a representative ground state, and a parent Hamiltonian for each of the six cases in Table~\ref{Table:phases}.

\begin{table*}[h!]
\centering
\begingroup
\renewcommand{\arraystretch}{1.35}
\begin{tabular}{|c|c|c|c|}
\hline
\begin{tabular}[c]{@{}c@{}}Conjugacy class of\\ unbroken symmetry $H$\end{tabular}
& \begin{tabular}[c]{@{}c@{}}$\mathrm{GSD}=|D_8/H|$\end{tabular}
& \begin{tabular}[c]{@{}c@{}}Representative GS $|\Omega_H\rangle$\end{tabular}
& \begin{tabular}[c]{@{}c@{}}Description\end{tabular} \\
\hline
\hline
$1$ 
& $8$ 
& $\displaystyle \otimes_j |\psi_{\mathrm e}\rangle_{2j}
|\psi_{\mathrm o}\rangle_{2j+1}$ & fully SSB = generic even/odd-site direct product state \\
\hline
$\mathbb Z_2=\langle a^2\rangle$ 
& $4$ 
& $\displaystyle \otimes_j
\left(
|\!\uparrow\uparrow\rangle
+\lambda |\!\downarrow\downarrow\rangle
\right)_{2j,2j+1}$ & generic dimer state \\
\hline
$\mathbb Z_2=\langle x\rangle$ or $\langle a^2x\rangle$ 
& $4$ 
& $\displaystyle \otimes_i |\psi\rangle_i$ & generic direct product state \\
\hline
$\mathbb Z_2^2=\langle a^2,x\rangle$ 
& $2$ 
& $\displaystyle\otimes_i |\!\uparrow\rangle_i$ & $z$FM = special direct product state \\
\hline
$\mathbb Z_2=\langle ax\rangle$ or $\langle a^3x\rangle$
& $4$ 
& $\displaystyle \otimes_j |+x\rangle_{2j}|+y\rangle_{2j+1}$ & special two-site product state \\
\hline
$\Z_2^{ax}\times \Z_2^{a^3x}=\langle ax,a^3x\rangle$ 
& $2$ 
& $\displaystyle \otimes_j (|\!\uparrow\uparrow\rangle+i|\!\downarrow\downarrow\rangle)_{2j,2j+1}$ & special dimer state \\
\hline
\end{tabular}
\endgroup
\caption{Gapped phases for the anomalous symmetry $\mathcal C=\Vec_{D_8}^{\omega_{\mathrm{LSM}}}$ with $\omega_{\mathrm{LSM}}=2\zeta$, labeled by conjugacy classes of anomaly-free unbroken subgroups $H\subseteq D_8$. Mathematically, each gapped phase is represented by module-category data $(H,[\gamma]=0)$, where $H$ is anomaly-free. Physically, $H$ is the unbroken symmetry of the representative state $|\Omega_H\rangle$, and the $D_8$ orbit of $|\Omega_H\rangle$ spans the ground-state subspace, giving $\mathrm{GSD}=|D_8/H|$. Generic choices of $|\psi_{\mathrm e}\rangle$, $|\psi_{\mathrm o}\rangle$, $|\psi\rangle$, and $\lambda$ avoid accidental enlargement of $H$.}
\label{Table:phases}
\end{table*}

\subsubsection{Unbroken symmetry $H=1$ (fully SSB)}
We first consider the fully symmetry-broken phase, for which the unbroken
subgroup is trivial, $H=1$.  A representative ground state can be chosen as a
generic period-two product state
\begin{equation}
|\Omega_1\rangle
=
\bigotimes_j
|\psi_{\mathrm e}\rangle_{2j}
|\psi_{\mathrm o}\rangle_{2j+1}.
\end{equation}
Here ``generic'' means that the two-site pattern has no accidental stabilizer
inside $D_8$.  Explicitly, we require
$|\psi_{\mathrm o}\rangle\not\propto|\psi_{\mathrm e}\rangle$,
$|\psi_{\mathrm o}\rangle\not\propto Z|\psi_{\mathrm e}\rangle$, and we
exclude the case in which both single-site states are $Z$-eigenstates.  We also
exclude the special choices
$|\psi_{\mathrm e}\rangle\in\{|\!+\!x\rangle,|\!-\!x\rangle\}$ with
$|\psi_{\mathrm o}\rangle\in\{|\!+\!y\rangle,|\!-\!y\rangle\}$, and the
corresponding choices with $x$ and $y$ interchanged.  These conditions remove
all nontrivial stabilizers, so that
$\mathrm{Stab}_{D_8}(|\Omega_1\rangle)=1$ and the $D_8$ orbit contains
$|D_8|=8$ ground states.

For concreteness, we now choose an explicit orbit on a single even--odd bond.
Starting from $|\psi_{1,1}\rangle=|0\rangle|\!+\!x\rangle$, the internal
symmetry operations generate
\begin{equation}
\begin{aligned}
|\psi_{1,1}\rangle &= |0\rangle|\!+\!x\rangle,\\
|\psi_{1,2}\rangle &= (X\otimes Y)|\psi_{1,1}\rangle
= -i|1\rangle|\!-\!x\rangle,\\
|\psi_{1,3}\rangle &= (Y\otimes X)|\psi_{1,1}\rangle
= i|1\rangle|\!+\!x\rangle,\\
|\psi_{1,4}\rangle &= (Z\otimes Z)|\psi_{1,1}\rangle
= |0\rangle|\!-\!x\rangle.
\end{aligned}
\end{equation}
On this bond, $X\otimes Y$ and $Y\otimes X$ represent the two generators of
$\mathbb{Z}_2^{ax}\times\mathbb{Z}_2^{a^3x}$, and their product is
$Z\otimes Z$.  The four states above are permuted freely by this subgroup and
therefore realize its regular representation.  The eight ground states in the
fully symmetry-broken phase are obtained by forming period-two products of
these bond states and their one-site translates:
\begin{equation}
\begin{aligned}
|\Omega_{1,i}\rangle
&= \bigotimes_j |\psi_{1,i}\rangle_{2j,2j+1},\\
T|\Omega_{1,i}\rangle
&= \bigotimes_j |\psi_{1,i}\rangle_{2j+1,2j+2},
\qquad i = 1,2,3,4.
\end{aligned}
\end{equation}
Together, these eight ground states furnish the regular representation of $D_8$. 

A convenient $D_8$-symmetric parent Hamiltonian is obtained by projecting
locally onto the allowed four-site product patterns.  We take
\begin{equation}
    \hat H_1 = -\sum_j P^{1}_{j,j+1,j+2,j+3}
\end{equation}
Here and below, the hat distinguishes the Hamiltonian $\hat H$ from the
unbroken subgroup $H$.  The local projector $P^1$ has image equal to the
seven-dimensional subspace spanned by
\begin{equation}
\begin{aligned}
    \{
&|0\rangle|\!+\!x\rangle|0\rangle|\!+\!x\rangle,
|0\rangle|\!-\!x\rangle|0\rangle|\!-\!x\rangle,
|1\rangle|\!+\!x\rangle|1\rangle|\!+\!x \rangle,
|1\rangle|\!-\!x\rangle|1\rangle|\!-\!x\rangle,\\
&|\!+\!x\rangle|0\rangle|\!+\!x\rangle|0\rangle,
|\!-\!x\rangle|0\rangle|\!-\!x\rangle|0\rangle, 
|\!+\!x\rangle|1\rangle|\!+\!x\rangle|1\rangle,
|\!-\!x\rangle|1\rangle|\!-\!x\rangle|1\rangle
    \}.
\end{aligned}
\end{equation}
An explicit expression for this projector is
\begin{equation}
\begin{aligned}
P^1= & \frac{7}{16} I I II 
 +\frac{3}{16}(X I X I+Z I Z I+I X I X+Z X Z X+I Z I Z+X Z X Z) \\
& +\frac{1}{16}(Y I Y I+Y X Y X+I Y I Y+X Y X Y+Z Y Z Y+Y Z Y Z) 
 -\frac{1}{16}(X X X X+Y Y Y Y+Z Z Z Z)\\
=& P_{Z X}+P_{X Z}-P_{Z X} P_{X Z} ,
\end{aligned}
\end{equation}
where $P_{ZX} = \frac{1+Z I Z I}{2} \frac{1+I X I X}{2}$ and $P_{XZ} = \frac{1+X I X I}{2} \frac{1+I Z I Z}{2}$.

\subsubsection{Unbroken symmetry $H=\mathbb Z_2^{a^2}$ (generic dimer state)}

Next, we consider the phase in which the unbroken subgroup is
$H=\mathbb{Z}_2^{a^2}$.  A representative ground state can be chosen as an
even-bond dimer product state
\begin{equation}
|\Omega_{a^2}\rangle
=
\bigotimes_j
|D_\lambda\rangle_{2j,2j+1},
\end{equation}
where
\begin{equation}
|D_\lambda\rangle
=
\frac{
|00\rangle
+
\lambda|11\rangle
}{
\sqrt{1+|\lambda|^2}
}.
\end{equation}
We take $\lambda$ to be finite and nonzero, and impose
$\lambda\neq \pm i$.  The excluded limits $\lambda=0,\infty$ give the
$z$-ferromagnetic product states, for which the one-site translation $x=T$ is
restored.  The values $\lambda=\pm i$ give special dimer states with enlarged
stabilizer $\Z_2^{ax}\times\Z_2^{a^2}$.

Each dimer is fixed by $Z\otimes Z$, so the chain preserves
$a^2=U_{XY}U_{YX}$.  For generic $\lambda$, it preserves neither
$ax=U_{XY}$ nor $a^3x=U_{YX}$, and it also breaks the one-site translation.
Thus $\mathrm{Stab}_{D_8}(|\Omega_{a^2}\rangle)=\langle a^2\rangle$, and the
$D_8$ orbit contains $\mathrm{GSD}=|D_8/\langle a^2\rangle|=4$ ground states.

For concreteness, choose $\lambda=1$.  On an even--odd bond, define
\begin{equation}
\begin{aligned}
|\psi_{a^2,1}\rangle
&= \frac{1}{\sqrt{2}}(|00\rangle + |11\rangle),\\
|\psi_{a^2,2}\rangle
&= (X\otimes Y)|\psi_{a^2,1}\rangle
= (Y\otimes X)|\psi_{a^2,1}\rangle
= \frac{-i}{\sqrt{2}} (|00\rangle - |11\rangle).
\end{aligned}
\end{equation}
Both states are fixed by $a^2$, represented on the bond by $Z\otimes Z$, while
$ax$ and $a^3x$ exchange them.  Equivalently, they form the regular
representation of
$(\mathbb{Z}_2^{ax}\times\mathbb{Z}_2^{a^3x})/\mathbb{Z}_2^{a^2}$.
The four ground states are obtained by forming dimer products and their
one-site translates:
\begin{equation}
\begin{aligned}
|\Omega_{a^2,i}\rangle
&= \bigotimes_j |\psi_{a^2,i}\rangle_{2j,2j+1},\\
T|\Omega_{a^2,i}\rangle
&= \bigotimes_j |\psi_{a^2,i}\rangle_{2j+1,2j+2},
\qquad i = 1,2.
\end{aligned}
\end{equation}

A convenient $D_8$-symmetric parent Hamiltonian is obtained by projecting
locally onto the allowed five-site dimer patterns.  We take
\begin{equation}
    \hat H_{a^2}  = -\sum_j P^{a^2}_{j,j+1,j+2,j+3,j+4}
\end{equation}
where $P^{a^2}$ is the projector onto the $8$-dimensional subspace spanned by
\begin{equation}
\big\{|\psi_{a^2,i}\rangle|\psi_{a^2,i}\rangle|0\rangle,|\psi_{a^2,i}\rangle|\psi_{a^2,i}\rangle|1\rangle,|0\rangle|\psi_{a^2,i}\rangle|\psi_{a^2,i}\rangle,|1\rangle|\psi_{a^2,i}\rangle|\psi_{a^2,i}\rangle\ |\ i=1,2\big\}.
\end{equation}
The corresponding projector can be written explicitly as
\[
\begin{aligned}
P^{a^2}={}&
\frac14\,IIIII
\\
&+\frac{5}{48}(
ZZIII
+XXXXI
-YYXXI
-XXYYI
+YYYYI
+ZZZZI
\\
&\qquad
+IXXXX
-IYYXX
-IXXYY
+IYYYY
+IIIZZ
+IZZZZ
)
\\
&+\frac18(
IZZII
+IIZZI
) 
-\frac{1}{48}(
ZIZII
+ZIIZI
+IZIIZ
+IIZIZ
)
\\
&+\frac{1}{48}(
YXYXI
+XYYXI
+YXXYI
+XYXYI
+IYXYX
+IXYYX
+IYXXY
+IXYXY
)
\\
&-\frac{1}{24}(XIIIX+YIIIY+ZIIIZ)(IIIII+IZZII+IZIZI+IIZZI).
\end{aligned}
\]

\subsubsection{Unbroken symmetry $H=\mathbb Z_2^x$ or $\mathbb{Z}_2^{a^2x}$ (generic product state)}

We next consider the conjugacy class represented by
$H=\mathbb{Z}_2^x$.  The conjugate subgroup
$\mathbb{Z}_2^{a^2x}$ is obtained from the same $D_8$ orbit.  A representative
ground state with unbroken $x=T$ is a uniform product state
\begin{equation}
|\Omega_x\rangle
=
\bigotimes_j |\psi_x\rangle_j .
\end{equation}
We choose $|\psi_x\rangle$ generically so that it has no accidental internal
stabilizer.  In particular, $|\psi_x\rangle$ is not a $Z$-eigenstate, or
equivalently $Z|\psi_x\rangle\not\propto|\psi_x\rangle$, so the diagonal
element $a^2$ is broken.  The state is uniform and therefore preserves
$x=T$, while a generic choice of $|\psi_x\rangle$ preserves neither
$ax=U_{XY}$ nor $a^3x=U_{YX}$.  Hence
$\mathrm{Stab}_{D_8}(|\Omega_x\rangle)=\langle x\rangle$, and the orbit has
$\mathrm{GSD}=|D_8/\langle x\rangle|=4$.

For concreteness, we choose an explicit orbit on an even--odd bond:
\begin{equation}
\begin{aligned}
|\psi_{x,1}\rangle &= |\!+\!x\rangle|\!+\!x\rangle,\\
|\psi_{x,2}\rangle &= |\!+\!x\rangle|\!-\!x\rangle,\\
|\psi_{x,3}\rangle &= |\!-\!x\rangle|\!+\!x\rangle,\\
|\psi_{x,4}\rangle &= |\!-\!x\rangle|\!-\!x\rangle.
\end{aligned}
\end{equation}
Up to unimportant phases, the internal generators $X\otimes Y$ and
$Y\otimes X$ permute these four states freely.  They therefore realize the
regular representation of
$\mathbb{Z}_2^{ax}\times\mathbb{Z}_2^{a^3x}$.  The corresponding chain ground
states are
\begin{equation}
|\Omega_{x,i}\rangle
=
\bigotimes_j |\psi_{x,i}\rangle_{2j,2j+1},
\qquad i=1,2,3,4.
\end{equation}
The states $|\Omega_{x,1}\rangle$ and $|\Omega_{x,4}\rangle$ are uniform and
preserve $\mathbb{Z}_2^x$, whereas $|\Omega_{x,2}\rangle$ and
$|\Omega_{x,3}\rangle$ are staggered and preserve the conjugate subgroup
$\mathbb{Z}_2^{a^2x}$.

A convenient $D_8$-symmetric parent Hamiltonian may first be written in
projector form as $\hat H_x^{\mathrm{proj}}=-P^x$, with
$P^x=\prod_j(1+X_{2j}X_{2j+2})/2\prod_j(1+X_{2j+1}X_{2j+3})/2$.  Equivalently,
we use the simpler local form
\begin{equation}
    \hat H_x = -\sum_j X_jX_{j+2}.
\end{equation}
This Hamiltonian locks the $X$ polarization independently on the even and odd sublattices,
so its ground space is spanned by the four product states above.

\subsubsection{Unbroken symmetry $H=\mathbb Z_2^{a^2}\times \mathbb Z_2^x$ ($z$FM)}

The $z$-ferromagnetic phase is obtained as a special product-state limit in
which the unbroken subgroup is
$H=\mathbb{Z}_2^{a^2}\times\mathbb{Z}_2^x=\langle a^2,x\rangle$.  A
representative $D_8$ orbit is
\begin{equation}
|\Omega_{z\mathrm{FM},\uparrow}\rangle
=
\bigotimes_j |\!\uparrow\rangle_j,
\qquad
|\Omega_{z\mathrm{FM},\downarrow}\rangle
=
\bigotimes_j |\!\downarrow\rangle_j .
\end{equation}
Since each site is a $Z$ eigenstate, the state preserves the diagonal internal
element $a^2=U_{XY}U_{YX}$.  It is also uniform, and hence preserves the
one-site translation $x=T$.  The remaining internal generators
$ax=U_{XY}$ and $a^3x=U_{YX}$ exchange the two states above, up to phases.
Thus the stabilizer of either representative is
$\mathrm{Stab}_{D_8}(|\Omega_{z\mathrm{FM},\uparrow}\rangle)
=\mathrm{Stab}_{D_8}(|\Omega_{z\mathrm{FM},\downarrow}\rangle)
=\langle a^2,x\rangle$,
and the orbit contains $\mathrm{GSD}=|D_8/\langle a^2,x\rangle|=2$ ground
states.

A convenient $D_8$-symmetric parent Hamiltonian may be written in projector
form as
$\hat H_{z\mathrm{FM}}^{\mathrm{proj}}=-\sum_j(1+Z_jZ_{j+1})/2$.  Equivalently,
up to an additive constant, we use the ferromagnetic Ising form
\begin{equation}
\hat H_{z\mathrm{FM}}
=
-\sum_j Z_jZ_{j+1}.
\label{H1}
\end{equation}
Its ground space is spanned by the two $z$-ferromagnetic states
$|\Omega_{z\mathrm{FM},\uparrow}\rangle$ and
$|\Omega_{z\mathrm{FM},\downarrow}\rangle$.

\subsubsection{Unbroken symmetry $H=\mathbb Z_2^{ax}$ (or $\mathbb{Z}_2^{a^3x}$) (special two-site product state)}

We next consider the conjugacy class represented by
$H=\mathbb{Z}_2^{ax}=\langle ax\rangle$; the conjugate representative is
$\mathbb{Z}_2^{a^3x}=\langle a^3x\rangle$.  A ground state with unbroken
$\langle ax\rangle$ can be chosen as the period-two product state
\begin{equation}
|\Omega_{ax}\rangle
=
\bigotimes_j
|\!+\!x\rangle_{2j}
|\!+\!y\rangle_{2j+1}.
\end{equation}
Indeed, each even--odd bond is fixed by $X\otimes Y$, so the chain preserves
$U_{XY}=ax$.  For this representative, $a^2$, $x=T$, and $a^3x=U_{YX}$ are
broken.  Thus
$\mathrm{Stab}_{D_8}(|\Omega_{ax}\rangle)=\langle ax\rangle$, and the orbit
contains $\mathrm{GSD}=|D_8/\langle ax\rangle|=4$ ground states.

For concreteness, take the two $ax$-invariant bond states
$|\psi_{ax,1}\rangle=|\!+\!x\rangle|\!+\!y\rangle$ and
$|\psi_{ax,2}\rangle=|\!-\!x\rangle|\!-\!y\rangle$.  The four ground states
are obtained by forming products of these bond states and their one-site
translates:
\begin{equation}
\begin{aligned}
|\Omega_{ax,i}\rangle
&= \bigotimes_j|\psi_{ax,i}\rangle_{2j,2j+1},\\
T|\Omega_{ax,i}\rangle
&= \bigotimes_j|\psi_{ax,i}\rangle_{2j+1,2j+2},
\qquad i=1,2.
\end{aligned}
\end{equation}
The first two states preserve $\mathbb{Z}_2^{ax}$, while the translated pair
preserve the conjugate subgroup $\mathbb{Z}_2^{a^3x}$.

A convenient $D_8$-symmetric parent Hamiltonian is
\begin{equation}
    \hat H_{ax} = -\sum_j P^{ax}_{j,j+1,j+2,j+3}.
\end{equation}
Here $P^{ax}$ is the projector onto the ten-dimensional subspace
\begin{equation}
\begin{aligned}
\mathrm{span}\,\{
&|\psi_{ax,i}\rangle|\psi_{ax,i}\rangle,
    |0\rangle|\psi_{ax,i}\rangle|0\rangle,
    |0\rangle|\psi_{ax,i}\rangle|1\rangle,
    |1\rangle|\psi_{ax,i}\rangle|0\rangle,
    |1\rangle|\psi_{ax,i}\rangle|1\rangle
    \;|\; i=1,2\}.
\end{aligned}
\end{equation}
For the representative above, this projector can be written explicitly as
\begin{equation}
    \begin{aligned}
P^{ax}&=  \frac{5}{8} I I I I+\frac{3}{8} I X Y I+\frac{1}{8} I Y X I-\frac{1}{8} I Z Z I 
 +\frac{1}{8} X I I Y+\frac{1}{8} X Y X Y-\frac{1}{8} X X Y Y-\frac{1}{8} X Z Z Y \\
 &=\frac{1+I X Y I}{2}+\frac{1-I X Y I}{2} \frac{1+I Y X I}{2} \frac{1+X I I Y}{2}
\end{aligned}
\end{equation}

Another convenient choice of $ax$-invariant bond states is
\begin{equation}
\begin{aligned}
|\psi_{ax,1}\rangle
&= \frac12 (|00\rangle+|01\rangle-i|10\rangle+i|11\rangle),\\
|\psi_{ax,2}\rangle
&= \frac12 (|00\rangle-|01\rangle+i|10\rangle+i|11\rangle).
\end{aligned}
\end{equation}
The corresponding projector is
\begin{equation}
    \begin{aligned}
P^{ax}&= \frac{5}{8} I I II+\frac{3}{8} I X Y I+\frac{1}{8} Y Y X X-\frac{1}{8} Y Z Z X+\frac{1}{8} X Y X Y 
 -\frac{1}{8} X Z Z Y+\frac{1}{8} Z I I Z-\frac{1}{8} Z X Y Z\\
 &=\frac{1+I X Y I}{2}+\frac{1-I X Y I}{2} \frac{1+Y Y X X}{2} \frac{1+X Y X Y}{2} .
\end{aligned}
\end{equation}

\subsubsection{Unbroken symmetry $H=\Z_2^{ax}\times \Z_2^{a^3x}$ (special dimer state)}

Finally, we discuss the special dimer phase, for which the unbroken subgroup is
$H=\Z_2^{ax}\times \Z_2^{a^3x}=\langle ax,a^3x\rangle$.  A representative
ground state can be chosen as an even-bond dimer product state
\begin{equation}
\begin{aligned}
|\Omega_{\mathrm{dimer}}\rangle
&=
\bigotimes_j
|\psi_{++}\rangle_{2j,2j+1},\\
|\psi_{++}\rangle
&=
\frac{1}{\sqrt{2}}(|00\rangle+i|11\rangle),
\qquad
(X\otimes Y)|\psi_{++}\rangle
=(Y\otimes X)|\psi_{++}\rangle
=|\psi_{++}\rangle .
\end{aligned}
\end{equation}
On each dimer bond, $|\psi_{++}\rangle$ is invariant under both $X\otimes Y$
and $Y\otimes X$.  The full chain therefore preserves $ax=U_{XY}$,
$a^3x=U_{YX}$, and $a^2=U_{XY}U_{YX}$.  It breaks the one-site translation
$x=T$, because the representative chooses the even-bond dimerization pattern.
Thus
$\mathrm{Stab}_{D_8}(|\Omega_{\mathrm{dimer}}\rangle)
=\langle a^2,ax\rangle=\langle ax,a^3x\rangle$, and the $D_8$ orbit contains
$\mathrm{GSD}=|D_8/\langle a^2,ax\rangle|=2$ ground states.  These are the
even- and odd-bond dimer patterns
$|\Omega_{\mathrm{dimer}}\rangle$ and $T|\Omega_{\mathrm{dimer}}\rangle$.

A convenient $D_8$-symmetric parent Hamiltonian is obtained by
projecting locally onto the allowed three-site dimer patterns.  We take
\begin{equation}
    \hat H_{\mathrm{dimer}} = -\sum_j P^{\langle ax,a^3x\rangle}_{j,j+1,j+2}.
\end{equation}
Here $P^{\langle ax,a^3x\rangle}$ is the projector onto the four-dimensional
subspace
\begin{equation}
\mathrm{span}\,\{
|\psi_{++}\rangle|0\rangle,
|\psi_{++}\rangle|1\rangle,
|0\rangle|\psi_{++}\rangle,
|1\rangle|\psi_{++}\rangle
\}.
\end{equation}
An explicit expression for this projector is
\begin{equation}
    P^{\langle ax,a^3x \rangle}
    =\frac{1}{2} I I I
    +\frac{1}{6}\!\left[
    I(X Y+Y X+Z Z)
    -(X I X+Y I Y+Z I Z)
    +(X Y+Y X+Z Z) I
    \right].
\end{equation}
Equivalently, after shifting and rescaling the local projector, the parent
Hamiltonian can be written as
\begin{align}\label{H2}
\hat H_\mathrm{dimer} = - \sum_j 3\left(P^{\langle ax,a^3x \rangle}_{j-1,j,j+1} - \frac{1}{2}\right)
= \sum_j \left[ -(X_j Y_{j+1} + Y_j X_{j+1} + Z_j Z_{j+1}) + \frac{1}{2}(X_{j-1} X_{j+1} + Y_{j-1} Y_{j+1} + Z_{j-1} Z_{j+1}) \right].
\end{align}

This Hamiltonian preserves the full symmetry group $G$.  Its two ground states
spontaneously break translation symmetry from $\mathbb{Z}_\mathrm{trans}$ to
$2\mathbb{Z}_\mathrm{trans}$ while preserving the full internal symmetry group
$\mathbb{Z}_2^{XY} \times \mathbb{Z}_2^{YX}$.  In terms of $D_8$ symmetry, the
remaining symmetry is the subgroup $\Z_2^{ax}\times\Z_2^{a^3x}$ of $D_8$.
Therefore this gapped phase corresponds to the module category
$(\mathrm{Vec}_{D_8}^{2\zeta})_{A_\mathrm{dimer}}$ with algebra
$A_\mathrm{dimer}=\mathbb C[\Z_2^{ax}\times\Z_2^{a^3x}]$.

\subsection{The dimer-to-ferromagnet DQCP}

We now construct an explicit spin-chain realization of a direct transition between the special dimer phase and the $z$FM phase identified above. The model below provides a concrete DQCP candidate for this dimer-to-ferromagnet transition. After a simple sublattice unitary transformation, it becomes a nearest-neighbor-$z$-anisotropic Majumdar-Ghosh chain.

\subsubsection{Interpolation between parent Hamiltonians}

With this motivation, we interpolate between the dimer and zFM parent Hamiltonians \eq{H2} and \eq{H1}:
\begin{align}
\hat H_{z\mathrm{MG}}(\lambda)
&=\hat H_\mathrm{dimer}+\lambda \hat H_{z\mathrm{FM}}\nonumber\\
&=-\sum_i \left[ X_iY_{i+1}+Y_iX_{i+1}+(1+\lambda)Z_iZ_{i+1} \right]
+\frac{1}{2}\sum_i
\left[
X_{i-1}X_{i+1}+Y_{i-1}Y_{i+1}+Z_{i-1}Z_{i+1}
\right].
\label{H}
\end{align}
The endpoint $\lambda=0$ is the special dimer parent Hamiltonian, while the large positive $\lambda$ limit is governed by the $z$FM Ising interaction. Since both terms preserve the full microscopic symmetry $G$, the entire interpolation preserves $G$.

These two phases preserve different order-four stabilizer subgroups of the same anomalous $D_8$ symmetry:
\begin{equation}
H_{\mathrm{dimer}}
=\langle ax,a^3x\rangle
=\langle U_{XY},U_{YX}\rangle,
\qquad
H_{z\mathrm{FM}}
=\langle a^2,x\rangle
=\langle U_{XY}U_{YX},T\rangle.
\end{equation}
They share the unbroken symmetry $a^2=U_{XY}U_{YX} = \prod_i Z_i$, but neither unbroken subgroup is contained in the other. Thus a direct transition is not naturally described by a conventional Landau theory based on a single order parameter.

\subsubsection{Anisotropic Majumdar-Ghosh model}

It is useful to perform a unitary transformation on the odd sites,
\begin{equation}
    U_{\mathrm{MG}}=\prod_j \frac{1}{\sqrt{2}}\left(X_{2j+1}-Y_{2j+1}\right).
\end{equation}
Its action on local operators is
\begin{align}
U_\mathrm{MG}:
\begin{cases}
X_{2i+1}\rightarrow -Y_{2i+1},\\
Y_{2i+1}\rightarrow -X_{2i+1},\\
Z_{2i+1}\rightarrow -Z_{2i+1},\\
\sigma_{2i}^{\alpha}\rightarrow \sigma_{2i}^{\alpha},\quad \alpha=x,y,z .
\end{cases}
\end{align}
Under this transformation, the dimer parent Hamiltonian \eq{H2} is mapped to the Majumdar-Ghosh (MG) Hamiltonian \cite{MajumdarGhosh1969I,MajumdarGhosh1969II}
\begin{align}
\tilde H_\mathrm{dimer}
=U_{\mathrm{MG}}\hat H_\mathrm{dimer}U_{\mathrm{MG}}^\dagger
=\sum_i \left[
\vec{\sigma}_i\cdot\vec{\sigma}_{i+1}
+\frac{1}{2}\vec{\sigma}_i\cdot\vec{\sigma}_{i+2}
\right],
\end{align}
and the special dimer state is mapped to the standard singlet dimer state. The $z$FM parent Hamiltonian \eq{H1} is mapped to the antiferromagnetic Ising interaction
\begin{align}
\tilde H_{z\mathrm{FM}}
&=U_{\mathrm{MG}}\hat H_{z\mathrm{FM}}U_{\mathrm{MG}}^\dagger
=\sum_i Z_iZ_{i+1},
\label{H1tilde}
\end{align}
with antiferromagnetic N\'eel ground state. Therefore the interpolating Hamiltonian becomes
\begin{align}\nonumber
\tilde H_{z\mathrm{MG}}(\lambda)
&=U_{\mathrm{MG}}\hat H_{z\mathrm{MG}}(\lambda)U_{\mathrm{MG}}^\dagger\\
&=\sum_i
\left[
X_iX_{i+1}+Y_iY_{i+1}+(1+\lambda)Z_iZ_{i+1}
\right]
+\frac{1}{2}\sum_i
\left[
X_{i-1}X_{i+1}+Y_{i-1}Y_{i+1}+Z_{i-1}Z_{i+1}
\right].
\label{Htilde}
\end{align}
Thus $\lambda$ adds an easy-axis anisotropy only to the nearest-neighbor part of the MG Hamiltonian. At $\lambda=0$, \eq{Htilde} is the MG model; as $\lambda\rightarrow\infty$, it approaches the antiferromagnetic Ising limit \eq{H1tilde}.

In the transformed basis, the symmetry generators become
\begin{align}
\tilde U_{XY}
&=U_{\mathrm{MG}}U_{XY}U_{\mathrm{MG}}^\dagger
=(-1)^{L/2}\prod_i X_i,\nonumber\\
\tilde U_{YX}
&=U_{\mathrm{MG}}U_{YX}U_{\mathrm{MG}}^\dagger
=(-1)^{L/2}\prod_i Y_i,\nonumber\\
\tilde T
&=U_{\mathrm{MG}}TU_{\mathrm{MG}}^\dagger
=\left[\prod_j \frac{1}{\sqrt{2}}\left(X_j-Y_j\right)\right]T .
\end{align}
The first two are the usual $\pi$ rotations about the $x$ and $y$ axes, whereas translation is dressed by an on-site rotation. This dressed translation is the remnant of the original anomalous lattice symmetry and is essential for identifying the transition in \eq{H} as the original dimer-to-$z$FM transition rather than an ordinary dimer-to-N\'eel transition with conventional translation symmetry.

\subsubsection{Symmetry-breaking pattern and dual interpretation}

These two phases preserve different order-four stabilizer subgroups of the same anomalous $D_8$ symmetry:
\begin{equation}
H_{\mathrm{dimer}}
=\langle ax,a^3x\rangle
=\langle U_{XY},U_{YX}\rangle,
\qquad
H_{z\mathrm{FM}}
=\langle a^2,x\rangle
=\langle U_{XY}U_{YX},T\rangle.
\end{equation}
They share the unbroken symmetry $a^2=U_{XY}U_{YX}=\prod_i Z_i$, but neither unbroken subgroup is contained in the other. Thus a direct transition is not naturally described by a conventional Landau theory based on a single order parameter. In the transformed MG representation, this same obstruction appears through the nontrivial action of $\tilde T$: the N\'eel order parameter of the anisotropic MG chain is the image of the original $z$FM order parameter, while the dimer order parameter remains tied to translation breaking. The numerical results below therefore test whether the direct interpolation realizes a continuous deconfined transition rather than a first-order transition or an intervening phase.

This interpretation is also reflected in the gauged description. As shown later in Sec.~\ref{sec:latticegauging}, gauging the symmetry $\Z_2^{XY}=\langle ax\rangle$ maps the one-parameter Hamiltonian to a dual spin chain with non-invertible $\Rep(H_8)$ symmetry. Under this gauging, the dimer phase maps to a partially symmetry-broken gapped phase with unbroken $\Z_2^{\eta_1}$ (or $\Z_2^{\eta_2}$ for the translated dimer), while the $z$FM phase maps to the fully $\Rep(H_8)$-symmetric gapped phase. Therefore the dimer-to-ferromagnet DQCP candidate is dual to a transition between partial symmetry breaking and full symmetry restoration for the non-invertible $\Rep(H_8)$ symmetry.

\subsubsection{Numerical evidence}

To further investigate the DQCP candidate described above, we employ the variational uniform matrix product state (VUMPS) method \cite{PhysRevB.97.045145} in the thermodynamic limit. For each bond dimension $\chi$, we perform two independent VUMPS optimizations starting from the two sides of the transition, representing the two competing MPS ansatz. On the dimer side, the initial state is chosen to be the exact ground state of $\hat H_{\mathrm{dimer}}$. On the $z$FM side, the initial state is chosen to be the exact ground state of $\hat H_{z\mathrm{FM}}$. 
We then sweep the tuning parameter $\lambda$ toward the transition from both sides. At each value of $\lambda$, the converged MPS from the previous parameter point is used as the initial state for the next optimization. For a fixed bond dimension $\chi$, we determine the pseudocritical point $\lambda_c(\chi)$ by comparing the variational energies of these two branches\cite{PhysRevB.99.165143}. This provides a stable operational definition of $\lambda_c(\chi)$ for the subsequent finite-entanglement scaling analysis.

\begin{figure}[htbp]
\centering    
\includegraphics[width=0.45\textwidth]{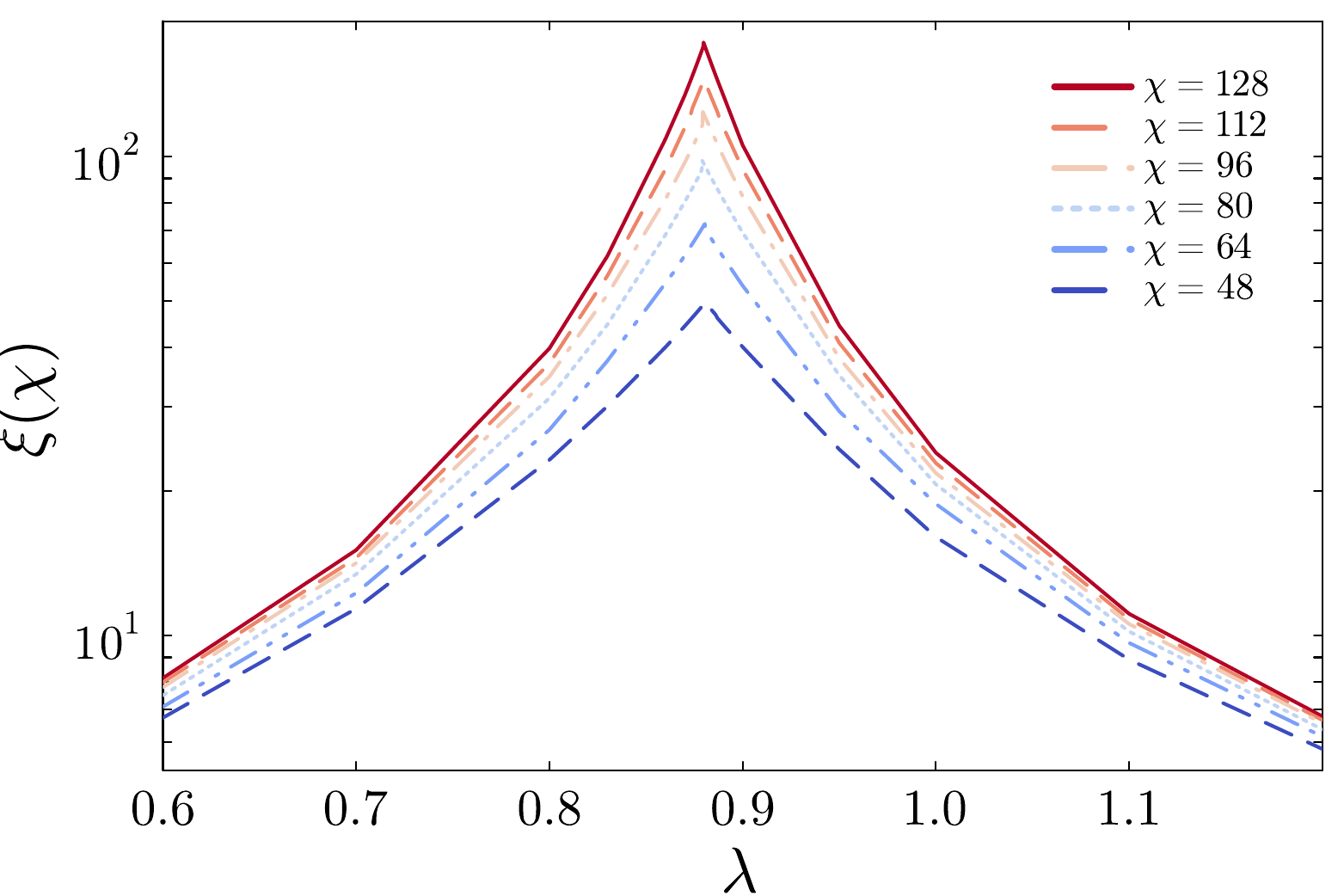}
\caption{The peak of the MPS correlation length $\xi(\chi)$ grows with increasing bond dimension $\chi$, indicating a continuous transition between the dimer phase near $\lambda=0$ and the $z$FM phase at large positive $\lambda$. As shown in Fig.~\ref{fig:corscaling}, the peak height follows a power-law scaling with $\chi$.}
\label{fig:cor}
\end{figure}

At a continuous phase transition, the correlation length $\xi(\chi)$ and the entanglement entropy $S(\chi)$ of an MPS with bond dimension $\chi$ at $\lambda_c(\chi)$ diverge in the $\chi\rightarrow \infty$ limit, and $\xi(\chi)$ should exhibit a peak with $\chi$-dependent height, as shown in Fig.~\ref{fig:cor}. The pronounced correlation-length peak and its growth with bond dimension provide evidence for a continuous phase transition.

\begin{figure}[htbp]
    \centering
    \begin{subfigure}{0.45\textwidth}
        \centering
        \includegraphics[width=\linewidth]{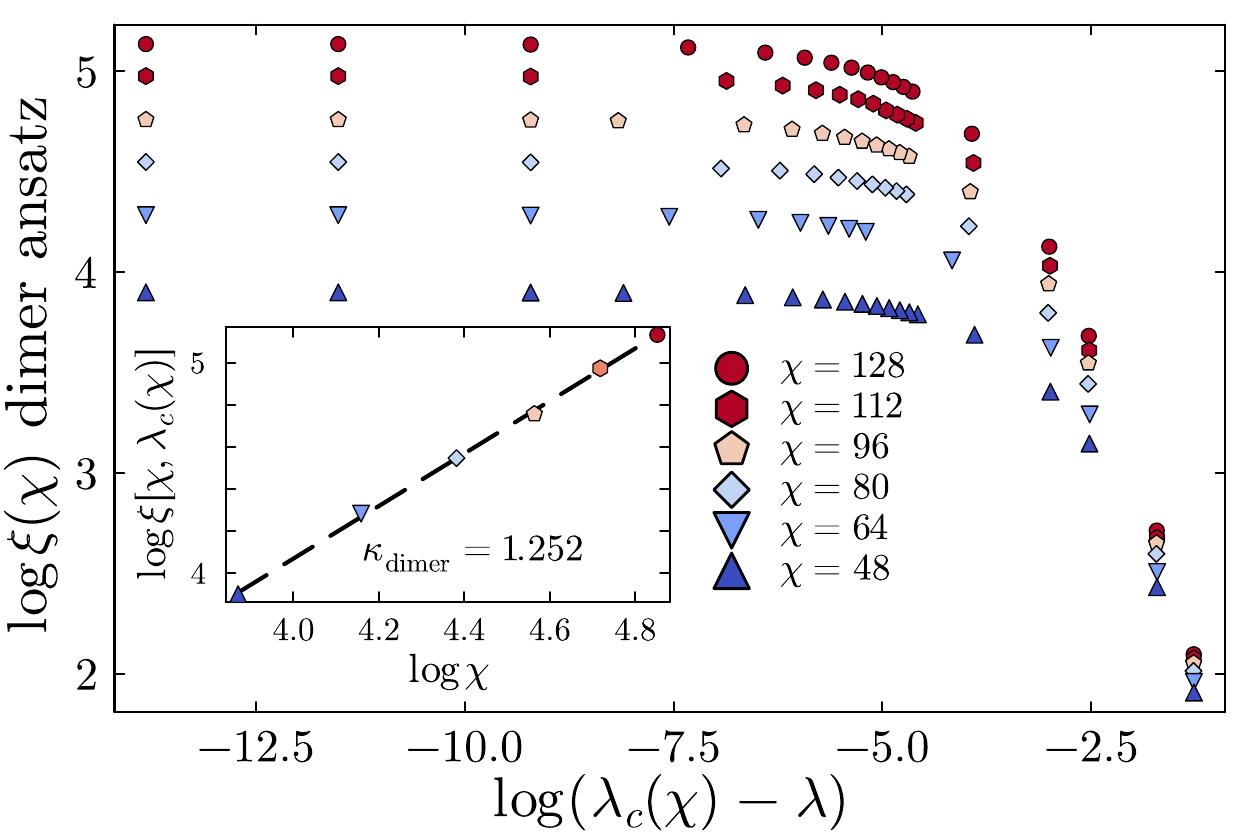}
    \end{subfigure}
    \begin{subfigure}{0.45\textwidth}
        \centering
        \includegraphics[width=\linewidth]{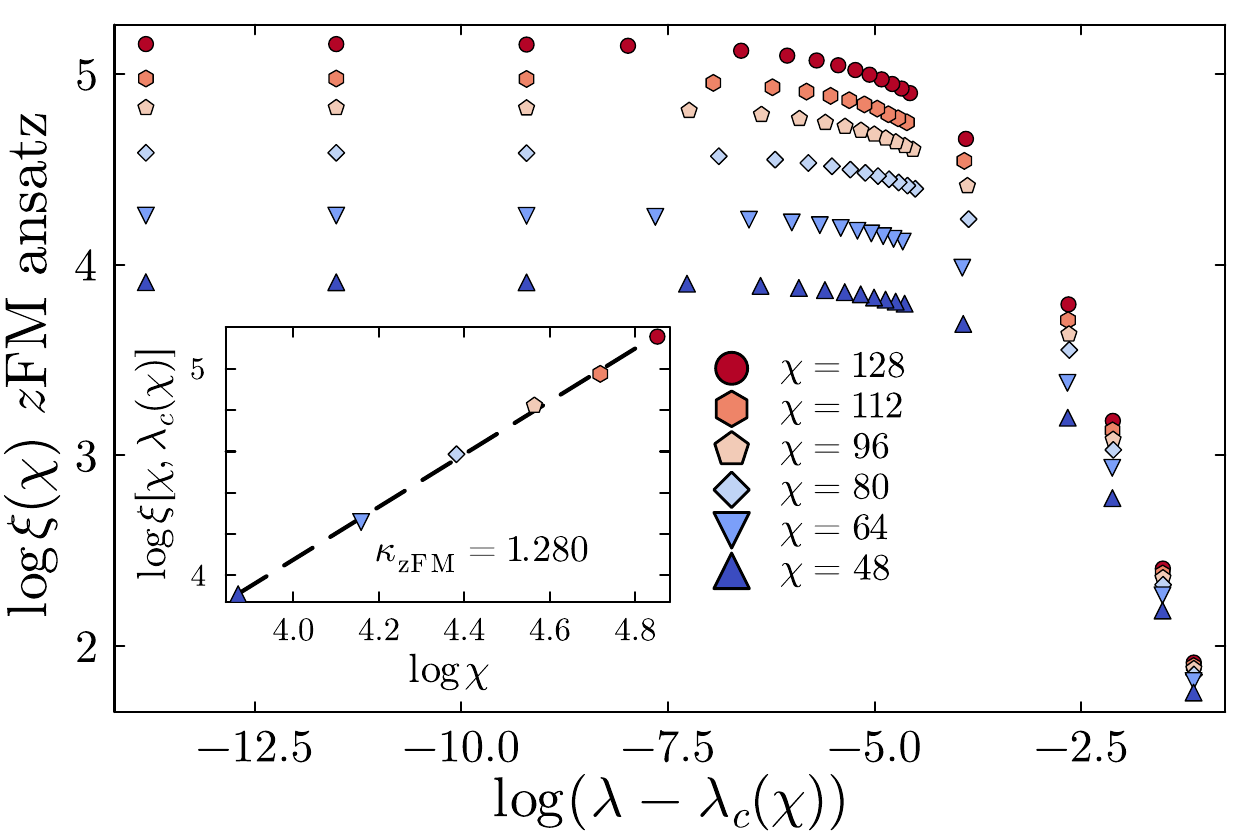}
    \end{subfigure}
    \caption{The MPS correlation length $\xi(\chi)$ exhibits power-law scaling with $\chi$ in the critical regime. In the insets, the peak value $\xi[\chi,\lambda_c(\chi)]$ is plotted as a function of $\chi$, from which the finite-entanglement exponent $\kappa$ is extracted. The estimate from the dimer ansatz is shown in the left panel, while that from the $z$FM ansatz is shown in the right panel.}
    \label{fig:corscaling}
\end{figure}

To substantiate the critical nature of this transition, we turn to finite-entanglement scaling analysis for a more quantitative characterization. For each bond dimension $\chi$, we determine the peak height of the MPS correlation length $\xi(\chi)$ and the entanglement entropy $S(\chi)$ from their values at the pseudocritical point $\lambda_c(\chi)$. According to finite-entanglement scaling theory \cite{PhysRevLett.102.255701}, in a critical regime these peak values scale as
\begin{align}
    \xi(\chi) &\sim \chi^\kappa,\\\label{eq:entropyscaling}
    S(\chi) &\sim \frac{c}{6}\log \xi(\chi),
\end{align}
where the exponent $\kappa$ depends on the central charge $c$ of the underlying conformal field theory.
As shown in Fig.~\ref{fig:corscaling}, we find that $\xi(\chi)$ saturates to its finite peak value $\xi[\chi,\lambda_c(\chi)]$ as $|\lambda-\lambda_c(\chi)|\rightarrow0$ because of the finite-entanglement cutoff effect. In the insets of Fig.~\ref{fig:corscaling}, we plot $\xi[\chi,\lambda_c(\chi)]$ as a function of $\chi$ on a log-log scale for both ansatzes. The data are well described by a power law in $\chi$, yielding the finite-entanglement exponents $\kappa_{\mathrm{dimer}} = 1.252$ and $\kappa_{z\mathrm{FM}} = 1.280$. The divergent trend further supports a continuous phase transition.

\begin{figure}[htbp]
\centering    
\includegraphics[width=0.45\textwidth]{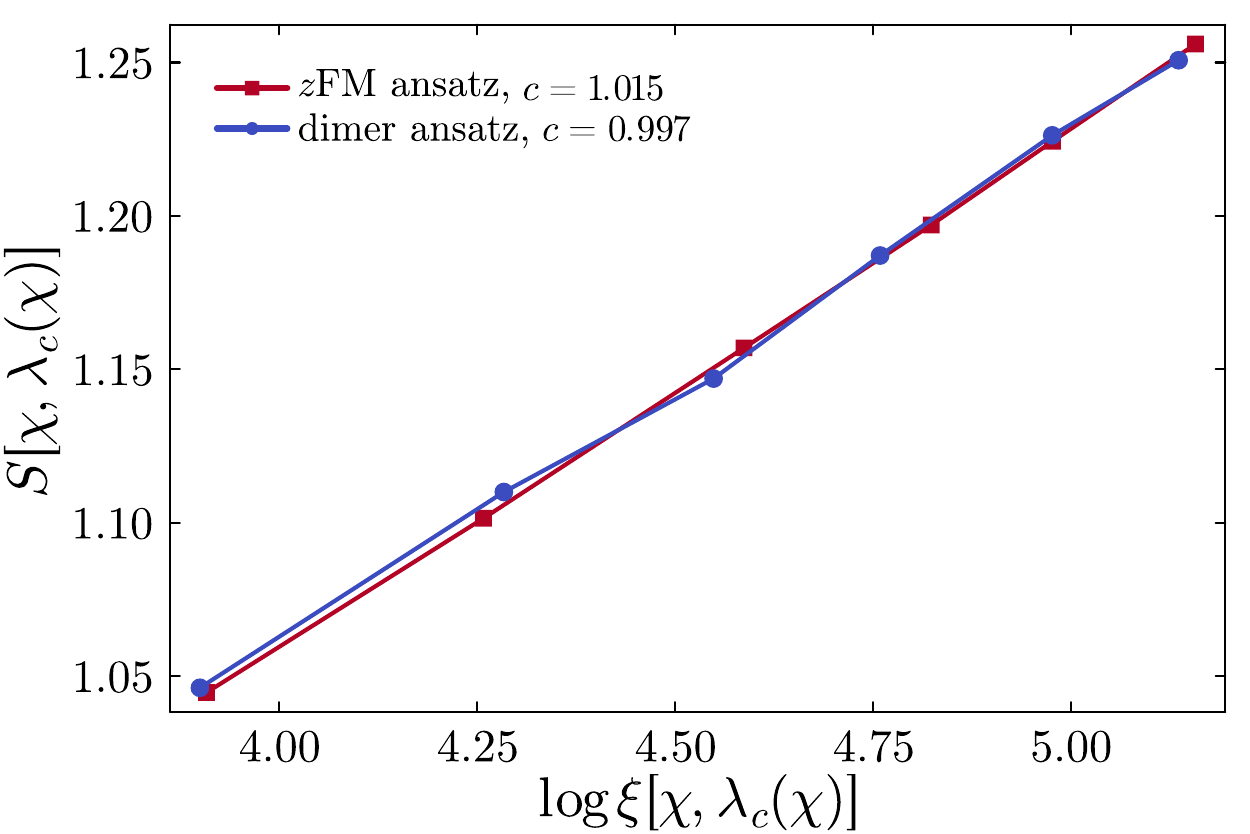}
\caption{The MPS entanglement entropy $S(\chi)$ is plotted as a function of the logarithm of the MPS correlation length $\xi(\chi)$ at the pseudocritical points $\lambda_c(\chi)$. The linear dependence is consistent with entropy scaling $S(\chi)\sim \frac{c}{6}\log\xi(\chi)$, yielding the central charge $c \approx 1$.}
\label{fig:entropyscaling}
\end{figure}

While the fitted value of $\kappa$ is broadly consistent with $\kappa=1.344$ expected for $c=1$, we do not use it as a precise determination of $c$, since $\kappa$ is sensitive to finite-$\chi$ corrections. Instead, we extract $c$ directly from the entropy scaling in Eq.~\ref{eq:entropyscaling}. Figure~\ref{fig:entropyscaling} shows $S(\chi)$ plotted as a function of $\log\xi(\chi)$ at the pseudocritical points $\lambda_c(\chi)$ for different bond dimensions. The resulting fits from the two ansatzes give $c_{z\mathrm{FM}} = 1.015$ and $c_{\mathrm{dimer}} = 0.997$, consistent with a $c=1$ critical theory. Together with the power-law growth of the correlation length, this entanglement scaling gives strong evidence that the transition is continuous and governed by a conformal field theory.

\subsection{Category theory of subgroup gauging, dual symmetries, and gapped-phase correspondences}

In this section, we use the method described in Appendix~\ref{App:A2} to determine the correspondence between the gapped phases of the original system and those of the theory obtained by generalized gauging. In Appendix~\ref{App:D8ACA}, we classify the condensable algebras in $\mathrm{Vec}_{D_8}^{\omega_{\mathrm{LSM}}=2\zeta}$, which are labeled by certain subgroups of $D_8$. We now work out this correspondence for the gaugings associated with these subgroups.

Physically, the computation should be read as a change of duality frame for the same infrared phases. A condensable algebra specifies which collection of topological symmetry lines can be absorbed by a given gapped phase. Generalized gauging by an algebra $A$ inserts an $A$-condensation interface, and the internal-Hom construction asks how the same phase is described on the other side of that interface. Thus the tables below track how order variables, disorder variables, and condensed symmetry defects are relabeled after gauging.

\begin{center}
\begin{table}[h]
        \centering
        
\begin{tabular}{|c|c|}
\hline 
 \multicolumn{2}{|c|}{$\displaystyle ( G,\omega_3) =( D_{8} ,\omega_\mathrm{LSM}=2\zeta )$} \\
\hline \hline 
 $\displaystyle A=( H,\gamma )$ & $\displaystyle _{A}\mathcal{C}_{A}$ \\
\hline 
 $\displaystyle ( 1,0)$ & $\displaystyle \mathrm{Vec}_{D_{8}}^{2\zeta }$ \\
\hline 
 $\displaystyle ( \langle x\rangle ,1)$ $\overset{\mathrm{M}}{\sim}$ $ $$\displaystyle \left( \langle xa^{2} \rangle =\mathbb{Z}_{2} ,0\right)$ & $\displaystyle \mathrm{Rep} H_{8}$ \\
\hline 
 $\displaystyle ( \langle xa\rangle ,1)$ $\overset{\mathrm{M}}{\sim}$ $ $$\displaystyle \left( \langle xa^{3} \rangle =\mathbb{Z}_{2} ,0\right)$ & $\displaystyle \mathrm{Rep} H_{8}$ \\
\hline 
 $\displaystyle \left( \langle a^{2} \rangle =\mathbb{Z}_{2} ,0\right)$  & $\displaystyle \mathrm{Vec}_{D_{8}}^{2\zeta }$ \\
\hline 
 $\displaystyle \left( \langle x,a^{2} \rangle =\mathbb{Z}_{2} \times \mathbb{Z}_{2} ,0\right)\overset{\mathrm{M}}{\sim} \left( \langle x,a^{2} \rangle =\mathbb{Z}_{2} \times \mathbb{Z}_{2} ,\gamma\right)$ & $\displaystyle \mathrm{Rep} H_{8}$ \\
\hline 
 $\displaystyle \left( \langle xa,a^{2} \rangle =\mathbb{Z}_{2} \times \mathbb{Z}_{2} ,0\right) \overset{\mathrm{M}}{\sim} \left( \langle xa,a^{2} \rangle =\mathbb{Z}_{2} \times \mathbb{Z}_{2} ,\gamma\right)$  & $\displaystyle \mathrm{Rep} H_{8}$ \\
\hline 
\end{tabular}
\caption{Morita equivalence classes of condensable algebras in $\mathrm{Vec}_{D_8}^{2\zeta}$ and the corresponding dual categories}
\label{Table:D82zeta}
        \end{table}
\end{center}

The two possible dual categories in Table~\ref{Table:D82zeta} have distinct physical meanings. When the dual category is again $\mathrm{Vec}_{D_8}^{2\zeta}$, all simple symmetry defects remain invertible, so the gauging acts as a generalized Kramers-Wannier transformation inside a pointed symmetry theory. When the dual category is $\mathrm{Rep}(H_8)$, the gauged theory contains a genuinely non-invertible defect, represented by the two-dimensional simple object $X_0$. In the lattice systems discussed in this paper, this non-invertible sector is the dual-symmetry remnant of the projective LSM data carried by a unit cell.

\subsubsection{Subgroup gauging in $\mathrm{Vec}_{D_8}^{\omega_\mathrm{LSM}}$}

As shown in Appendix~\ref{App:D8ACA}, the Morita equivalence classes of condensable algebras in $\mathrm{Vec}_{D_8}^{2\zeta}$ and their corresponding dual categories are summarized in Table~\ref{Table:D82zeta}. Since the condensable algebras of $\mathrm{Vec}_{D_8}^{2\zeta}$ are labeled by subgroups of $D_8$, we denote these algebras directly by the corresponding subgroups.

First consider the condensable algebra $A$ labeled by the subgroup $H=\langle xa,a^2\rangle$. Then ${}_{H}(\mathrm{Vec}_{D_8}^{2\zeta})_{H} \simeq \mathrm{Rep}(H_8)$. Following Appendix~\ref{App:method}, the simple objects in ${}_{H}(\mathrm{Vec}_{D_8}^{2\zeta})_{H}$ can be represented explicitly as
$$
Y_1:=(H1H,\rho^H_{\mathrm{1,1,1,1}}), \  
Y_2:=(H1H,\rho^H_{\mathrm{1,1,-1,-1}}), \ 
Y_3:=(H1H,\rho^H_{\mathrm{1,-1,1,-1}}), \ 
Y_4:=(H1H,\rho^H_{\mathrm{1,-1,-1,1}}), \ 
Y_0:=(Ha^2H,\rho^H_{\mathrm{proj}})
$$
where $\rho^H_{a,b,c,d}$ denotes the representation of $H$ whose character takes the values $a, b, c$, and $d$ on the elements $1, a, xa$, and $xa^3$, respectively. The symbol $\rho^H_{\mathrm{proj}}$ denotes the irreducible projective representation of $H$. Thus the first four objects are one-dimensional, whereas the last one is two-dimensional. 

We denote the simple objects of $\mathrm{Rep}(H_8)$ by $X_1, X_2, X_3, X_4$, and $X_0$. The first four objects are one-dimensional, with $X_1$ serving as the tensor unit, while $X_0$ is two-dimensional. The category $\mathrm{Rep}(H_8)$ has a nontrivial tensor auto-equivalence \cite{tambara2000representations} that permutes two of the one-dimensional objects; we choose $X_3$ and $X_4$ to be the two objects exchanged by this auto-equivalence. According to Ref.~\cite{etingof2021tensor}, the condensable algebras in $\mathrm{Rep}(H_8)$ are classified as follows:
$$
X_1, \quad X_1 \oplus X_2, \quad X_1 \oplus X_3 \stackrel{\text{Morita}}{\sim} X_1 \oplus X_4, \quad X_1 \oplus X_2 \oplus X_0, \quad X_{1} \oplus X_{2} \oplus X_{3} \oplus X_{4}, \quad X_{1} \oplus X_{2} \oplus X_{3} \oplus X_{4} \oplus X_{0}^{\oplus 2}.
$$

We fix the equivalence ${}_{H}(\mathrm{Vec}_{D_8}^{2\zeta})_{H} \simeq \mathrm{Rep}(H_8)$ by sending $Y_i$ to $X_i$ for $i = 0, 1, 2, 3, 4$. A different choice of equivalence would permute $X_3$ and $X_4$. However, this tensor auto-equivalence preserves the Morita classes of condensable algebras in $\mathrm{Rep}(H_8)$, and therefore does not change the resulting correspondence between gapped phases.

The following internal-Hom calculation has a direct physical interpretation. The object $M$ represents the gauging interface with the original gapped phase, labeled by $L$, attached to one side. Acting on $M$ by a dual defect $Y_i$ tests whether that defect can end on, or be absorbed by, the gauged phase without changing its infrared boundary condition. The summands of $[M,M]$ are precisely the dual symmetry defects that are condensed in the image phase.

For each subgroup $L \le D_8$ satisfying $2\zeta|_L = 1 \in H^3(L,U(1))$, we follow the computational framework of Ref.~\cite{etingof2021tensor} and choose the object $M \in {}_{H}(\mathrm{Vec}_{D_8}^{2\zeta})_{L}$ represented by $(H1L,\rho^{H\cap L}_{\mathrm{triv}})$. We then compute the internal Hom $[M,M]$, which is a condensable algebra in ${}_{H}(\mathrm{Vec}_{D_8}^{2\zeta})_{H} \simeq \mathrm{Rep}(H_8)$; see Definition~\ref{internalhom}. It is determined by the natural isomorphism
$$
\mathrm{Hom}_{{}_{H}(\mathrm{Vec}_{D_8}^{2\zeta})_{H}}(Y_i,[M,M]) \simeq \mathrm{Hom}_{{}_{H}(\mathrm{Vec}_{D_8}^{2\zeta})_{L}}(Y_i\odot M,M).
$$
The results are listed in Table~\ref{Table:D8 gauge xa a2}. As an illustration, let us spell out the case $L=\langle xa\rangle$. The relevant double-coset decomposition is $H \backslash D_8 / L = \{H1L, HaL\}$. In this case $M=(H1L,\rho^L_{\mathrm{triv}})$. For $i=1,2,3,4$, the support of $Y_i\odot M$ is $H1H \otimes_H H1L \simeq H1L$. Requiring the action of $\langle xa\rangle$ to be trivial, we find that $Y_i\odot M \simeq M$ holds only for $i=1,3$. On the other hand, $Y_0\odot M$ is supported on $HaH \otimes_H H1L \simeq HaL$. Hence $[M,M] \simeq Y_1 \oplus Y_3$, which corresponds to the condensable algebra $X_1 \oplus X_3$ in $\mathrm{Rep}(H_8)$.

\begin{table}[!h]
        \centering
\begin{tabular}{|c|c|}
\hline 
	 \multicolumn{2}{|c|}{Gauging $\displaystyle \langle xa,a^{2} \rangle$ in $\displaystyle \mathrm{Vec}_{D_8}^{2\zeta}$} \\
\hline \hline
 condensable algebras in $\displaystyle \mathrm{Vec}_{D_{8}}^{2\zeta }$ & condensable algebras in $\displaystyle \mathrm{Rep}( H_{8})$ \\
\hline 
 $\displaystyle 1$ & $\displaystyle X_{1} \oplus X_{2} \oplus X_{3} \oplus X_{4}$ \\
\hline 
 $\displaystyle \langle xa,a^{2} \rangle $ & $\displaystyle X_{1}$ \\
\hline 
 $\displaystyle \langle a^{2} \rangle $ & $\displaystyle X_{1} \oplus X_{2}$ \\
\hline 
 $\displaystyle \langle xa \rangle $ & $\displaystyle X_{1} \oplus X_{3}$ \\
\hline 
 $\displaystyle \langle x \rangle $ & $\displaystyle X_{1} \oplus X_{2} \oplus X_{3} \oplus X_{4} \oplus X_{0}^{\oplus 2}$ \\
\hline 
 $\displaystyle \langle x,a^{2} \rangle $ & $\displaystyle X_{1} \oplus X_{2} \oplus X_{0}$ \\
 \hline
\end{tabular}
\caption{Correspondence between gapped phases of $\mathrm{Vec}_{D_8}^{2\zeta}$ and $\mathrm{Rep}(H_8)$ after gauging the subgroup $\langle ax, a^2\rangle$}
\label{Table:D8 gauge xa a2}
        \end{table}

Next consider the condensable algebra $A$ labeled by the subgroup $H=\langle ax\rangle$. Then ${}_{\langle ax\rangle}(\mathrm{Vec}_{D_8}^{2\zeta})_{\langle ax\rangle} \simeq \mathrm{Rep}(H_8)$. Following Appendix~\ref{App:method}, the simple objects in ${}_{\langle ax\rangle}(\mathrm{Vec}_{D_8}^{2\zeta})_{\langle ax\rangle}$ can be represented as
$$
Z_1:=(H1H,\rho^H_{\mathrm{triv}}), \quad 
Z_2:=(H1H,\rho^H_{\mathrm{sign}}), \quad
Z_3:=(Ha^2H,\rho^H_{\mathrm{triv}}), \quad
Z_4:=(Ha^2H,\rho^H_{\mathrm{sign}}), \quad
Z_0:=(HxH,\rho^1_{\mathrm{triv}})
$$
where $\rho^H_{\mathrm{triv}}$ and $\rho^H_{\mathrm{sign}}$ denote the trivial and sign representations of $H$, respectively. The first four objects are one-dimensional, whereas the last one is two-dimensional. We fix the equivalence ${}_{\langle ax\rangle}(\mathrm{Vec}_{D_8}^{2\zeta})_{\langle ax\rangle} \simeq \mathrm{Rep}(H_8)$ by sending $Z_i$ to $X_i$ for $i=0,1,2,3,4$.

For each subgroup $L\leqslant D_8$ satisfying $2\zeta|_L = 1 \in H^3(L,U(1))$, we again choose the object $M \in {}_{H}(\mathrm{Vec}_{D_8}^{2\zeta})_{L}$ represented by $(H1L,\rho^{H\cap L}_{\mathrm{triv}})$. The resulting internal Homs $[M,M]$, for all allowed choices of $L$, are summarized in Table~\ref{Table:D8 gauge xa}. These computations also appear in Ref.~\cite{etingof2021tensor}.

\begin{table}[!h]
        \centering
\begin{tabular}{|c|c|}
\hline 
 \multicolumn{2}{|c|}{
Gauging $\displaystyle \langle xa\rangle$ in $\displaystyle \mathrm{Vec}_{D_8}^{2\zeta}$
} \\
\hline \hline
 condensable algebras in $\displaystyle \mathrm{Vec}_{D_{8}}^{2\zeta }$ & condensable algebras in $\displaystyle \mathrm{Rep}( H_{8})$ \\
\hline 
 $\displaystyle 1$ & $\displaystyle X_{1} \oplus X_{2}$ \\
\hline 
 $\displaystyle \langle xa,a^{2} \rangle $ & $\displaystyle X_{1} \oplus X_{3}$ \\
\hline 
 $\displaystyle \langle a^{2} \rangle $ & $\displaystyle X_{1} \oplus X_{2} \oplus X_{3} \oplus X_{4}$ \\
\hline 
 $\displaystyle \langle xa \rangle $ & $\displaystyle X_{1}$ \\
\hline 
 $\displaystyle \langle x \rangle $ & $\displaystyle X_{1} \oplus X_{2} \oplus X_{0}$ \\
\hline 
 $\displaystyle \langle x,a^{2} \rangle $ & $\displaystyle X_{1} \oplus X_{2} \oplus X_{3} \oplus X_{4} \oplus X_{0}^{\oplus 2}$ \\
 \hline
\end{tabular}
\caption{Correspondence between gapped phases of $\mathrm{Vec}_{D_8}^{2\zeta}$ and $\mathrm{Rep}(H_8)$ after gauging the subgroup $\langle xa\rangle$}
\label{Table:D8 gauge xa}
        \end{table}

The cases $H=\langle x \rangle$ and $H=\langle x,a^2\rangle$ are listed in Tables~\ref{Table:D8 gauge x a2} and \ref{Table:D8 gauge x}. They can be obtained from Tables~\ref{Table:D8 gauge xa a2} and \ref{Table:D8 gauge xa} by applying an outer automorphism of $\mathrm{Vec}_{D_8}^{2\zeta}$. The underlying group automorphism is $\phi: D_8 \to D_8$, with $\phi(x) = xa$ and $\phi(a) = a$.

\begin{table}[!h]
        \centering
        
\begin{tabular}{|c|c|}
\hline 
	 \multicolumn{2}{|c|}{Gauging $\displaystyle \langle x,a^{2} \rangle$ in $\displaystyle \mathrm{Vec}_{D_8}^{2\zeta}$} \\
\hline \hline
 condensable algebras in $\displaystyle \mathrm{Vec}_{D_{8}}^{2\zeta }$ & condensable algebras in $\displaystyle \mathrm{Rep}( H_{8})$ \\
\hline 
 $\displaystyle 1$ & $\displaystyle X_{1} \oplus X_{2} \oplus X_{3} \oplus X_{4}$ \\
\hline 
 $\displaystyle \langle xa,a^{2} \rangle $ & $\displaystyle X_{1} \oplus X_{2} \oplus X_{0}$ \\
\hline 
 $\displaystyle \langle a^{2} \rangle $ & $\displaystyle X_{1} \oplus X_{2}$ \\
\hline 
 $\displaystyle \langle xa \rangle $ & $\displaystyle X_{1} \oplus X_{2} \oplus X_{3} \oplus X_{4} \oplus X_{0}^{\oplus 2}$ \\
\hline 
 $\displaystyle \langle x \rangle $ & $\displaystyle X_{1} \oplus X_{3}$ \\
\hline 
 $\displaystyle \langle x,a^{2} \rangle $ & $\displaystyle X_{1}$ \\
 \hline
\end{tabular}
\caption{Correspondence between gapped phases of $\mathrm{Vec}_{D_8}^{2\zeta}$ and $\mathrm{Rep}(H_8)$ after gauging the subgroup $\langle x,a^2\rangle$}
\label{Table:D8 gauge x a2}
        \end{table}

\begin{table}[!h]
        \centering
        
\begin{tabular}{|c|c|}
\hline 
 \multicolumn{2}{|c|}{Gauging $\displaystyle \langle x\rangle$ in $\displaystyle \mathrm{Vec}_{D_8}^{2\zeta}$} \\
\hline \hline
 condensable algebras in $\displaystyle \mathrm{Vec}_{D_{8}}^{2\zeta }$ & condensable algebras in $\displaystyle \mathrm{Rep}( H_{8})$ \\
\hline 
 $\displaystyle 1$ & $\displaystyle X_{1} \oplus X_{2}$ \\
\hline 
 $\displaystyle \langle xa,a^{2} \rangle $ & $\displaystyle X_{1} \oplus X_{2} \oplus X_{3} \oplus X_{4} \oplus X_{0}^{\oplus 2}$ \\
\hline 
 $\displaystyle \langle a^{2} \rangle $ & $\displaystyle X_{1} \oplus X_{2} \oplus X_{3} \oplus X_{4}$ \\
\hline 
 $\displaystyle \langle xa \rangle $ & $\displaystyle X_{1} \oplus X_{2} \oplus X_{0}$ \\
\hline 
 $\displaystyle \langle x \rangle $ & $\displaystyle X_{1}$ \\
\hline 
 $\displaystyle \langle x,a^{2} \rangle $ & $\displaystyle X_{1} \oplus X_{3}$ \\
 \hline
\end{tabular}
\caption{Correspondence between gapped phases of $\mathrm{Vec}_{D_8}^{2\zeta}$ and $\mathrm{Rep}(H_8)$ after gauging the subgroup $\langle x\rangle$}
\label{Table:D8 gauge x}       
        \end{table}

For $A=\langle a^2\rangle$, we have ${}_{\langle a^2\rangle} (\mathrm{Vec}_{D_8}^{2\zeta})_{\langle a^2\rangle} \simeq \mathrm{Vec}_{D_8}^{2\zeta}$. There is one subtlety in determining the correspondence between the algebras before and after gauging. We must first fix a particular equivalence ${}_{\langle a^2\rangle} (\mathrm{Vec}_{D_8}^{2\zeta})_{\langle a^2\rangle} \simeq \mathrm{Vec}_{D_8}^{2\zeta}$; only after making this choice can we identify, inside ${}_{\langle a^2\rangle} (\mathrm{Vec}_{D_8}^{2\zeta})_{\langle a^2\rangle}$, the algebras labeled by subgroups of $D_8$. Different choices of this equivalence differ by monoidal auto-equivalences of $\mathrm{Vec}_{D_8}^{2\zeta}$. After fixing our choice, we obtain the correspondence shown in Table~\ref{Table:gauge a2}. These auto-equivalences are discussed in more detail in Appendix~\ref{App:A3}.

\begin{table}[h]
        \centering
        
\begin{tabular}{|c|c|}
\hline 
 \multicolumn{2}{|c|}{
Gauging $\displaystyle \langle a^{2} \rangle$ in $\displaystyle \mathrm{Vec}_{D_8}^{2\zeta}$
} \\
\hline \hline
 condensable algebras in $\displaystyle \mathrm{Vec}_{D_{8}}^{2\zeta }$ & condensable algebras in the dual $\displaystyle \mathrm{Vec}_{D_{8}}^{2\zeta }$ \\
\hline 
 $\displaystyle 1$ & $\displaystyle \langle a^{2} \rangle $ \\
\hline 
 $\displaystyle \langle xa,a^{2} \rangle $ & $\displaystyle \langle xa \rangle $ \\
\hline 
 $\displaystyle \langle a^{2} \rangle $ & $\displaystyle 1$ \\
\hline 
 $\displaystyle \langle xa \rangle $ & $\displaystyle \langle xa,a^{2} \rangle $ \\
\hline 
 $\displaystyle \langle x \rangle $ & $\displaystyle \langle x,a^{2} \rangle $ \\
\hline 
 $\displaystyle \langle x,a^{2} \rangle $ & $\displaystyle \langle x \rangle $ \\
 \hline
\end{tabular}
\caption{Correspondence between gapped phases of $\mathrm{Vec}_{D_8}^{2\zeta}$ and its dual after gauging the subgroup $\langle a^2 \rangle$} 
\label{Table:gauge a2}
        \end{table}

This last case is the pointed self-dual gauging. Unlike the gaugings that produce $\mathrm{Rep}(H_8)$, it does not turn the symmetry into a non-invertible one; instead, it exchanges the subgroup labels that play the roles of order and disorder data. Table~\ref{Table:gauge a2} should therefore be viewed as a relabeling of the same $D_8$-type symmetry-breaking patterns in a dual description.

\subsubsection{Generalized gauging in the dual $\Rep(H_8)$}

We next determine the correspondence between gapped phases obtained by gauging symmetries in $\mathrm{Rep}(H_8)$. According to Table~\ref{Table:D82zeta}, there are four condensable algebras in $\mathrm{Rep}(H_8)$ whose gauging produces a dual symmetry category that is again equivalent to $\mathrm{Rep}(H_8)$. These four algebras are $X_1$, $X_1\oplus X_3$, $X_1\oplus X_2\oplus X_0$, and $X_1\oplus X_2\oplus X_3\oplus X_4\oplus X_0^{\oplus 2}$.

In the dual $\mathrm{Rep}(H_8)$ frame, a gapped phase is no longer described only by a subgroup of an ordinary symmetry group. Instead, its condensable algebra records which invertible and non-invertible dual defects are invisible in that phase. Gauging one of the four self-dual algebras above is therefore a further change of duality frame inside the non-invertible symmetry theory. Since the output category is again $\mathrm{Rep}(H_8)$, these operations permute the possible $\mathrm{Rep}(H_8)$-symmetric gapped phases rather than changing the abstract symmetry type.

We determine this correspondence systematically by composing sequential gauging operations. Consider, for example, the condensable algebra $X_1\oplus X_3$. Starting from $\mathrm{Vec}_{D_8}^{2\zeta}$, first gauge the subgroup $\langle xa \rangle$. Under this gauging, the gapped phase labeled by $\langle xa,a^2 \rangle$ maps to the gapped phase labeled by $X_1\oplus X_3$ in $\mathrm{Rep}(H_8)$. Gauging $X_1\oplus X_3$ within $\mathrm{Rep}(H_8)$ then maps this phase to the one labeled by $X_1$ in the resulting dual $\mathrm{Rep}(H_8)$ category. The composition of these two gauging operations therefore gives a direct map from $\mathrm{Vec}_{D_8}^{2\zeta}$ to the dual $\mathrm{Rep}(H_8)$ in which $\langle xa,a^2\rangle$ maps to $X_1$. This composed operation is equivalent to directly gauging the subgroup $\langle xa,a^2\rangle$ in $\mathrm{Vec}_{D_8}^{2\zeta}$. By comparing Tables~\ref{Table:D8 gauge xa} and \ref{Table:D8 gauge xa a2}, we obtain the full correspondence between gapped phases of $\mathrm{Rep}(H_8)$ and its dual after gauging $X_1\oplus X_3$, as summarized in Table~\ref{Table:H8 gauge X1X3}.

The gauging procedures for $X_1\oplus X_2\oplus X_0$ and $X_1\oplus X_2\oplus X_3\oplus X_4\oplus X_0^{\oplus 2}$ are analogous. The corresponding results are listed in Tables~\ref{Table:H8 gauge X1X2X0} and \ref{Table:H8 gauge X1X2X3X4X0}. These four self-dual gaugings form a group isomorphic to $\Z_2\times \Z_2$, in agreement with Ref.~\cite{marshall2018brauer}.

\begin{table}[!h]
        \centering
        
\begin{tabular}{|c|c|}
\hline 
 \multicolumn{2}{|c|}{Gauging $\displaystyle X_{1} \oplus X_{3}$ in $\displaystyle \mathrm{Rep}(H_8)$} \\
\hline \hline
 condensable algebras in $\displaystyle \mathrm{Rep}( H_{8})$ & condensable algebras in the dual $\displaystyle \mathrm{Rep}( H_{8})$ \\
\hline 
 $\displaystyle X_{1} \oplus X_{2}$ & $\displaystyle X_{1} \oplus X_{2} \oplus X_{3} \oplus X_{4}$ \\
\hline 
 $\displaystyle X_{1} \oplus X_{3}$ & $\displaystyle X_{1}$ \\
\hline 
 $\displaystyle X_{1} \oplus X_{2} \oplus X_{3} \oplus X_{4}$ & $\displaystyle X_{1} \oplus X_{2}$ \\
\hline 
 $\displaystyle X_{1}$ & $\displaystyle X_{1} \oplus X_{3}$ \\
\hline 
 $\displaystyle X_{1} \oplus X_{2} \oplus X_{0}$ & $\displaystyle X_{1} \oplus X_{2} \oplus X_{3} \oplus X_{4} \oplus X_{0}^{\oplus 2}$ \\
\hline 
 $\displaystyle X_{1} \oplus X_{2} \oplus X_{3} \oplus X_{4} \oplus X_{0}^{\oplus 2}$ & $\displaystyle X_{1} \oplus X_{2} \oplus X_{0}$ \\
 \hline
\end{tabular}
\caption{Correspondence between gapped phases of $\mathrm{Rep}(H_8)$ and its dual after gauging $X_1\oplus X_3$}
\label{Table:H8 gauge X1X3}     
        \end{table}

\begin{table}[!h]
        \centering
        
\begin{tabular}{|c|c|}
\hline 
 \multicolumn{2}{|c|}{Gauging $\displaystyle X_{1} \oplus X_{2} \oplus X_{0}$ in $\displaystyle \mathrm{Rep}(H_8)$} \\
\hline \hline
 condensable algebras in $\displaystyle \mathrm{Rep}( H_{8})$ & condensable algebras in the dual $\displaystyle \mathrm{Rep}( H_{8})$ \\
\hline 
 $\displaystyle X_{1} \oplus X_{2}$ & $\displaystyle X_{1} \oplus X_{2}$ \\
\hline 
 $\displaystyle X_{1} \oplus X_{3}$ & $\displaystyle X_{1} \oplus X_{2} \oplus X_{3} \oplus X_{4} \oplus X_{0}^{\oplus 2}$ \\
\hline 
 $\displaystyle X_{1} \oplus X_{2} \oplus X_{3} \oplus X_{4}$ & $\displaystyle X_{1} \oplus X_{2} \oplus X_{3} \oplus X_{4}$ \\
\hline 
 $\displaystyle X_{1}$ & $\displaystyle X_{1} \oplus X_{2} \oplus X_{0}$ \\
\hline 
 $\displaystyle X_{1} \oplus X_{2} \oplus X_{0}$ & $\displaystyle X_{1}$ \\
\hline 
 $\displaystyle X_{1} \oplus X_{2} \oplus X_{3} \oplus X_{4} \oplus X_{0}^{\oplus 2}$ & $\displaystyle X_{1} \oplus X_{3}$ \\
 \hline
\end{tabular}
\caption{Correspondence between gapped phases of $\mathrm{Rep}(H_8)$ and its dual after gauging $X_1\oplus X_2\oplus X_0$}
\label{Table:H8 gauge X1X2X0}    
        \end{table}

\begin{table}[!h]
        \centering
        
\begin{tabular}{|c|c|}
\hline 
 \multicolumn{2}{|c|}{Gauging $\displaystyle X_{1} \oplus X_{2} \oplus X_{3} \oplus X_{4} \oplus X_{0}^{\oplus 2}$ in $\displaystyle \mathrm{Rep}(H_8)$} \\
\hline \hline
 condensable algebras in $\displaystyle \mathrm{Rep}( H_{8})$ & condensable algebras in the dual $\displaystyle \mathrm{Rep}( H_{8})$ \\
\hline 
 $\displaystyle X_{1} \oplus X_{2}$ & $\displaystyle X_{1} \oplus X_{2} \oplus X_{3} \oplus X_{4}$ \\
\hline 
 $\displaystyle X_{1} \oplus X_{3}$ & $\displaystyle X_{1} \oplus X_{2} \oplus X_{0}$ \\
\hline 
 $\displaystyle X_{1} \oplus X_{2} \oplus X_{3} \oplus X_{4}$ & $\displaystyle X_{1} \oplus X_{2}$ \\
\hline 
 $\displaystyle X_{1}$ & $\displaystyle X_{1} \oplus X_{2} \oplus X_{3} \oplus X_{4} \oplus X_{0}^{\oplus 2}$ \\
\hline 
 $\displaystyle X_{1} \oplus X_{2} \oplus X_{0}$ & $\displaystyle X_{1} \oplus X_{3}$ \\
\hline 
 $\displaystyle X_{1} \oplus X_{2} \oplus X_{3} \oplus X_{4} \oplus X_{0}^{\oplus 2}$ & $\displaystyle X_{1}$ \\
 \hline
\end{tabular}
\caption{Correspondence between gapped phases of $\mathrm{Rep}(H_8)$ and its dual after gauging $X_1\oplus X_2\oplus X_3\oplus X_4 \oplus X_0^{\oplus 2}$}
\label{Table:H8 gauge X1X2X3X4X0} 
        \end{table}

\subsubsection{Composition of generalized gaugings}
\label{section:BrPic}

The general Brauer-Picard groupoid input is summarized in Appendix~\ref{App:A3}. Here we apply it to $\mathrm{Vec}_{D_8}^{\omega_{\text{LSM}}}$.

The role of the Brauer-Picard groupoid is to organize these duality frames and their compositions. Physically, two successive gaugings may pass through different-looking symmetry descriptions, but the combined operation is equivalent to condensing a single effective algebra in the starting category. The composition tables below are a compact record of this network of dual descriptions: their entries tell us which final gapped phase label is obtained after the full two-step gauging process.

As recalled in Appendix~\ref{App:A3}, for a fusion category $\C$ one has $\text{Out}(\C) \backslash \text{BrPic}(\C) \simeq \Gamma(\C)$. For $\C = \mathrm{Vec}_{D_8}^{\omega_{\text{LSM}}}$, the outer autoequivalence group is $\text{Out}(\mathrm{Vec}_{D_8}^{\omega_{\text{LSM}}}) = \Z_2 \times \Z_2$, while the Brauer-Picard group is $\text{BrPic}(\mathrm{Vec}_{D_8}^{\omega_{\text{LSM}}}) = \Z_2^3$~\cite{marshall2018brauer}. Therefore the coset space $\Gamma(\mathrm{Vec}_{D_8}^{\omega_{\text{LSM}}})$ has exactly two elements, corresponding to the two indecomposable module categories labeled by $1$ and $\langle a^2 \rangle$. 

For $\mathrm{Rep}(H_8)$, one has $\text{Out}(\mathrm{Rep}(H_8))=\Z_2$~\cite{tambara2000representations} and $\text{BrPic}(\mathrm{Rep}(H_8))=\Z_2^3$~\cite{marshall2018brauer}. Hence $\Gamma(\mathrm{Rep}(H_8))\simeq \Z_2\times \Z_2$. The four corresponding module categories are represented by the algebras $X_1$, $X_1\oplus X_3$, $X_1\oplus X_2\oplus X_0$, and $X_1\oplus X_2\oplus X_3\oplus X_4\oplus X_0^{\oplus 2}$. The group structure of $\Gamma(\mathrm{Rep}(H_8))$ was described above.

After fixing category equivalences ${}_{\langle a^2 \rangle}(\mathrm{Vec}_{D_8}^{\omega_{\text{LSM}}})_{\langle a^2 \rangle} \simeq \mathrm{Vec}_{D_8}^{\omega_{\text{LSM}}}$ and ${}_A(\mathrm{Vec}_{D_8}^{\omega_{\text{LSM}}})_A \simeq \mathrm{Rep}(H_8)$, we can compose the gauging maps listed above. Cross-referencing Tables~\ref{Table:D8 gauge xa a2}, \ref{Table:D8 gauge xa}, \ref{Table:D8 gauge x a2}, \ref{Table:D8 gauge x}, \ref{Table:gauge a2}, \ref{Table:H8 gauge X1X3}, \ref{Table:H8 gauge X1X2X0}, and \ref{Table:H8 gauge X1X2X3X4X0} gives the two-step composition rules in Tables~\ref{Table:compose gauge1}, \ref{Table:compose gauge2}, \ref{Table:compose gauge3}, and \ref{Table:compose gauge4}.

\begin{table}[!h]
        \centering
        
\begin{tabular}{|c|c|c|c|c|}
\hline 
  & $\displaystyle \langle xa\rangle $ & $\displaystyle \langle xa,a^{2} \rangle $ & $\displaystyle \langle x\rangle $ & $\displaystyle \langle x,a^{2} \rangle $ \\
\hline 
 $\displaystyle 1$ & $\displaystyle \langle xa\rangle $ & $\displaystyle \langle xa,a^{2} \rangle $ & $\displaystyle \langle x\rangle $ & $\displaystyle \langle x,a^{2} \rangle $ \\
\hline 
 $\displaystyle \langle a^{2} \rangle $ & $\displaystyle \langle xa,a^{2} \rangle $ & $\displaystyle \langle xa\rangle $ & $\displaystyle \langle x,a^{2} \rangle $ & $\displaystyle \langle x\rangle $ \\
 \hline
\end{tabular}
\caption{\label{Table:compose gauge1}  The row headers record the first gauging step from $\mathrm{Vec}_{D_8}^{\omega_{\text{LSM}}}$, and the column headers record the second gauging step in the dual $\mathrm{Vec}_{D_8}^{\omega_{\text{LSM}}}$. The entries in the body of the table give the final effective condensable algebras in $\mathrm{Vec}_{D_8}^{\omega_{\text{LSM}}}$ for the full two-step composition.}
\end{table}

\begin{table}[!h]
        \centering
        
\begin{tabular}{|c|c|c|}
\hline 
  & $\displaystyle 1$ & $\displaystyle \langle a^{2} \rangle $ \\
\hline 
 $\displaystyle X_{1} \oplus X_{2}$ & $\displaystyle X_{1} \oplus X_{2}$ & $\displaystyle X_{1} \oplus X_{2} \oplus X_{3} \oplus X_{4}$ \\
\hline 
 $\displaystyle X_{1} \oplus X_{2} \oplus X_{3} \oplus X_{4}$ & $\displaystyle X_{1} \oplus X_{2} \oplus X_{3} \oplus X_{4}$ & $\displaystyle X_{1} \oplus X_{2}$ \\
 \hline
\end{tabular}
\caption{\label{Table:compose gauge2}  The row headers record the first gauging step from $\mathrm{Rep}(H_8)$, and the column headers record the second gauging step in $\mathrm{Vec}_{D_8}^{\omega_{\text{LSM}}}$. The entries in the body of the table give the final effective condensable algebras in $\mathrm{Rep}(H_8)$ for the full two-step composition.}         
\end{table}

\begin{table}[!h]
        \centering
        
\begin{tabular}{|c|c|c|c|c|}
\hline 
  & $\displaystyle X_{1}$ & $\displaystyle X_{1} \oplus X_{3}$ & $\displaystyle X_{1} \oplus X_{2} \oplus X_{0}$ & $\displaystyle X_{1} \oplus X_{2} \oplus X_{3} \oplus X_{4} \oplus X_{0}^{\oplus 2}$ \\
\hline 
 $\displaystyle \langle xa\rangle $ & $\displaystyle \langle xa\rangle $ & $\displaystyle \langle xa,a^{2} \rangle $ & $\displaystyle \langle x\rangle $ & $\displaystyle \langle x,a^{2} \rangle $ \\
\hline 
 $\displaystyle \langle xa,a^{2} \rangle $ & $\displaystyle \langle xa,a^{2} \rangle $ & $\displaystyle \langle xa\rangle $ & $\displaystyle \langle x,a^{2} \rangle $ & $\displaystyle \langle x\rangle $ \\
\hline 
 $\displaystyle \langle x\rangle $ & $\displaystyle \langle x\rangle $ & $\displaystyle \langle x,a^{2} \rangle $ & $\displaystyle \langle xa\rangle $ & $\displaystyle \langle xa,a^{2} \rangle $ \\
\hline 
 $\displaystyle \langle x,a^{2} \rangle $ & $\displaystyle \langle x,a^{2} \rangle $ & $\displaystyle \langle x\rangle $ & $\displaystyle \langle xa,a^{2} \rangle $ & $\displaystyle \langle xa\rangle $ \\
 \hline
\end{tabular}
\caption{\label{Table:compose gauge3}  The row headers record the first gauging step from $\mathrm{Vec}_{D_8}^{\omega_{\text{LSM}}}$, while the column headers record the second gauging step in $\mathrm{Rep}(H_8)$. The entries in the body of the table give the final effective condensable algebras in $\mathrm{Vec}_{D_8}^{\omega_{\text{LSM}}}$ for the full two-step composition.}      
\end{table}

\begin{table}[!h]
        \centering
\begin{tabular}{|c|c|c|}
\hline 
  & $\displaystyle X_{1} \oplus X_{2}$ & $\displaystyle X_{1} \oplus X_{2} \oplus X_{3} \oplus X_{4}$ \\
\hline 
 $\displaystyle X_{1}$ & $\displaystyle X_{1} \oplus X_{2}$ & $\displaystyle X_{1} \oplus X_{2} \oplus X_{3} \oplus X_{4}$ \\
\hline 
 $\displaystyle X_{1} \oplus X_{3}$ & $\displaystyle X_{1} \oplus X_{2} \oplus X_{3} \oplus X_{4}$ & $\displaystyle X_{1} \oplus X_{2}$ \\
\hline 
 $\displaystyle X_{1} \oplus X_{2} \oplus X_{0}$ & $\displaystyle X_{1} \oplus X_{2}$ & $\displaystyle X_{1} \oplus X_{2} \oplus X_{3} \oplus X_{4}$ \\
\hline 
 $\displaystyle X_{1} \oplus X_{2} \oplus X_{3} \oplus X_{4} \oplus X_{0}^{\oplus 2}$ & $\displaystyle X_{1} \oplus X_{2} \oplus X_{3} \oplus X_{4}$ & $\displaystyle X_{1} \oplus X_{2}$ \\
 \hline
\end{tabular}
\caption{\label{Table:compose gauge4}  The row headers record the first gauging step from $\mathrm{Rep}(H_8)$, while the column headers record the second gauging step in the dual $\mathrm{Rep}(H_8)$. The entries in the body of the table give the final effective condensable algebras in $\mathrm{Rep}(H_8)$ for the full two-step composition.}      
        \end{table}

\subsection{Lattice construction of subgroup gaugings and the dual symmetries}\label{sec:latticegauging}

In this section, we implement lattice gauging explicitly using the bond-algebra approach. Instead of tracking the symmetry operators themselves, we consider the algebra $\mathfrak{B}_G$ generated by local operators that are invariant under the internal symmetry. This algebra is called the local symmetric-operator algebra, or bond algebra~\cite{Cobanera_2010, 10.21468/SciPostPhys.17.4.115}. 
For an internal symmetry, the symmetry operators can be recovered from the commutant of the bond algebra, namely from operators that commute with all local symmetric operators. 

Physically, this formulation keeps the local Hamiltonian data in the foreground. The bond algebra is generated by the local terms that can appear in a symmetric Hamiltonian, while its commutant records the global charges that cannot be detected by any local symmetric probe. Gauging can therefore be described as a transformation of local observable algebras, without first choosing a special Hamiltonian.

Spatial symmetries must be treated differently. A translation does not commute with each local symmetric operator; rather, it maps one local symmetric operator to another. Thus the translation $T$ should be regarded as a locality-preserving automorphism $\alpha_T \in \mathrm{Aut}(\mathfrak{B}_G)$. In this language, a translation-invariant Hamiltonian is an element of the fixed-point subalgebra under the translation automorphism.

This distinction is important in the present LSM setting. Translation is not an onsite charge; it moves each local bond to the neighboring bond, and the anomaly is encoded in how this motion intertwines with the internal symmetry action. After gauging, the dual translation must reproduce this same action on the gauged images of the local bonds.

This viewpoint is especially useful for gauging. The gauging procedure maps the original bond algebra to the dual bond algebra of gauge-invariant local operators, $\mathcal G:\mathfrak{B}_G\longrightarrow \mathfrak{B}_{G/A}$.
The dual internal symmetry is then identified as the commutant of $\mathfrak{B}_{G/A}$.
The dual translation automorphism $\alpha_{\widetilde{T}} \in \mathrm{Aut}(\mathfrak{B}_{G/A})$ is defined by requiring the gauging map $\mathcal{G}$ to be an equivariant homomorphism between the original and dual bond algebras equipped with their translation automorphisms, namely $\mathcal{G} \circ \alpha_T = \alpha_{\widetilde{T}}\circ \mathcal{G}$.
This condition is the algebraic analogue of the intertwining relation between two representations.
It also gives a practical recipe: once the gauged images of the local bonds are known, the dual symmetry is whatever commutes with them, and the dual translation is the locality-preserving operation that makes the above diagram commute.

In our case, the symmetry group is $G = \mathbb{Z}_2^{XY}\times \mathbb{Z}_2^{YX} \rtimes \mathbb{Z}_{\mathrm{trans}}$. It is generated by $U_{XY} = \prod_i X_{2i}Y_{2i+1}$, $U_{YX} = \prod_i Y_{2i}X_{2i+1}$, and $T$.
The bond algebra of $G$ is
\begin{equation}
    \mathfrak{B}_{G} = \langle X_{2i}Y_{2i+1}, Z_{2i}Z_{2i+1},  Y_{2i+1}X_{2i+2}, Z_{2i+1}Z_{2i+2} | i \in \Lambda \rangle,
\end{equation}
where $\Lambda$ denotes the set of lattice sites and $|\Lambda| = 2L$. These generators generate all local symmetric operators. The internal symmetry is identified with the commutant of $\mathfrak{B}_{G}$, namely $G_{\mathrm{int}}=\mathbb{Z}_2^{XY} \times \mathbb{Z}_2^{YX}$. The translation symmetry is instead identified with the automorphism $\alpha_T(O_i) = TO_iT^{-1} = O_{i+1}$ of $\mathfrak{B}_{G}$, for $O_i \in \mathfrak{B}_{G}$.

\subsubsection{Gauging $\Z_2^{XY}$: non-invertible dual symmetry}

We now gauge the subgroup $\mathbb{Z}_2^{XY}$. We first introduce a $\mathbb{Z}_2$ gauge field on the links, with Pauli operators $\tau^x_{i+\frac12}$ and $\tau^z_{i+\frac12}$, and impose the Gauss-law constraint $G_i = 1$, where $G_{2i} = \tau^x_{2i-\frac12}X_{2i}\tau^x_{2i+\frac12}$ and $G_{2i+1} = \tau^x_{2i+\frac12}Y_{2i+1}\tau^x_{2i+\frac32}$.
This gauging promotes the global $XY$ flip to a local redundancy. The link variables are the corresponding parallel transporters, and the Gauss law ties the gauge electric field to the local $XY$ charge on even and odd sites.

After introducing minimal couplings, the original bond algebra is mapped to
\begin{equation}
\mathfrak{B}_{G / \mathbb{Z}_2^{X Y}}^{\mathrm{mc}}= \left\langle X_{2 i} Y_{2 i+1},\right. 
Z_{2 i} \tau_{2 i+\frac{1}{2}}^z Z_{2 i+1}, 
Y_{2 i+1} X_{2 i+2},
Z_{2 i+1} \tau_{2 i+\frac{3}{2}}^z Z_{2 i+2},
\left|G_i=1, i \in \Lambda\right\rangle.
\end{equation}
To solve the Gauss-law constraint, we apply the unitary transformation
\begin{equation}
    V =
    \prod_i
    \left[
    \frac{1}{2}(1+Z_i)
    +
    \frac{1}{2}(1-Z_i)\tau^x_{i-\frac12}\tau^x_{i+\frac12}
    \right].
\end{equation}
This transformation sends
\begin{equation}
    X_{i}\rightarrow \tau^x_{i-\frac12}X_{i}\tau^x_{i+\frac12},\quad Y_{i}\rightarrow \tau^x_{i-\frac12}Y_{i}\tau^x_{i+\frac12},\quad \tau_{i+\frac12}^z\rightarrow Z_i\tau_{i+\frac12}^z  Z_{i+1},
\end{equation}
and therefore
$G_{2i}\rightarrow X_{2i}$ and $G_{2i+1} \rightarrow Y_{2i+1}$.
The dual bond algebra of the gauged theory is
\begin{equation}
    \mathfrak{B}_{G / \mathbb{Z}_2^{X Y}} = V\mathfrak{B}_{G / \mathbb{Z}_2^{X Y}}^{\mathrm{mc}}V^{\dagger}|_{X_{2i} =1, Y_{2i+1} =1} 
    =\langle \tau_{2 i-\frac{1}{2}}^x \tau_{2 i+\frac{3}{2}}^x,  \tau^z_{2i+\frac12},
    \tau_{2 i+\frac{1}{2}}^x \tau_{2 i+\frac{5}{2}}^x, \tau^z_{2i+\frac32}| i\in\Lambda\rangle.
\end{equation}

Thus the original matter spins have been traded for link spins. The remaining generators are precisely the gauge-invariant remnants of the original symmetric bonds: products of dual electric fields and local dual magnetic fields.

The internal symmetry of the gauged theory is identified with the commutant of $\mathfrak{B}_{G / \mathbb{Z}_2^{X Y}}$, giving $G_{\mathrm{int}}^\vee=\mathbb{Z}_2^{\eta_1} \times \mathbb{Z}_2^{\eta_2}$. It is generated by $\eta_1 = \prod_i  \tau^z_{2i-\frac12}$ and $\eta_2 = \prod_i  \tau^z_{2i+\frac12}$.
These two operators are the quantum symmetries of the gauged theory. They measure the two sublattice parities of the dual link variables and commute with every local dual bond.

The translation symmetry of the gauged theory is more subtle. The naive one-site shift of the dual variables does not implement the gauged image of the original translation automorphism. In $\mathfrak{B}_G$, $T X_{2i}Y_{2i+1}T^{-1} = (Y_{2i+1}X_{2i+2})(Z_{2i+1}Z_{2i+2})$. By contrast, in $\mathfrak{B}_{G/\mathbb{Z}_2^{XY}}$, $T \tau_{2 i-\frac{1}{2}}^x \tau_{2 i+\frac{3}{2}}^xT^{-1} \neq (\tau_{2 i+\frac{1}{2}}^x \tau_{2 i+\frac{5}{2}}^x)(\tau^z_{2i+\frac32})$.
This mismatch is the lattice manifestation of the LSM decoration. Translation carries a gauged $XY$ bond through a neighboring bond that also contains the anomaly-related $ZZ$ factor, so a plain shift of the dual variables misses part of the original automorphism.

Therefore the gauged Hamiltonian obtained from a translation-invariant Hamiltonian in the original theory is not invariant under the one-site shift of the dual variables. Instead, it preserves a symmetry $\widetilde{T}$ that induces the same automorphism structure,
$\alpha_{\widetilde{T}}\left(\widetilde{O}_i^{(a)}\right) =\widetilde{O}_{i+1}^{(a)}$, with $\widetilde{O}_i^{(a)} = \mathcal{G} (O_i)$,
so that the equivariance condition above holds. Explicitly, this condition is
\begin{equation}
\begin{aligned}
    \widetilde{T} \tau_{2 i-\frac{1}{2}}^x \tau_{2 i+\frac{3}{2}}^x &= \tau_{2 i+\frac{1}{2}}^x \tau^z_{2i+\frac32} \tau_{2 i+\frac{5}{2}}^x \widetilde{T}, \\
    \widetilde{T} \tau^z_{2i+\frac12}&=\tau^z_{2i+\frac32}\widetilde{T},\\
    \widetilde{T} \tau_{2 i+\frac{1}{2}}^x \tau_{2 i+\frac{5}{2}}^x&=\tau^x_{2i+\frac32}\tau^z_{2i+\frac52}\tau^x_{2i+\frac72}\widetilde{T},\\
    \widetilde{T} \tau^z_{2i+\frac32}&=\tau^z_{2i+\frac52}\widetilde{T}.\label{eq:NItranslation}
\end{aligned}
\end{equation}
Therefore the new translation symmetry is $\widetilde{T} = TW$, where $W$ denotes the $\mathbf{TST}$ transformation~\cite{PhysRevLett.133.116601}. It acts on the spin chain with onsite symmetry $\mathbb{Z}_2^{\eta_1}\times\mathbb{Z}_2^{\eta_2}$.
Concretely, $W$ is obtained by stacking the corresponding SPT phase, performing a Kramers-Wannier transformation, and then stacking the same SPT phase again. Applying these three transformations sequentially gives the following maps:
\begin{equation}
\begin{aligned}
    \tau_{i+\frac{1}{2}}^z \rightarrow \tau_{i-\frac{1}{2}}^x\tau_{i+\frac{1}{2}}^z \tau_{i+\frac{3}{2}}^x \rightarrow \tau_{i-\frac{1}{2}}^x\tau_{i+\frac{1}{2}}^z \tau_{i+\frac{3}{2}}^x \rightarrow \tau_{i+\frac{1}{2}}^z,
    \\
    \tau_{i-\frac{1}{2}}^x\tau_{i+\frac{3}{2}}^x \rightarrow\tau_{i-\frac{1}{2}}^x\tau_{i+\frac{3}{2}}^x \rightarrow \tau_{i+\frac{1}{2}}^z\rightarrow \tau_{i-\frac{1}{2}}^x\tau_{i+\frac{1}{2}}^z \tau_{i+\frac{3}{2}}^x.
\end{aligned}
\end{equation}
After these three steps, applying the one-site translation gives Eq.~\ref{eq:NItranslation}. This is the desired non-invertible translation operator.
The key point is that the physical translation defect in the gauged theory is not an invertible group element by itself. It is a translation decorated by a duality operation, and its fusion necessarily produces a sum over the dual quantum-symmetry sectors.

On the $T^2=1$ subspace, the dual symmetry of the gauged theory satisfies the fusion rules of $\mathrm{Rep}(H_8)$:
\begin{equation}
    \widetilde{T}^2 = (1+\eta_1+\eta_2+\eta_1\eta_2)T^2,\qquad
    \widetilde{T} \eta_{i} = \eta_{i} \widetilde{T} = \widetilde{T}.
\end{equation}

\subsubsection{Gauging $\Z_2^{a^2}$: self-dual invertible symmetry}
We next gauge the central subgroup $\mathbb{Z}_2^{a^2}\subset G_{\mathrm{int}}$, which is generated by $U_{X Y} U_{Y X}=\prod_i Z_i$.
This gauging is qualitatively different from gauging $\mathbb{Z}_2^{XY}$. The subgroup is central and diagonal in the internal symmetry, so the operation is closer to an ordinary Kramers-Wannier relabeling of order and disorder variables. There is no additional translation decoration to generate a non-invertible defect.

We first introduce a $\mathbb{Z}_2$ gauge field on the links, with Pauli operators $\sigma_{i+\frac{1}{2}}^x$ and $\sigma_{i+\frac{1}{2}}^z$, and impose the Gauss-law constraint $G_i = 1$, where $G_{i} = \sigma_{i-\frac{1}{2}}^x Z_i \sigma_{i+\frac{1}{2}}^x$.
After introducing minimal couplings, the original bond algebra is mapped to 
\begin{equation}
    \mathfrak{B}_{G/\mathbb{Z}_2^{a^2}}^{\mathrm{mc}}=
    \langle
    X_{2 i}\sigma^z_{i+\frac12} Y_{2 i+1}, Z_{2 i} Z_{2 i+1}, Y_{2 i+1}\sigma^z_{i+\frac32} X_{2 i+2}, Z_{2 i+1} Z_{2 i+2}|G_i=1, i\in\Lambda
    \rangle.
\end{equation}
To solve the Gauss-law constraint, we apply the unitary transformation
\begin{equation}
    V =
    \prod_i
    \left[
    \frac{1}{2}(1+X_i)
    +
    \frac{1}{2}(1-X_i)\sigma_{i-\frac{1}{2}}^x \sigma_{i+\frac{1}{2}}^x
    \right].
\end{equation}
This transformation sends
\begin{equation}
    Z_i \rightarrow \sigma_{i-\frac{1}{2}}^xZ_i\sigma_{i+\frac{1}{2}}^x, \quad 
    Y_i \rightarrow \sigma_{i-\frac{1}{2}}^xY_i\sigma_{i+\frac{1}{2}}^x, \quad
    \sigma_{i+\frac{1}{2}}^z \rightarrow X_i \sigma_{i+\frac{1}{2}}^z X_{i+1},
\end{equation}
and therefore
$G_i \rightarrow Z_i$.
The dual bond algebra of the gauged theory is
\begin{equation}
    \mathfrak{B}_{G/\mathbb{Z}_2^{a^2}} = V \mathfrak{B}_{G/\mathbb{Z}_2^{a^2}}^{\mathrm{mc}} V^{\dagger}|_{Z_i = 1}= 
    \langle
    -\sigma_{2 i+\frac{1}{2}}^y \sigma_{2 i+\frac{3}{2}}^x, 
    \sigma_{2 i-\frac{1}{2}}^x \sigma_{2 i+\frac{3}{2}}^x, -\sigma_{2 i+\frac{1}{2}}^x \sigma_{2 i+\frac{3}{2}}^y, 
    \sigma_{2 i+\frac{1}{2}}^x \sigma_{2 i+\frac{5}{2}}^x
    |i\in\Lambda
    \rangle.
\end{equation}

This dual bond algebra is isomorphic to the original bond algebra $\mathfrak{B}_G$. Moreover, the naive shift of the dual variables implements the gauged image of the original translation automorphism. Therefore the dual symmetry is again group-like,
$G^\vee \simeq D_8$, with generators $\widetilde{U}_{XY} = \prod_i\sigma^x_{2i-\frac12}\sigma^y_{2i+\frac12}$, $\widetilde{U}_{YX} = \prod_i\sigma^y_{2i-\frac12}\sigma^x_{2i+\frac12}$, and $\widetilde{T} = T$.
Thus this gauging channel provides a useful check on the construction. It exchanges the central charge with the dual quantum symmetry, but it does not change the symmetry from an invertible group symmetry into a non-invertible one.

\subsubsection{Gauging the full internal symmetry $\Z_2^{XY}\times\Z_2^{YX}$}
Gauging the full internal symmetry means gauging the normal Abelian subgroup
$G_{\mathrm{int}}=\langle U_{XY},U_{YX}\rangle=\langle ax,a^2\rangle$.
It can be implemented sequentially by first gauging the diagonal subgroup
$\mathbb{Z}_2^{a^2}$ and then gauging the remaining $\mathbb{Z}_2^{XY}$ subgroup. The resulting dual symmetry is $\Rep(H_8)$.
This sequential implementation is an instance of the composition of generalized gaugings reviewed in Appendix~\ref{App:A3}; the corresponding $D_8$-specific composition tables are summarized in Sec.~\ref{section:BrPic}.

Physically, this operation gauges all onsite internal charges carried by the primitive unit cell. Before gauging, the two internal generators act projectively on each unit cell, and this projective representation is the microscopic LSM index. After gauging, the projective internal charge is no longer a global charge label; instead, it is converted into the endpoint data carried by internal gauge-flux defects. Thus the anomaly is not removed by gauging. It is transferred from an onsite projective representation into the fusion and translation properties of the dual defects.

The sequential description separates two physical effects. Gauging the central subgroup $\mathbb{Z}_2^{a^2}$ is essentially a self-dual Kramers-Wannier-type transformation: it exchanges the diagonal internal charge with a dual quantum symmetry, but the dual symmetry remains an ordinary invertible $D_8$ symmetry. The second step, gauging $\mathbb{Z}_2^{XY}$, is where the $E_\infty^{1,2}$ LSM decoration becomes visible. Translation carries a domain wall through a projective internal edge mode, so after the internal symmetry has been gauged the translated defect must be accompanied by a duality operation. Its fusion no longer closes onto a single group element, but instead splits over the quantum-symmetry sectors. This is the lattice origin of the non-invertible $\Rep(H_8)$ dual symmetry.

This full internal gauging is therefore the most direct way to see the type-II character of the transition. The dimer and $z$-ferromagnetic phases are not merely relabeled by subgroups of another ordinary group; they become phases distinguished by which condensable algebra, or equivalently which collection of $\Rep(H_8)$ defects, is invisible in the infrared. In the dual description developed below, the dimer side is partially symmetry-breaking, while the $z$FM side is fully $\Rep(H_8)$ symmetric. The original DQCP is thus mapped to a non-invertible symmetry-breaking transition.

\subsubsection{Dual model of the dimer-to-ferromagnet DQCP}
In the preceding discussion, we derived the lattice gauging procedure in detail. We now consider the dual Hamiltonian of Eq.~\ref{H} after gauging anomaly-free subgroups of $D_8$. The dual Hamiltonian is obtained by tracking the gauge image of each local symmetric operator. 

The advantage of the bond-algebra construction is that the Hamiltonian follows mechanically. Each term in the original Hamiltonian is replaced by its image under the appropriate gauging map, while the endpoint phases can be read from the exactly solvable projector limits of the resulting dual Hamiltonians.

We first gauge the central subgroup $A=\mathbb{Z}_2^{a^2}$.
Because this gauging is self-dual at the level of the symmetry category, it should not turn the transition into a non-invertible symmetry-restoration problem. Instead, it relabels the two endpoint phases within an ordinary dual $D_8$ symmetry frame.

The local symmetric-operator algebra of the original theory is
\begin{equation}
    \mathfrak{B}_G=\left\langle X_{2 i} Y_{2 i+1}, Z_{2 i} Z_{2 i+1}, Y_{2 i+1} X_{2 i+2}, Z_{2 i+1} Z_{2 i+2} | i \in \Lambda\right\rangle.
\end{equation}
After gauging, it is mapped to the gauge-invariant local-operator algebra
\begin{equation}
    \mathfrak{B}_{G / \mathbb{Z}_2^{a^2}}=\langle-\sigma_{2 i+\frac{1}{2}}^y \sigma_{2 i+\frac{3}{2}}^x, \sigma_{2 i-\frac{1}{2}}^x \sigma_{2 i+\frac{3}{2}}^x,-\sigma_{2 i+\frac{1}{2}}^x \sigma_{2 i+\frac{3}{2}}^y, \sigma_{2 i+\frac{1}{2}}^x \sigma_{2 i+\frac{5}{2}}^x | i \in \Lambda\rangle.
\end{equation}
Here the gauge-field degrees of freedom are $\{\sigma_{i+\frac 12}\}$. We have localized the Gauss-law constraint onto the original matter degrees of freedom by a unitary transformation. 

The dual symmetry is isomorphic to the original group, $G^\vee\simeq D_8$, with generators $\widetilde{U}_{X Y}=\prod_i \sigma_{2 i-\frac{1}{2}}^x \sigma_{2 i+\frac{1}{2}}^y$, $\widetilde{U}_{Y X}=\prod_i \sigma_{2 i-\frac{1}{2}}^y \sigma_{2 i+\frac{1}{2}}^x$, and $\widetilde{T}=T$.

The dual Hamiltonian is
\begin{align}
    \hat H_{G/\mathbb{Z}_2^{a^2}}(\lambda) = 
    \sum_i&[\sigma_{i+\frac12}^{y}\sigma_{i+\frac32}^{x} + 
    \sigma_{i-\frac12}^{x}\sigma_{i+\frac12}^{y} - 
    (1+\lambda)\sigma_{i-\frac12}^{x}\sigma_{i+\frac32}^{x}] 
    +\\ \frac12&[\sigma^z_{i+\frac12}\sigma^z_{i+\frac32} + 
    \sigma^x_{i-\frac12}\sigma^y_{i+\frac12}\sigma^y_{i+\frac32}\sigma^x_{i+\frac52} + 
    \sigma^x_{i-\frac12}\sigma^x_{i+\frac12}\sigma^x_{i+\frac32}\sigma^x_{i+\frac52}].
\end{align}

At $\lambda = 0$, the Hamiltonian can be written in projector form as $-\sum_i3(P^{G/\mathbb{Z}_2^{a^2}}_{i-\frac12,i+\frac12,i+\frac32,i+\frac52}-\frac12)$, where
\begin{align}
    P^{G/\mathbb{Z}_2^{a^2}}_{i-\frac12,i+\frac12,i+\frac32,i+\frac52} =& 
    \frac16[1-\sigma_{i-\frac{1}{2}}^x \sigma_{i+\frac{1}{2}}^y- \sigma_{i+\frac{1}{2}}^y \sigma_{i+\frac{3}{2}}^x+ \sigma_{i-\frac{1}{2}}^x \sigma_{i+\frac{3}{2}}^x] +
    \frac16[1-\sigma_{i+\frac{1}{2}}^x \sigma_{i+\frac{3}{2}}^y- \sigma_{i+\frac{3}{2}}^y \sigma_{i+\frac{5}{2}}^x+ \sigma_{i+\frac{1}{2}}^x \sigma_{i+\frac{5}{2}}^x]+\\
    &\frac16[1-\sigma_{i+\frac{1}{2}}^z \sigma_{i+\frac{3}{2}}^z-\sigma_{i-\frac{1}{2}}^x \sigma_{i+\frac{1}{2}}^y \sigma_{i+\frac{3}{2}}^y \sigma_{i+\frac{5}{2}}^x-\sigma_{i-\frac{1}{2}}^x \sigma_{i+\frac{1}{2}}^x \sigma_{i+\frac{3}{2}}^x \sigma_{i+\frac{5}{2}}^x].
\end{align}
The eigenspace $P^{G/\mathbb{Z}_2^{a^2}}=1$ is eight-dimensional and is spanned by
\begin{align}
    \{&
    |0\rangle|\!+\!x\rangle|\!-\!y\rangle|\!+\!x\rangle, 
    |1\rangle|\!+\!x\rangle|\!-\!y\rangle|\!+\!x\rangle,
    |0\rangle|\!-\!x\rangle|\!+\!y\rangle|\!-\!x\rangle,
    |1\rangle|\!-\!x\rangle|\!+\!y\rangle|\!-\!x\rangle,\\&
    |\!+\!x\rangle|\!-\!y\rangle|\!+\!x\rangle|0\rangle,
    |\!+\!x\rangle|\!-\!y\rangle|\!+\!x\rangle|1\rangle,
    |\!-\!x\rangle|\!+\!y\rangle|\!-\!x\rangle|0\rangle,
    |\!-\!x\rangle|\!+\!y\rangle|\!-\!x\rangle|1\rangle
    \}.
\end{align}
Therefore the Hamiltonian is frustration-free, and the four degenerate ground states are
\begin{equation}
    \bigotimes_i |\!+\!x\rangle_{i-\frac12}|\!-\!y\rangle_{i+\frac12},\quad
    \bigotimes_i |\!-\!x\rangle_{i-\frac12}|\!+\!y\rangle_{i+\frac12},\quad
    \bigotimes_i |\!+\!x\rangle_{i+\frac12}|\!-\!y\rangle_{i+\frac32},\quad
    \bigotimes_i |\!-\!x\rangle_{i+\frac12}|\!+\!y\rangle_{i+\frac32},
\end{equation}
where the first two states are symmetric under $\widetilde{U}_{XY}$ while the last two states are symmetric under $\widetilde{U}_{YX}$. Thus the unbroken symmetry is $H_{\lambda=0}^{G/\mathbb{Z}_2^{a^2}}=\mathbb{Z}_2^{ax}$ or $\mathbb{Z}_2^{a^3x}$.

When $\lambda\rightarrow \infty$, the Hamiltonian is dominated by $-\sum_i \sigma_{i-\frac{1}{2}}^x \sigma_{i+\frac{3}{2}}^x$.
Therefore the four degenerate ground states are
\begin{equation}
    \bigotimes_i |\!+\!x\rangle_{i-\frac12}|\!+\!x\rangle_{i+\frac12}, \quad 
    \bigotimes_i |\!-\!x\rangle_{i-\frac12}|\!-\!x\rangle_{i+\frac12}, \quad
    \bigotimes_i |\!+\!x\rangle_{i-\frac12}|\!-\!x\rangle_{i+\frac12}, \quad
    \bigotimes_i |\!-\!x\rangle_{i-\frac12}|\!+\!x\rangle_{i+\frac12},
\end{equation}
where the first two states are symmetric under $\widetilde{T}$. The last two states are symmetric under $\widetilde{U}_{XY}\widetilde{U}_{YX}\widetilde{T}$. The unbroken symmetry is $H_{\lambda\to\infty}^{G/\mathbb{Z}_2^{a^2}}=\mathbb{Z}_2^{x}$ or $\mathbb{Z}_2^{a^2x}$.
Combining the two endpoints of the interpolation, gauging $\mathbb{Z}_2^{a^2}$ maps the original $D_8$ symmetry to a dual $\hat{D}_8$ symmetry. The endpoint phases are mapped as
\begin{align}
    \mathbb{Z}_2^{ax}\times \mathbb{Z}_2^{a^2}
    &\longmapsto
    \mathbb{Z}_2^{a^2},\\
    \mathbb{Z}_2^{x}\times \mathbb{Z}_2^{a^2}
    &\longmapsto
    \mathbb{Z}_2^{x}.
\end{align}
Thus the dimer-$z$FM DQCP is mapped, as expected, to a dual DQCP between the two phases on the right-hand side.
This confirms that the central gauging leaves us in the familiar setting of ordinary symmetry breaking: both sides remain described by residual subgroups of an invertible dual group.

Next we gauge the subgroup $A=\mathbb{Z}_2^{ax} = \mathbb{Z}_2^{XY}$.
This is the more useful duality frame for the type-II interpretation of the transition. The gauged subgroup is precisely the one for which translation must be decorated by the duality operator $W$, so the dual theory naturally organizes the endpoint phases using $\mathrm{Rep}(H_8)$ rather than an ordinary group.

The local symmetric-operator algebra of the original theory is mapped to the gauge-invariant local-operator algebra
\begin{equation}
    \mathfrak{B}_{G / \mathbb{Z}_2^{X Y}}=\langle\tau_{2 i-\frac{1}{2}}^x \tau_{2 i+\frac{3}{2}}^x, \tau_{2 i+\frac{1}{2}}^z, \tau_{2 i+\frac{1}{2}}^x \tau_{2 i+\frac{5}{2}}^x, \tau_{2 i+\frac{3}{2}}^z | i \in \Lambda\rangle.
\end{equation}
Here the gauge-field degrees of freedom are $\{\tau_{i+\frac 12}\}$. Again, we have localized the Gauss-law constraint onto the original matter degrees of freedom by a unitary transformation. 

The dual symmetry is non-invertible, $\C^\vee=\mathrm{Rep}(H_8)$. Its generators are $\eta_1=\prod_i \tau_{2 i-\frac{1}{2}}^z$, $\eta_2=\prod_i \tau_{2 i+\frac{1}{2}}^z$, and $\widetilde{T} = TW$.
Here $\eta_1$ and $\eta_2$ are the invertible quantum-symmetry lines, while $\widetilde{T}$ is the non-invertible translation defect obtained above. The endpoint analysis below identifies which of these dual defects remain unbroken in each gapped phase.

The dual Hamiltonian is
\begin{align}
\hat H_{G/\mathbb{Z}_2^{XY}}(\lambda)=& -\sum_i[\tau_{i-\frac{1}{2}}^x \tau_{i+\frac{3}{2}}^x+\tau_{i-\frac{1}{2}}^x \tau_{i+\frac{1}{2}}^z \tau_{i+\frac{3}{2}}^x+(1+\lambda)\tau_{i+\frac{1}{2}}^z]+ \\
 & \frac{1}{2} \sum_i[\tau_{i-\frac{3}{2}}^x \tau_{i-\frac{1}{2}}^x \tau_{i+\frac{1}{2}}^x \tau_{i+\frac{3}{2}}^x+\tau_{i-\frac{3}{2}}^x \tau_{i-\frac{1}{2}}^y \tau_{i+\frac{1}{2}}^y \tau_{i+\frac{3}{2}}^x+\tau_{i-\frac{1}{2}}^z \tau_{i+\frac{1}{2}}^z].
\end{align}

At $\lambda = 0$, the Hamiltonian can be written in projector form as $-\sum 3(P_{i-\frac{1}{2}, i+\frac{1}{2}, i+\frac{3}{2}, i+\frac{5}{2}}^{G / \mathbb{Z}_2^{XY}}-\frac{1}{2})$, where
\begin{align}
    P_{i-\frac{1}{2}, i+\frac{1}{2}, i+\frac{3}{2}, i+\frac{5}{2}}^{G / \mathbb{Z}_2^{XY}}= 
    &\frac16[1+\tau_{i-\frac{3}{2}}^x \tau_{i+\frac{1}{2}}^x+\tau_{i-\frac{3}{2}}^x \tau_{i-\frac{1}{2}}^z \tau_{i+\frac{1}{2}}^x+\tau_{i-\frac{1}{2}}^z] +
    \frac16 [1+\tau_{i-\frac{1}{2}}^x \tau_{i+\frac{3}{2}}^x+\tau_{i-\frac{1}{2}}^x \tau_{i+\frac{1}{2}}^z \tau_{i+\frac{3}{2}}^x+\tau_{i+\frac{1}{2}}^z] +
    \\&\frac16[\tau_{i-\frac{3}{2}}^x \tau_{i-\frac{1}{2}}^x \tau_{i+\frac{1}{2}}^x \tau_{i+\frac{3}{2}}^x+\tau_{i-\frac{3}{2}}^x \tau_{i-\frac{1}{2}}^y \tau_{i+\frac{1}{2}}^y \tau_{i+\frac{3}{2}}^x+\tau_{i-\frac{1}{2}}^z \tau_{i+\frac{1}{2}}^z]
\end{align}
The eigenspace $P^{G/\mathbb{Z}_2^{XY}}=1$ is eight-dimensional and is spanned by
\begin{align}
    \{
    |0\rangle|\!+\!x\rangle|0\rangle|\!+\!x\rangle,
    |1\rangle|\!+\!x\rangle|0\rangle|\!+\!x\rangle, 
    |0\rangle|\!-\!x\rangle|0\rangle|\!-\!x\rangle, 
    |1\rangle|\!-\!x\rangle|0\rangle|\!-\!x\rangle, \\
    |\!+\!x\rangle|0\rangle|\!+\!x\rangle|0\rangle, 
    |\!+\!x\rangle|0\rangle|\!+\!x\rangle|1\rangle, 
    |\!-\!x\rangle|0\rangle|\!-\!x\rangle|0\rangle, 
    |\!-\!x\rangle|0\rangle|\!-\!x\rangle|1\rangle\}.
\end{align}
Therefore the Hamiltonian is frustration-free, and the four degenerate ground states are
\begin{equation}
    \bigotimes_i|0\rangle_{i-\frac12}|\!+\!x\rangle_{i+\frac12}, \quad
    \bigotimes_i|0\rangle_{i-\frac12}|\!-\!x\rangle_{i+\frac12}, \quad
    \bigotimes_i|\!+\!x\rangle_{i-\frac12}|0\rangle_{i+\frac12}, \quad
    \bigotimes_i|\!+\!x\rangle_{i-\frac12}|0\rangle_{i+\frac12},
\end{equation}
where the first two states are symmetric under $\eta_1$ and the last two states are symmetric under $\eta_2$. The unbroken symmetry is $H_{\lambda=0}^{G/\mathbb{Z}_2^{XY}}=\mathbb{Z}_2^{\eta_1}$ or $\mathbb{Z}_2^{\eta_2}$.

When $\lambda\rightarrow\infty$, the Hamiltonian is dominated by $-\sum_i \tau_{i+\frac12}^{z}$. Therefore the unique $\mathrm{Rep}(H_8)$-symmetric ground state is $\bigotimes_i |0\rangle_{i+\frac12}$. This gapped phase has fully unbroken $\mathrm{Rep}(H_8)$ symmetry, $\mathcal C_{\mathrm{unbroken}}^{G/\mathbb{Z}_2^{XY}}=\mathrm{Rep}(H_8)$. Combining the two endpoints of the interpolation, gauging $\mathbb{Z}_2^{ax}$ maps $D_8 \longmapsto \mathrm{Rep}(H_8)$.
The corresponding endpoint phases transform as
\begin{align}
    \mathbb{Z}_2^{ax}\times \mathbb{Z}_2^{a^2}
    &\longmapsto
    \mathbb{Z}_2^{\eta_1},\\
    \mathbb{Z}_2^{x}\times \mathbb{Z}_2^{a^2}
    &\longmapsto
    \mathrm{Rep}(H_8).
\end{align}
Thus the dimer-$z$FM DQCP is mapped to a dual DQCP between a partially symmetry-broken phase and a fully $\mathrm{Rep}(H_8)$-symmetric phase. In this dual frame, the ferromagnetic side becomes the symmetric phase of the non-invertible category, while the dimer side breaks only part of the dual symmetry.

\section{Summary and outlook}
\label{sec:dis}

In this work we identified a general mechanism for DQCPs associated with $E_\infty^{1,2}$-type LSM anomalies. The anomaly is the projective representation of the internal symmetry carried by each primitive unit cell, or equivalently a translation defect decorated by a one-dimensional internal-symmetry SPT phase:
\begin{align}
\omega_{\mathrm{LSM}}\in E_\infty^{1,2}= H^1(\mathbb Z_{\mathrm{trans}},H^2(G_{\mathrm{int}},\U))\subseteq H^3(G_{\mathrm{int}}\rtimes_{\rho}\mathbb Z_{\mathrm{trans}},\U)
\end{align}
or, after restricting to a finite translation quotient, an $E_\infty^{1,2}$ component of $H^3(G,\U)$. The key result is that this component forces non-invertibility after gauging: for an Abelian normal gaugeable subgroup $A$, a nontrivial $E_\infty^{1,2}$ component makes some dual defect sector carry a nontrivial projective representation of $A$, producing simple objects with quantum dimension larger than one in ${}_A(\Vec_G^\omega)_A$. This is the categorical mechanism for type-II DQCPs, which are dual to symmetry breaking of a non-invertible fusion-category symmetry rather than to ordinary group-like symmetry breaking.

We realized this mechanism in a spin-$1/2$ chain with effective anomalous $D_8$ symmetry. The LSM anomaly is the class $\omega_{\mathrm{LSM}}=2\zeta\in H^3(D_8,\U)$. We classified the gapped phases of $\Vec_{D_8}^{\omega_{\mathrm{LSM}}}$, constructed representative ground states and parent Hamiltonians, and focused on the direct transition between the special dimer phase with $H_{\mathrm{dimer}}=\langle ax,a^3x\rangle$ and the $z$-ferromagnetic phase with $H_{z\mathrm{FM}}=\langle a^2,x\rangle$. These unbroken subgroups are not nested, so the transition is not naturally Landau-like. The interpolating Hamiltonian is mapped by a sublattice unitary transformation to a $z$-anisotropic Majumdar-Ghosh-type chain, and VUMPS finite-entanglement scaling gives a growing correlation length and central charge $c\simeq 1$, supporting a continuous DQCP. Finally, both categorical and explicit lattice gauging show that gauging $\langle ax\rangle$ or the full internal subgroup $\langle ax,a^2\rangle$ produces the non-invertible dual symmetry $\Rep(H_8)$; in this dual description, the dimer-to-$z$FM transition becomes a transition between a partially symmetry-broken phase and the fully $\Rep(H_8)$-symmetric phase.

There are several natural directions for future work. First, it would be useful to identify the continuum field theory of the critical point more explicitly. The numerical central charge suggests a $c=1$ CFT, but a sharper description should determine the compactification radius or orbifold data, the symmetry action on primary fields, and the role of marginal perturbations. This would also clarify the relation between the lattice $\Rep(H_8)$ symmetry and the non-invertible topological lines appearing in $c=1$ orbifold CFTs, including the known $\Rep(H_8)$ symmetry on the orbifold branch, and related generalized orbifold constructions \cite{FrohlichFuchsRunkelSchweigert2004,FuchsRunkelSchweigert2009DefectLines,ChangLinShaoWangYin2019TopologicalDefectLines,diatlyk2024gauging,ChoiLuSun2024SelfDuality,PerezLonaRobbinsSharpeVandermeulenYu2024GaugingPartI}.

Second, it would be interesting to extend the type-I/type-II distinction beyond the present finite-group examples and compare it systematically with other one-dimensional DQCP constructions, including the Ising-ferromagnet-to-VBS transition of Jiang and Motrunich and the exactly solvable model of Zhang and Levin \cite{JiangMotrunich2019,ZhangLevin2023}. For general internal symmetry $G_{\mathrm{int}}$, nontrivial translation action, non-Abelian or non-normal gaugeable subgroups, and possibly continuous symmetries, one would like a systematic classification of which LHS components lead to pointed dual symmetries and which force non-invertible ones. Such a classification would give a more general map between anomaly data, allowed symmetry-breaking patterns, and possible DQCPs.

Finally, it would be useful to develop a more intrinsic Landau-like description for the non-invertible symmetry-breaking transition that appears in the gauged theory. In ordinary symmetry-breaking transitions, local order parameters and their transformation properties organize the low-energy theory. For a fusion-category symmetry such as $\Rep(H_8)$, the corresponding order parameters are naturally categorical objects, including condensable algebras, module categories, and non-invertible order/disorder operators. A systematic dictionary between these categorical data and continuum critical fields would make the dual non-invertible description of type-II DQCPs more predictive.\\

\paragraph*{Note added.}
When finishing this work, we became aware of Ref.~\onlinecite{ChenLiuLuTantivasadakarn2026BeyondLandau}, which also studies spontaneous breaking of non-invertible symmetries and its duality to beyond-Landau transitions in an internal-symmetry setting, without the lattice-translation LSM anomaly considered here. We also note the related work of Ref.~\onlinecite{Hiromi:2026}, which discusses similar non-invertible-symmetry-breaking viewpoints on DQCPs.

\acknowledgments
We thank Weiguang Cao for bringing Ref.~\onlinecite{Furukawa2025SubsystemWeakNISPT} to our attention. This work is supported by the National Natural Science Foundation of China (NSFC) under Grants No. 12274250 and 12475022.

\appendix


\section{Categorical tools for $\C$-symmetric gapped phases and generalized gauging}
\label{App:A}

The purpose of this appendix is to summarize the categorical language used to understand the constructions in the main text better.

We use \emph{fusion categories} to describe anomalous and non-invertible symmetry lines. A $\C$-symmetric gapped phase is described by a \emph{module category} over $\C$. A \emph{condensable algebra} enters in a related but conceptually different way: it can be used as an algebraic representative of a module category, and it can also be chosen as the concrete input of a \emph{generalized gauging} operation. Throughout this appendix, we denote algebra representatives of gapped phases by $B$, as in $\M\simeq \C_B$, and reserve $A$ for the algebra that is actually gauged, as in the dual symmetry ${}_A\C_A$. \emph{Internal Homs} track the correspondence between module-category phases under gauging. The discussion below is intentionally brief and physics-oriented: we emphasize the dictionary and the computations used in this paper, rather than the most general mathematical framework. For systematic mathematical treatments of fusion categories, module categories, Morita theory, and the Brauer-Picard groupoid, see Refs.~\onlinecite{etingof2016tensor,ostrik2002module,ostrik2003module,etingof2010fusion,etingof2011weakly,galindo2017tensor}; for condensable algebras, gapped boundaries, and anyon condensation, see Refs.~\onlinecite{muger2003subfactors,muger2003subfactorsII,kitaev2012models,kong2014anyon,kong2022invitation,davydov2010modular,davydov2017lagrangian}; for group-theoretical fusion categories and Morita duals of pointed categories, see Refs.~\onlinecite{ostrik2002module,Naidu2007CategoricalMorita,gelaki2009some,natale2017equivalence,uribe2017classification}; and for generalized gauging, condensation defects, and topological interfaces in two-dimensional QFT, see Refs.~\onlinecite{bhardwaj2018finite,tachikawa2020gauging,diatlyk2024gauging,RoumpedakisSeifnashriShao2023HigherGauging,ChoiCordovaHsinLamShao2022DualityDefects,ChoiCordovaHsinLamShao2023CondensationDefects,PerezLonaRobbinsSharpeVandermeulenYu2024GaugingPartI,SeifnashriShaoYang2025LatticeGauging}.

\begin{table}[!h]
        \centering
        \begingroup
        \renewcommand{\arraystretch}{1.35}
        \setlength{\tabcolsep}{8pt}
        \resizebox{0.98\linewidth}{!}{%

\begin{tabular}{|c|c|}
\hline 
 Mathematical concept & Physical meaning \\
\hline 
\hline 
 fusion category $\displaystyle \mathcal{C}$ & global symmetry of 1+1D system $\displaystyle \mathcal{T}$ \\
 objects in $\displaystyle \mathcal{C}$ & topological defect lines \\
\hline 
 gapped-phase algebra $\displaystyle B\in \mathcal{C}$ & choice presenting a $\displaystyle \mathcal{C}$-symmetric gapped phase as $\displaystyle \mathcal{M}\simeq \mathcal{C}_{B}$ \\
 gapped-phase module category $\displaystyle \mathcal{M}\simeq \mathcal{C}_{B}$ & gapped phase with $\displaystyle \mathcal{C}$ symmetry \\
 simple objects in $\displaystyle \mathcal{M}$ & in bijection with the ground states of gapped phases \\
\hline 
 gauging algebra $\displaystyle A\in \mathcal{C}$ & gaugeable network of defect lines summed over in generalized gauging \\
 interface module category $\displaystyle \mathcal{C}_{A}$ & topological interface between $\displaystyle \mathcal{T}$ and $\displaystyle \mathcal{T}/A$ \\
 fusion category $\displaystyle {}_{A}\mathcal{C}_{A}$ & global symmetry of the gauged theory $\displaystyle \mathcal{T}/A$ \\
 \hline
\end{tabular}%
}
        \endgroup
\caption{Categorical dictionary between mathematics and physics. The notation separates the two roles of algebras in $\C$: $B$ presents a $\C$-symmetric gapped phase, whereas $A$ specifies a generalized gauging procedure.}
\end{table}

The basic dictionary used most often in the main text is:
\begin{itemize}
\item A $\C$-symmetric gapped phase is an indecomposable semisimple left $\C$-\emph{module category} $\M$;
\item after choosing an algebra representative $B\in\C$ for this module category, one may write $\M\simeq \C_B$;
\item A \emph{generalized gauging} operation is specified by a concrete condensable algebra $A\in\C$, and changes the symmetry category from $\C$ to the \emph{Morita dual} category ${}_A\C_A$;
\item The correspondence between gapped phases before and after this gauging is computed by an \emph{internal-Hom} construction.
\end{itemize}
Thus, when the main text labels a gapped phase by a condensable algebra $B$, this means that we have chosen an algebra representative of the underlying module category. By contrast, when the main text says that we gauge a condensable algebra $A$, the algebra $A$ is part of the gauging interface itself. These statements explain why the $D_8$ spin chain can be discussed in terms of $\Vec_{D_8}^{\omega_{\mathrm{LSM}}}$, why gauging certain subgroups produces $\Rep(H_8)$, and how the correspondence tables for gapped phases in Sec.~\ref{section:BrPic} are obtained.

The appendix is organized as follows. Appendix~\ref{App:A1} gives the basic dictionary for fusion-category symmetries and the module-category description of gapped phases. Appendix~\ref{App:A1gauging} explains the separate role of gauging algebras and the Morita dual symmetry after gauging. Appendix~\ref{App:A1pointed} specializes this dictionary to pointed categories $\Vec_G^\omega$, where algebra representatives and gauging algebras are described by subgroup data. Appendix~\ref{App:A2} explains how internal Homs compute the correspondence between gapped phases before and after gauging. Appendix~\ref{App:A3} explains how the Brauer-Picard groupoid packages the composition of generalized gaugings. Finally, Appendix~\ref{App:method} records the group-theoretical category computation used for the $D_8$ example.

\subsection{Fusion-category symmetries and gapped phases}
\label{App:A1}

Let $\C$ be a fusion category over $\mathbb C$. Physically, in a 1+1D system $\T$ with fusion-category symmetry $\C$, objects of $\C$ are topological defect lines, the tensor product $X\otimes Y$ is the fusion of two defect lines, and the associator is the $F$-symbol that tells us how three lines reassociate \cite{FrohlichFuchsRunkelSchweigert2004,FuchsRunkelSchweigert2009DefectLines,thorngren1912fusion,etingof2016tensor,BhardwajBottiniSchaferNamekiTiwari2023HigherCategorical}. A simple object is an elementary defect line. If every simple object has a tensor inverse, then $\C$ is called \emph{pointed}; this is the categorical version of an ordinary invertible symmetry.

Gapped phases with symmetry $\C$ are classified by indecomposable semisimple left $\C$-module categories \cite{ostrik2002module,ostrik2003module,kitaev2012models,kong2014anyon,kong2022invitation}. The physical meaning is that symmetry lines in $\C$ can act on the ground-state sectors, and the module category records this action. For an ordinary broken finite-group symmetry, this reduces to the familiar statement that the ground states form a $G$-set such as $G/H$.

A condensable algebra gives a convenient way to present a module category, but it is not the primary phase datum. In the categorical setting, Ostrik's theorem says that every indecomposable semisimple left $\C$-module category can be written as $\C_B$, the category of right modules over an algebra object $B\in\C$ \cite{ostrik2003module,etingof2016tensor}. The relevant algebra objects are connected separable algebras; in a unitary theory these are the same objects usually called 1d condensable algebras, or connected symmetric normalized-special $\ast$-Frobenius algebras \cite{muger2003subfactors,muger2003subfactorsII,kong2014anyon,davydov2010modular,davydov2017lagrangian}. Two algebras $B$ and $B'$ in $\C$ are said to be \emph{Morita equivalent} if their respective right module categories $\C_B$ and $\C_{B'}$ are equivalent as left $\C$-module categories. Consequently, gapped phases with symmetry $\C$ can be equivalently classified by the \emph{Morita equivalence classes} of condensable algebras in $\C$. For brevity, unless explicitly noted otherwise, for the purpose of classification, by ``condensable algebras'' we always mean their Morita equivalence classes in this work.

\subsection{Gauging algebras and dual symmetries}
\label{App:A1gauging}

Generalized gauging is a different use of condensable algebras. To gauge, one chooses a condensable algebra $A\in\C$, or equivalently the corresponding interface module category $\C_A$. This is operational data, not merely a name for a gapped phase: the network to be summed over is built from the multiplication and unit of $A$. Physically, gauging $A$ means summing over networks made from the defect lines contained in $A$. A defect line that remains after the gauging must be compatible with the $A$ network on both sides, so it becomes an $A$-$A$ bimodule. Therefore the dual symmetry category becomes
\begin{align}
\C \quad \longrightarrow \quad {}_A\C_A .
\end{align}
The category $\C_A$ is the topological interface between the original theory $\T$ and the gauged theory $\T/A$; it is a $\C$-${}_A\C_A$ bimodule category. This is the formulation of generalized gauging used throughout the main text \cite{bhardwaj2018finite,tachikawa2020gauging,diatlyk2024gauging,ChoiCordovaHsinLamShao2022DualityDefects,ChoiCordovaHsinLamShao2023CondensationDefects}.

Two fusion categories $\C$ and $\mathcal{D}$ are Morita equivalent if there is an invertible $\C$-$\mathcal{D}$ bimodule category between them \cite{muger2003subfactors,muger2003subfactorsII,etingof2010fusion,etingof2011weakly}. More precisely, there exist a finite semisimple $\C$-$\mathcal{D}$ bimodule category $\M$ and a finite semisimple $\mathcal{D}$-$\mathcal{C}$ bimodule category $\mathcal{N}$ such that $\M\boxtimes_{\mathcal{D}}\mathcal{N}\simeq \mathcal{C}$ as $\C$-$\C$ bimodule categories, and $\mathcal{N}\boxtimes_{\mathcal{C}}\M\simeq \mathcal{D}$ as $\mathcal{D}$-$\mathcal{D}$ bimodule categories. Here, $\boxtimes_{\C}$ denotes the relative Deligne tensor product over $\C$. The bimodule categories $\M$ and $\mathcal{N}$ satisfying these two conditions are said to be \textit{invertible}. Equivalently, $\mathcal{D}$ can be realized as ${}_A\C_A$ for some condensable algebra $A\in\C$. Hence generalized gauging does not produce an arbitrary new symmetry category; it moves inside the Morita class of the original symmetry. This is why the dual categories that appear below are Morita duals of $\Vec_{D_8}^{\omega_{\mathrm{LSM}}}$.

\subsection{Algebras in $\Vec_G^\omega$}
\label{App:A1pointed}

The basic pointed example is $\Vec_G^\omega$. Its simple objects are labeled by elements $g\in G$, and they fuse by group multiplication, $g\otimes h=gh$. The associator is multiplication by a normalized 3-cocycle $\omega(g,h,k)$. Thus $\Vec_G$ describes an ordinary non-anomalous $G$ symmetry, while $\Vec_G^\omega$ describes a $G$ symmetry with 't Hooft anomaly $[\omega]\in H^3(G,\U)$, as for the boundary of a Dijkgraaf-Witten SPT phase \cite{DijkgraafWitten1990,kapustin2014anomalies,etingof2016tensor,BhardwajBottiniSchaferNamekiTiwari2023HigherCategorical}. In the main text, the anomalous finite symmetry of the spin chain is precisely of this form, with $\C=\Vec_{D_8}^{\omega_{\mathrm{LSM}}}$.

For $\C=\Vec_G$, this dictionary reproduces the usual classification of one-dimensional gapped phases with finite group symmetry. The indecomposable module categories are represented by a subgroup $H\leq G$ and a class $\gamma\in H^2(H,\U)$: $H$ is the unbroken symmetry, while $\gamma$ is the $H$-SPT index of a symmetric ground state. Equivalently, the phase has $G/H$ symmetry-related ground states, each carrying the same $H$-SPT decoration. The corresponding algebra representative is supported on the lines $h\in H$ \cite{ostrik2002module,ChenGuWen2011,ChenGuWen2011Complete,SchuchPerezGarciaCirac2011}.

For the anomalous pointed category $\Vec_G^\omega$, the same statement holds with one important modification. If $(H,\gamma)$ is used as an algebra representative of a gapped phase, or if it is used as a concrete gauging algebra, then the anomaly restricted to $H$ must be canceled by the local counterterm $\gamma$. This gives the standard classification used in Appendix~\ref{App:D8ACA}.

\begin{proposition}\label{algebraMorita}\cite{ostrik2002module,Naidu2007CategoricalMorita,gelaki2009some,natale2017equivalence,uribe2017classification}
The condensable algebras in $\Vec_G^\omega$ are parameterized by pairs $(H,\gamma)$, where $H\leq G$ and $\gamma\in C^2(H,\U)$ satisfy
\begin{align}
\omega|_H=d\gamma .
\end{align}
The corresponding algebra object is supported on $\oplus_{h\in H}h$, with multiplication twisted by $\gamma$. Two pairs $(H,\gamma)$ and $(H',\gamma')$ define isomorphic algebras in $\Vec_G^\omega$ if and only if $H=H'$ and there exists a 1-cochain $c\in C^1(H,\U)$ such that $\gamma=\gamma'\cdot dc$.

Two condensable algebras $(H,\gamma)$ and $(H',\gamma')$ in $\Vec_G^\omega$ are Morita equivalent if and only if there exists $g\in G$ such that $H=gH'g^{-1}$, and the 2-cocycle class of $\gamma'^{-1}g^*(\gamma)\Omega_g|_{H'}$ is trivial in $H^2(H',\U)$. Here, the pullback 2-cochain $g^*(\gamma) \in C^2(H', U(1))$ is defined via
\begin{align}
(g^*\gamma)(x,y)&=\gamma(gxg^{-1},gyg^{-1}),
\end{align}
and the 2-cocycle $\Omega_g \in C^2(H', U(1))$ is given by
\begin{align}
\Omega_g(x,y)&=\frac{\omega(gxg^{-1},gyg^{-1},g)\omega(g,x,y)}
{\omega(gxg^{-1},g,y)} .
\end{align}
\end{proposition}

The condition $\omega|_H=d\gamma$ means that the anomaly restricted to $H$ can be canceled by the 1+1D counterterm $\gamma$. In the gapped-phase interpretation, one may denote this representative by $B=(H,\gamma)$, and it presents the module category of the phase. In the generalized-gauging interpretation, the same pair is now chosen as the concrete gauging algebra $A=(H,\gamma)$. In the $D_8$ example, this is exactly the criterion that selects the gaugeable subgroups listed in Appendix~\ref{App:D8ACA} and Table~\ref{Table:D82zeta}. After gauging $A=(H,\gamma)$, the dual category is usually denoted by
\begin{align}
\C(G,\omega,H,\gamma) = {}_A(\Vec_G^\omega)_A,
\end{align}
and is called a group-theoretical fusion category \cite{ostrik2002module,Naidu2007CategoricalMorita,gelaki2009some}. It need not be pointed. When it is not pointed, the gauged theory has a non-invertible dual symmetry.

\subsection{Gapped phases under gauging}
\label{App:A2}

Let $\C$ be a fusion category. For a chosen condensable algebra $A\in\C$, after gauging $A$, the new system, denoted by $\T/A$, has fusion-category symmetry ${}_A\C_A$. Categorically, the Morita context defined by the interface $\C_A$ gives a correspondence between indecomposable module categories of $\C$ and those of ${}_A\C_A$. Physically, this is the correspondence between gapped phases of $\T$ and gapped phases of $\T/A$. In terms of algebra representatives, it gives a map from $B$ in $\C$ to $B'$ in ${}_A\C_A$, understood up to Morita equivalence. We now explain how to determine this map.

By \cite[Proposition 2.1]{ostrik2002module} and the well-known equivalence $\mathrm{Fun}_{\C}(\C_A,\C_B) \simeq {}_A\C_B$, there exists a canonical bijection between the indecomposable semisimple (left) module categories of $\C$ and those of ${}_A\C_A$. Specifically, for a module category $\C_B$ of $\C$, the corresponding module category of ${}_A\C_A$ is given by ${}_A\C_B$, the category of $A$-$B$ bimodules in $\C$. Furthermore, according to \cite{ostrik2003module}, there exists an algebra $B' \in {}_A\C_A$ such that ${}_A\C_B \simeq ({}_A\C_A)_{B'}$ as left ${}_A\C_A$ module categories. Here, the object $B'$ can be explicitly determined via the internal Hom.

\begin{definition}\label{internalhom}

Let $\C$ be a monoidal category, and let $\M$ be a left $\C$-module category with the left $\C$-module action denoted by $\odot: \C \times \M \to \M$. For any objects $M, N \in \M$, the \textit{internal Hom} of $M$ and $N$, denoted by $[M,N]$, is an object in $\C$ satisfying the natural isomorphism
$$
\mathrm{Hom}_{\C}(X, [M,N]) \simeq \mathrm{Hom}_{\M}(X \odot M, N) \quad \forall X \in \C.
$$
\end{definition}
\begin{remark}
   The internal Hom $[M,M]$ naturally carries an algebra structure. Moreover, when $\C$ is a fusion category and $\M$ is an indecomposable semisimple left $\C$-module category, $[M,M]$ equips the structure of a condensable algebra for any simple object $M \in \M$.
\end{remark}

Following \cite{ostrik2003module}, the algebra $B' \in {}_A\C_A$ satisfying ${}_A\C_B \simeq ({}_A\C_A)_{B'}$ can be identified via the internal Hom $[M,M]$ for the left ${}_A\C_A$-module category ${}_A\C_B$, where $M \in {}_A\C_B$ is an arbitrarily chosen simple object. Physically, this implies that the gapped phase of $\C$ labeled by $B$ maps to the gapped phase of ${}_A\C_A$ labeled by $B'$ upon gauging the condensable algebra $A$.

\subsection{Composing gaugings}
\label{App:A3}

We use the Brauer-Picard groupoid to mathematically describe the composition of generalized gaugings. Similar discussions can be found in \cite{diatlyk2024gauging}. Here, we discuss the autoequivalences more precisely. 

For a 1+1D system $\T$ with a fusion category symmetry $\C$, gauging a condensable algebra $A\in \C$ yields a new system $\T/A$ with fusion category symmetry $_A\C_A$. The $\C$-$_A\C_A$ bimodule category $\C_A$ serves as the topological interface between $\T$ and $\T/A$. Next, considering a condensable algebra $A' \in {_A\C_A}$ and the subsequent gauged system $(\T/A)/A'$ with symmetry $_{A'}(_A\C_A)_{A'}$, the $_A\C_A$-$_{A'}(_A\C_A)_{A'}$ bimodule category $({_A\C_A})_{A'}$ becomes the topological interface between $\T/A$ and $(\T/A)/A'$. According to \cite{diatlyk2024gauging}, when these two gauging steps are composed, the resulting interface between $\T$ and $(\T/A)/A'$ is given by the fusion of the two individual interfaces. Mathematically, this combined interface is described by $\C_A\boxtimes_{_A\C_A}(_A\C_A)_{A'}$, where $\boxtimes_{_A\C_A}$ denotes the relative Deligne tensor product over $_A\C_A$.  

To determine the condensable algebra $A''$ in $\C$ such that $(\T/A)/A'=\T/A''$, we can compute the internal Hom for the left $\C$-module category $\C_A\boxtimes_{_A\C_A}(_A\C_A)_{A'}$. Alternatively, $A'$ can be identified through the correspondence of condensable algebras: the first step of gauging maps the condensable algebras of $\C$ to those of $_A\C_A$, and the preimage of $A'$ under this map is precisely the desired $A''$.

For a fusion category $\C$, we introduce the \emph{Brauer-Picard groupoid} $\mathbf{BrPic}(\C)$ \cite{etingof2010fusion} as follows: the objects of $\mathbf{BrPic}(\C)$ are the equivalence classes of fusion categories that are Morita equivalent to $\C$. The morphisms between two objects $\C$ and $\mathcal{D}$ are the equivalence classes of invertible $\C$-$\mathcal{D}$ bimodule categories. The composition of morphisms is given by the relative Deligne tensor product. Consequently, all morphisms in $\mathbf{BrPic}(\C)$ are invertible. In particular, $\text{BrPic}(\C) := \text{Hom}_{\mathbf{BrPic}(\C)}(\C,\C)$ forms a group, called the \textit{Brauer-Picard group} of $\C$.

The identity morphism in $\text{BrPic}(\C)$ is the category $\C$ itself, equipped with its canonical $\C$-$\C$ bimodule structure. Tensor autoequivalence in $\text{Aut}(\C)$ may induce another $\C$-$\C$ bimodule structure on $\C$ via its action. However, inner tensor autoequivalences do not alter this bimodule structure up to equivalence of bimodule categories. Let $\text{Inn}(\C)$ denote the normal subgroup of $\text{Aut}(\C)$ consisting of all inner tensor autoequivalences, and let $\text{Out}(\C) = \text{Aut}(\C) / \text{Inn}(\C)$ be the group of outer tensor autoequivalences. As shown in \cite{galindo2017tensor}, there is a subgroup embedding $\text{Out}(\C) \hookrightarrow \text{BrPic}(\C)$, where elements of $\text{Out}(\C)$ correspond to the category $\C$ equipped with  non-equivalent $\C$-$\C$ bimodule structures. Furthermore, let $\Gamma(\C)$ denote the set of right $\C$-module category equivalence classes of invertible $\C$-$\C$ bimodule categories. We have the following right $\text{BrPic}(\C)$-set isomorphism \cite{galindo2017tensor}:
$$
\text{Out}(\C) \backslash \text{BrPic}(\C) \simeq \Gamma(\C).
$$
When the gauging of $A \in \C$ yields ${}_A\C_A \simeq \C$, the gauging operation induces a permutation among the condensable algebras of $\C$. The above isomorphism implies that to explicitly determine this algebra permutation, one must choose a specific representative within the coset $\text{Out}(\C) \backslash \text{BrPic}(\C)$ to perform the relative Deligne tensor product. It is equivalent to saying that we should choose an equivalence ${}_A\C_A\simeq \C$ so that we can well-define the permutation of condensable algebras.

\subsection{Computing group-theoretical categories}
\label{App:method}

The simple objects in a group-theoretical fusion category can be represented in the following way:

\begin{proposition}\cite{ostrik2002module}
    Let $A=(H_1,\gamma_1),B=(H_2,\gamma_2)\in \mathrm{Vec}_G^\omega$ be two algebras. The simple objects of $A$-$B$ bimodule $_A(\mathrm{Vec}_G^\omega)_B$ are classified by the pairs $(Z,\pi)$, where $Z$ is a $H_1$-$H_2$ double coset in $G$, and $\pi$ is an irreducible projective representation of $H_1\cap gH_2g^{-1}$ with the Schur multiplication $\gamma^g\in Z^2(H_1\cap gH_2g^{-1},U(1))$. The Schur multiplication $\gamma^g$ is defined as follows: 
    $$
\gamma^g(h,h'):=\gamma_1(h,h')\gamma_2(g^{-1}h'^{-1}g,g^{-1}h^{-1}g)\omega(hh'g,g^{-1}h'^{-1},g^{-1}h^{-1}g)^{-1}\omega(h,h',g)\omega(h,h'g,g^{-1}h'^{-1}g)
$$

When $A = B = (H, \gamma)$, the pairs $(Z, \pi)$ classify the simple objects in the group-theoretical fusion category $\C(G, \omega, H, \gamma) = {}_A(\mathrm{Vec}_G^{\omega})_A$.
\end{proposition}

The double coset is the support of the bimodule category, and $\gamma^g$ characterizes the action of algebras in the bimodule.

For a given quadruple $(G,\omega,H,\gamma)$, the program \href{https://github.com/junkicide/fusionrules\#}{fusionrules} can compute the simple objects, fusion rules, dimensions of objects, and (higher) Frobenius-Schur indicators of the group-theoretical fusion category $\C(G,\omega,H,\gamma)$. While these data are not theoretically guaranteed to fully determine a fusion category up to equivalence in all general settings, they are practically sufficient to uniquely identify the category in most cases with small total quantum dimensions.

For dimension 8, the thesis referenced in \href{https://github.com/junkicide/fusionrules\#}{fusionrules} shows that the data output by the program can determine the fusion category. Using the dimension and fusion rule of the objects, we can determine if the fusion category is pointed or not. If the output fusion category is pointed, we can use Table \ref{Morita} to determine which pointed fusion category it is. If the output is a non-pointed fusion category, the thesis in \href{https://github.com/junkicide/fusionrules\#}{fusionrules} shows that there are only two types of dimension 8 non-pointed group-theoretical fusion categories: $\mathrm{TY}(\Z_2\times \Z_2)$ and $\mathrm{TY}(\Z_4)$. We can use the fusion rule and the higher Frobenius-Schur indicator to determine which specific dimension 8 fusion category it is, between the various types of $\mathrm{TY}(\Z_2\times \Z_2)$ and $\mathrm{TY}(\Z_4)$.

\section{Type-I and type-II DQCPs from $\mathbb{Z}_2^3$ with type-III anomaly}

We consider the symmetry group $\Z_2^a\times\Z_2^b\times\Z_2^c$. The associated 2+1D SPT phases, and hence their 1+1D boundary anomalies, are classified by
$H^3(\Z_2^a\times\Z_2^b\times\Z_2^c,\U)=\Z_2^7$. Among the seven generators, three type-I cocycles are associated with the individual $\Z_2$ factors, three type-II cocycles are associated with pairs of $\Z_2$ factors, and the remaining generator is the type-III cocycle involving all three $\Z_2$ factors. We choose the representative
\begin{align}
\omega_3^\mathrm{III}(g,h,k)=(-1)^{g_ah_bk_c},\quad \forall g,h,k\in \Z_2^a\times\Z_2^b\times\Z_2^c.
\end{align}
The important point for this appendix is that the same group $G=\Z_2^a\times\Z_2^b\times\Z_2^c$, with the same type-III anomaly, admits two natural choices of normal subgroup and hence two different LHS decompositions. Choosing a rank-one normal subgroup places the type-III class in the $E_\infty^{2,1}$ decomposition and leads to a type-I DQCP. Choosing a rank-two normal subgroup instead places the same class in the $E_\infty^{1,2}$ decomposition and leads to a type-II DQCP.

This mixed behavior is why we treat the $\Z_2^3$ type-III anomaly in an appendix rather than as the main example: in the $D_8$ spin chain discussed in the main text, apart from the pointed self-dual gauging, the relevant gaugings realize the $E_\infty^{1,2}$-type mechanism and hence lead to type-II DQCPs.

\subsection{From $E_\infty^{2,1}$ decomposition to type-I DQCP}

We first choose the rank-one normal subgroup
$A=\Z_2^a$ and quotient $Q=G/A=\Z_2^b\times \Z_2^c$, giving the short exact sequence
\begin{align}
1\rightarrow \Z_2^a
\rightarrow \Z_2^a\times\Z_2^b\times\Z_2^c
\rightarrow \Z_2^b\times \Z_2^c
\rightarrow 1.
\end{align}
In the corresponding LHS spectral sequence
\begin{align}
E_2^{p,q}=H^p(\Z_2^b\times\Z_2^c,H^q(\Z_2^a,\U))
\Rightarrow H^{p+q}(\Z_2^a\times\Z_2^b\times\Z_2^c,\U),
\end{align}
the type-III cocycle contributes to
\begin{align}
E_\infty^{2,1}
\subset
H^2(\Z_2^b\times\Z_2^c,H^1(\Z_2^a,\U)).
\end{align}
In the bulk decorated-domain-wall picture, this means that a 0+1D $\mathbb{Z}_2^a$ SPT, equivalently a $\mathbb{Z}_2^a$ charge, is placed at each codimension-two intersection of $\mathbb{Z}_2^b$ and $\mathbb{Z}_2^c$ domain walls. On the 1+1D boundary, this decoration implies that exchanging a $\mathbb{Z}_2^b$ domain wall with a $\mathbb{Z}_2^c$ domain wall creates a $\mathbb{Z}_2^a$ charge.

Thus, condensing $\mathbb{Z}_2^b\times \mathbb{Z}_2^c$ domain walls on the boundary necessarily produces fluctuating $\mathbb{Z}_2^a$ charges, which break $\mathbb{Z}_2^a$ symmetry and can drive a DQCP. Gauging the subgroup $A=\mathbb{Z}_2^a$ gives a $D_8$ dual symmetry. The DQCP between the $\mathbb{Z}_2^b\times \mathbb{Z}_2^c$-symmetric phase and the $\mathbb{Z}_2^a$-symmetric phase is then mapped to a conventional SSB critical point between a $D_8$-symmetric phase and a $D_8$-SSB phase. In this sense, the $E_\infty^{2,1}$ decomposition leads to a type-I DQCP: the $E_\infty^{2,1}$ data changes the dual group extension, but the dual defects remain invertible.

\subsection{From $E_\infty^{1,2}$ decomposition to type-II DQCP}

We now choose the rank-two normal subgroup
$A=\mathbb{Z}_2^a\times\mathbb{Z}_2^b$ and quotient $Q=G/A=\mathbb{Z}_2^c$, giving the short exact sequence
\begin{align}
1\rightarrow \Z_2^a\times\Z_2^b
\rightarrow \Z_2^a\times\Z_2^b\times\Z_2^c
\rightarrow \Z_2^c
\rightarrow 1.
\end{align}
This choice is the one most directly adapted to the lattice construction below. In the corresponding LHS spectral sequence
\begin{align}
E_2^{p,q}=H^p(\Z_2^c,H^q(\Z_2^a\times\Z_2^b,\U))
\Rightarrow H^{p+q}(\Z_2^a\times\Z_2^b\times\Z_2^c,\U),
\end{align}
the same type-III cocycle contributes to
\begin{align}
E_\infty^{1,2}
\subset
H^1(\Z_2^c,H^2(\Z_2^a\times\Z_2^b,\U)).
\end{align}
In the bulk decorated-domain-wall picture, this means that a 1+1D $\mathbb{Z}_2^a\times \mathbb{Z}_2^b$ SPT phase is placed on each codimension-one $\mathbb{Z}_2^c$ domain wall. On the 1+1D boundary, this decoration implies that a $\mathbb{Z}_2^c$ domain wall carries a projective representation of $\mathbb{Z}_2^a\times \mathbb{Z}_2^b$.

Thus, condensing $\mathbb{Z}_2^c$ domain walls on the boundary necessarily produces fluctuating $\mathbb{Z}_2^a\times \mathbb{Z}_2^b$ projective representations, which break $\mathbb{Z}_2^a\times \mathbb{Z}_2^b$ symmetry and can drive a DQCP. Gauging the subgroup $A=\mathbb{Z}_2^a\times \mathbb{Z}_2^b$ gives a $\mathrm{Rep}D_8$ dual symmetry (see Appendix~\ref{App:gaugingZ23}). The DQCP between the $\mathbb{Z}_2^c$-symmetric phase and the $\mathbb{Z}_2^a\times\mathbb{Z}_2^b$-symmetric phase is then mapped to an SSB critical point between a $\mathrm{Rep}D_8$-symmetric phase and a $\mathrm{Rep}D_8$-SSB phase. In this sense, the $E_\infty^{1,2}$ decomposition leads to a type-II DQCP: the $E_\infty^{1,2}$ data makes the dual symmetry non-invertible.

\subsection{A candidate type-II DQCP}

We start from the boundary-action formula with $\omega_3=\omega_3^\mathrm{III}$. In the group-variable basis $|\{g_i\}\rangle$, the anomalous actions of the generators $a$, $b$, and $c$ of $\Z_2^a\times\Z_2^b\times\Z_2^c$ are
\begin{align}
U_a |\{g_i\}\rangle 
&= \prod_{\langle ij\rangle} \omega_3^{s_{ij}}(g_i^{-1}g_j,g_j^{-1},a) 
|\{g_i+a\}\rangle
=|\{g_i+a\}\rangle,\\
U_b |\{g_i\}\rangle 
&= \prod_{\langle ij\rangle} \omega_3^{s_{ij}}(g_i^{-1}g_j,g_j^{-1},b) 
|\{g_i+b\}\rangle
=|\{g_i+b\}\rangle,\\
U_c |\{g_i\}\rangle 
&= \prod_{\langle ij\rangle} \omega_3^{s_{ij}}(g_i^{-1}g_j,g_j^{-1},c) 
|\{g_i+c\}\rangle
= \prod_{\langle ij\rangle} 
(-1)^{(g_i^a+g_j^a) g_j^b}
|\{g_i+c\}\rangle,
\end{align}
where $g_i=(g_i^a,g_i^b,g_i^c)\in \Z_2^a\times\Z_2^b\times\Z_2^c$ is the group variable at site $i$ of the spin chain.
For the lattice construction below, we encode each group variable $g_i\in\mathbb{Z}_2^3$ by three qubits per unit cell. In this qubit representation, the anomalous symmetry actions become
\begin{align}
U_a &= \prod_{i=2\, (\mathrm{mod}\, 3)}X_i,\\
U_b &= \prod_{i=1\, (\mathrm{mod}\, 3)}X_i,\\
U_c &= \prod_{i=0\, (\mathrm{mod}\, 3)}X_i \prod_{i=0\, (\mathrm{mod}\, 3)} (-1)^{(g_{i-1}^a+g_{i+2}^a)g_{i+1}^b}
= \prod_{i=0\, (\mathrm{mod}\, 3)} X_i \prod_{i=0\, (\mathrm{mod}\, 3)} CZ_{i-1,i+1}CZ_{i+1,i+2}.
\end{align}

The same anomalous symmetry actions can also be obtained from an anomaly-free $D_8 =\langle a,x|a^4 = x^2=1, xax=a^{-1}\rangle$ symmetry by gauging its central subgroup $\mathbb{Z}_2^{a^2}$:
\begin{equation}
    1\rightarrow \Z_2^{a^2}\rightarrow D_8 \rightarrow \Z_2^{x}\times \Z_2^{ax}\rightarrow 1.
\end{equation}
Here and below, we use a mild abuse of notation: $a$ denotes both the generator of $\mathbb{Z}_2^a \subseteq \mathbb{Z}_2^3$ and the order-four rotation generator of $D_8$.
On a spin-1/2 chain with three qubits per unit cell, the $D_8$ symmetry actions can be realized as
\begin{align}
    U_{a^2} &= \prod_i X_{3i+2},\\
    U_{x} &= \prod_i X_{3i+1},\\
    U_{ax} &= \prod_i \frac{1}{2}(1+X_{3i+2} +Z_{3i+1} - X_{3i+2}Z_{3i+1})X_{3i},\\
    U_x U_{ax} &= U_{ax} U_x U_{a^2}.
\end{align}
To gauge the subgroup $\mathbb{Z}_2^{a^2}$, introduce a $\mathbb{Z}_2$ gauge field $\sigma^x_{3i+2}, \sigma^z_{3i+2}$ and impose the Gauss-law constraint $G_{3i+2} = \sigma_{3i-1}^{x}X_{3i+2}\sigma^x_{3i+2} = 1$. The resulting dual symmetry is generated by
\begin{align}
    \widetilde{U}_{\hat{a}^2} &= \prod_i \sigma^z_{3i+2},\\
    \widetilde{U}_{x} &= \prod_i X_{3i+1},\\
    \widetilde{U}_{ax} &= \prod_i [\frac12 (1+Z_{3i+1})+\frac{1}{2}(1-Z_{3i+1})\sigma^x_{3i-1}\sigma^x_{3i+2}]X_{3i}.
\end{align}
After identifying $\sigma^x_{3i+2}$ with $Z_{3i+2}$ and $\sigma^z_{3i+2}$ with $X_{3i+2}$, these generators reproduce the type-III anomalous $\mathbb{Z}_2^3$ symmetry written above.

We now use this anomalous symmetry action to propose a candidate Hamiltonian for the type-II DQCP. The Hamiltonian interpolates between two symmetry-breaking limits. One endpoint preserves $\mathbb{Z}_2^c$ while breaking $\mathbb{Z}_2^a \times \mathbb{Z}_2^b$:
\begin{equation}
    \hat H_{\Z_2^c} = \sum_i -Z_{3i-1}Z_{3i+2} -Z_{3i-2}Z_{3i+1} - X_{3i},
    \label{Z2c}
\end{equation}
whereas the other endpoint preserves $\mathbb{Z}_2^a \times \mathbb{Z}_2^b$ while breaking $\mathbb{Z}_2^c$:
\begin{equation}
\begin{aligned}
    \hat H_{\Z_2^a\times\Z_2^b} = \sum_i &- \frac{1+Z_{3i}Z_{3i+3}}{2}(X_{3i+1}\frac{1+Z_{3i-1}Z_{3i+2}}{2} + X_{3i+2}\frac{1+Z_{3i+1}Z_{3i+4}}{2}) \\&- \frac{Z_{3i} + Z_{3i+3}}{2}(X_{3i+1}\frac{1-Z_{3i-1}Z_{3i+2}}{2} + X_{3i+2}\frac{1-Z_{3i+1}Z_{3i+4}}{2}),
\end{aligned}
\end{equation}
\begin{equation}
\hat H_{\mathbb{Z}_2^a\times\mathbb{Z}_2^b \rightarrow\mathbb{Z}_2^c}(\lambda) = \lambda\hat H_{\Z_2^c} + (1-\lambda)\hat H_{\Z_2^a \times \Z_2^b}.
\label{Z2abZ2c}
\end{equation}
The projector factors in $\hat H_{\Z_2^a\times\Z_2^b}$ have a simple anomaly-based explanation. When a $\mathbb{Z}_2^a$ domain wall is present, i.e. when $\frac{1-Z_{3 n-1} Z_{3 n+2}}{2} = 1$, the operator $X_{3n+1}$ flips a $\mathbb{Z}_2^b$ spin and moves a $\mathbb{Z}_2^b$ domain wall across it. This operation exchanges the relative positions of the $\mathbb{Z}_2^a$ and $\mathbb{Z}_2^b$ domain walls. Because of the type-III anomaly, the exchange creates a $\mathbb{Z}_2^c$ charge. Therefore, the corresponding spin-flip operator must be dressed by an operator, here $Z_{3i}$, that is odd under $\mathbb{Z}_2^c$.

The endpoint Hamiltonian $\hat H_{\Z_2^a\times\Z_2^b}$ is frustration-free, with two exact ground states
\begin{align}
    |\Omega_1\rangle &= [\otimes_i|0\rangle_{3i}]_c \otimes |\mathrm{trivial}\rangle_{ab},\\
    |\Omega_2\rangle &= [\otimes_i|1\rangle_{3i}]_c \otimes |\mathrm{cluster}\rangle_{ab},
\end{align}
where $|\mathrm{trivial}\rangle_{ab}$ and $|\mathrm{cluster}\rangle_{ab}$ denote the trivial product state and the cluster SPT state of the $\mathbb{Z}_2^a\times\mathbb{Z}_2^b$ spin chain on sites $\{3i+1, 3i+2\mid i\in \mathbb{Z}\}$. These two states form the regular representation of $\mathbb{Z}_2^c$:
\begin{equation}
    \prod_iX_{3i} CZ_{3i+1,3i-1}CZ_{3i+1,3i+2} |\Omega_1\rangle = |\Omega_2\rangle.
\end{equation}

At $\lambda = 0$, the ground states $|\Omega_1\rangle$ and $|\Omega_2\rangle$ spontaneously break $\mathbb{Z}_2^c$ symmetry while preserving $\mathbb{Z}_2^{a}\times \mathbb{Z}_2^b$ symmetry. Increasing $\lambda$ restores $\mathbb{Z}_2^c$ symmetry by condensing $\mathbb{Z}_2^c$ domain walls. The controlled-$Z$ dressing string in $U_c$ then forces these domain walls to carry $\mathbb{Z}_2^{a}\times \mathbb{Z}_2^b$ projective representations. This follows because the controlled-$Z$ circuit maps the trivial paramagnet to the nontrivial $\mathbb{Z}_2^a\times \mathbb{Z}_2^b$ SPT phase: a truncated controlled-$Z$ string creates the corresponding SPT edge mode at its endpoint, and this edge mode transforms projectively under $\mathbb{Z}_2^a\times \mathbb{Z}_2^b$. As a result, the proliferation of $\mathbb{Z}_2^c$ domain walls also proliferates $\mathbb{Z}_2^a\times \mathbb{Z}_2^b$ projective representations, which necessarily break $\mathbb{Z}_2^a\times \mathbb{Z}_2^b$ symmetry and can drive a DQCP.

Gauging the $\mathbb{Z}_2^a$ symmetry maps this $\Z_2^a\times\Z_2^b\rightarrow\Z_2^c \subseteq \mathbb{Z}_2^3$ DQCP to a $\Z_2^{x} \rightarrow \Z_2^{ax} \times \Z_2^{a^3x} \subseteq D_8$ critical point. Gauging the dual $D_8$ symmetry then maps the transition further to a $\mathrm{Rep}D_8$ SSB critical point.

\subsection{Numerical evidence}

\begin{figure}[htbp]
    \centering
    \begin{subfigure}{0.45\textwidth}
        \centering
        \includegraphics[width=\linewidth]{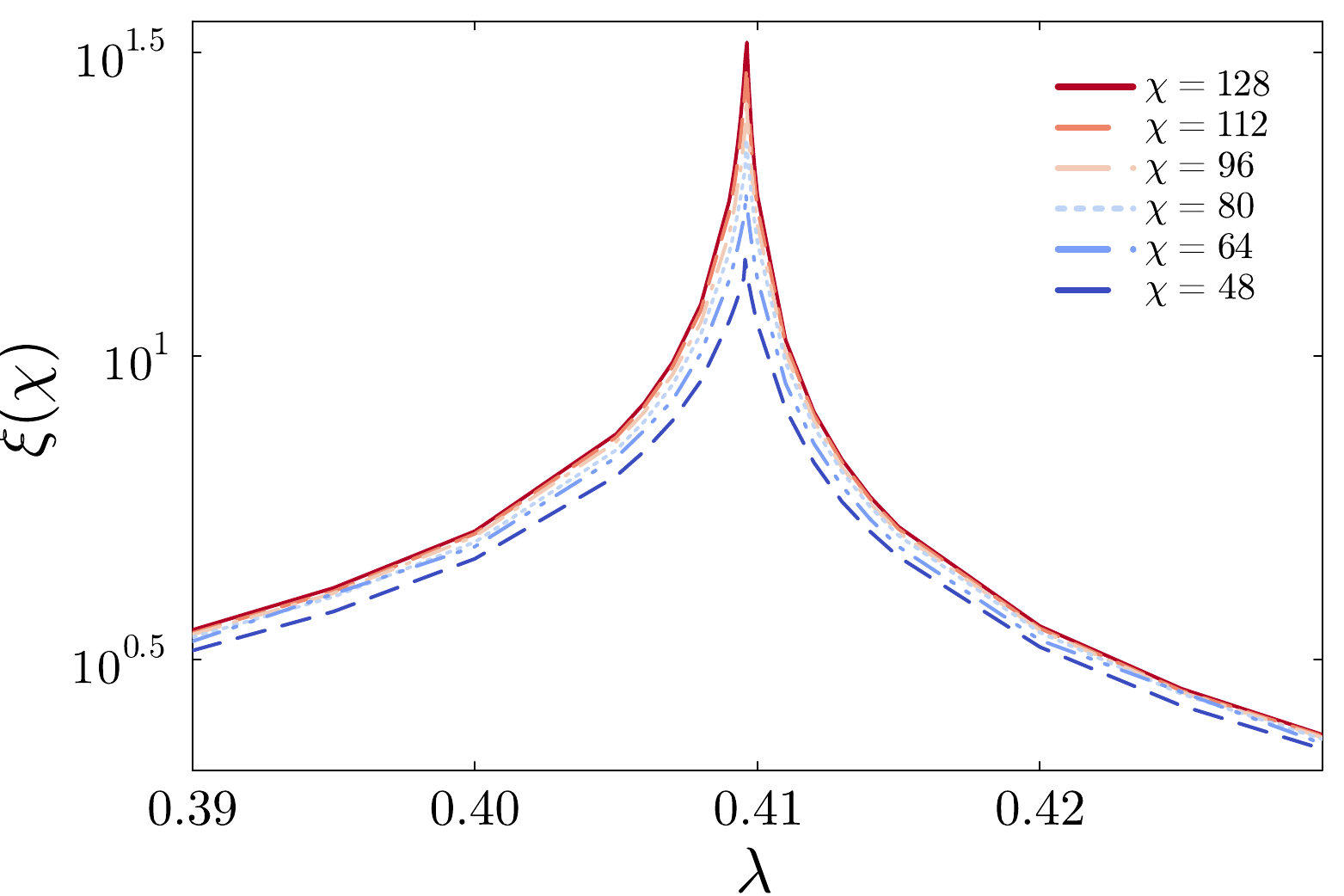}
    \end{subfigure}
    \begin{subfigure}{0.45\textwidth}
        \centering
        \includegraphics[width=\linewidth]{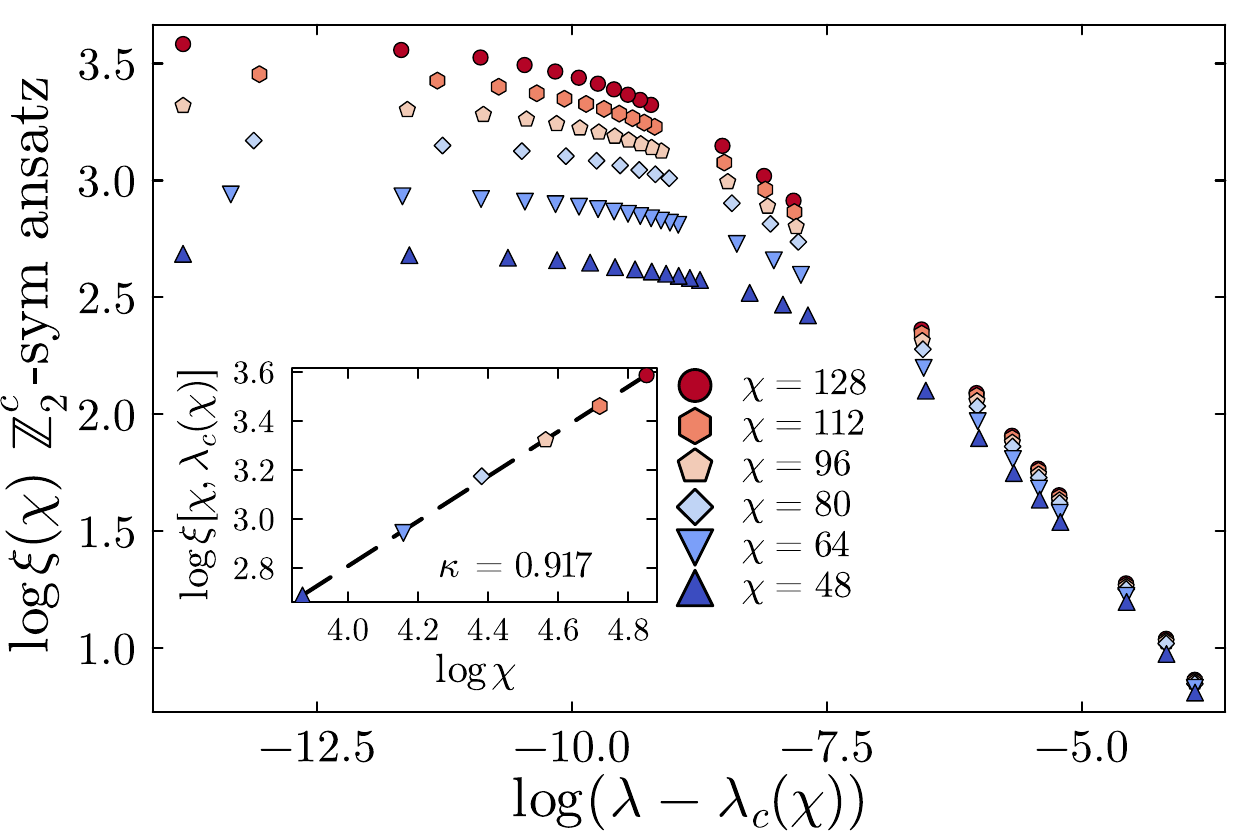}
    \end{subfigure}
    \caption{Growing peak of the MPS correlation length $\xi(\chi)$ in the $\mathbb{Z}_2^a\times\mathbb{Z}_2^b\rightarrow\mathbb{Z}_2^c$ transition. The peak value shows power-law scaling with the bond dimension $\chi$, suggesting a continuous phase transition.}
    \label{fig:Z23corscaling}
\end{figure}

We now provide numerical evidence that the interpolation in Eq.~\eqref{Z2abZ2c} realizes a continuous transition rather than a first-order level crossing. As shown in Fig.~\ref{fig:Z23corscaling}, the MPS correlation length develops a growing peak near the $\mathbb{Z}_2^a\times\mathbb{Z}_2^b\rightarrow\mathbb{Z}_2^c$ transition. The peak value shows power-law scaling with the bond dimension, providing suggestive evidence for a continuous DQCP.

This behavior is the standard finite-entanglement signature of criticality: a finite-$\chi$ MPS cuts off the correlation length, so a continuous transition appears as a peak whose height grows with $\chi$. By contrast, at a strongly first-order transition between two gapped phases, the correlation length is expected to saturate once the two competing states are well resolved. We therefore regard the data as evidence for criticality, although we do not attempt to determine the full universality class of this appendix example. The result is consistent with the type-II DQCP picture, in which proliferating $\mathbb{Z}_2^c$ domain walls carry $\mathbb{Z}_2^a\times\mathbb{Z}_2^b$ projective representations.

\subsection{Subgroup gauging in $\mathrm{Vec}_{\Z_2^3}^{\omega}$ for all $\omega\in H^3(\Z_2^3,\U)$}\label{App:gaugingZ23}

This subsection summarizes subgroup gaugings of $\mathrm{Vec}_{\Z_2^3}^{\omega}$ for all 3-cocycles $\omega\in H^3(\Z_2^3,\U)$. For each gaugeable algebra $A$, the gauged symmetry is the $A$-bimodule category $\ACA$, which is group-theoretical and Morita equivalent to $\mathrm{Vec}_{\Z_2^3}^{\omega}$. The tables below list one representative from each tensor-equivalence class of the twisted pointed categories $\mathrm{Vec}_{\mathbb{Z}_2^3}^\omega$. For a pair $(H,\gamma)$, we use $0$ for the trivial cohomology class in $H^2(H, U(1))$ and $\gamma$ for a nontrivial class. The Morita equivalence classes are determined using Proposition~\ref{algebraMorita}. In what follows, $A \overset{\mathrm{M}}{\sim} B$ means that the algebras $A$ and $B$ are Morita equivalent.

\begin{table}[h!]
        \centering
        
\begin{tabular}{|c|c|}
\hline 
 \multicolumn{2}{|c|}{$\displaystyle \mathcal C = ( G,\omega) =\left(\mathbb{Z}_{2}^{a} \times \mathbb{Z}_{2}^{b} \times \Z_{2}^{c} ,0\right)$} \\
\hline \hline 
 $\displaystyle A=( H,\gamma )$ & $ $$\displaystyle _{A}\mathcal{C}_{A}$ \\
\hline 
 $\displaystyle ( 1,0)$ & $\displaystyle \mathrm{Vec}_{\mathbb{Z}_{2}^{a} \times \mathbb{Z}_{2}^{b} \times \Z_{2}^{c}}$ \\
\hline 
 $\displaystyle (\mathbb{Z}_{2},0)$ & $\displaystyle \mathrm{Vec}_{\mathbb{Z}_{2}^{a} \times \mathbb{Z}_{2}^{b} \times \Z_{2}^{c}}$ \\
\hline 
 $\displaystyle (\mathbb{Z}_{2} \times \mathbb{Z}_{2} ,0), (\mathbb{Z}_{2} \times \mathbb{Z}_{2} ,\gamma) $ & $\displaystyle \mathrm{Vec}_{\mathbb{Z}_{2}^{a} \times \mathbb{Z}_{2}^{b} \times \Z_{2}^{c}}$ \\
\hline 
 $\displaystyle \left(\mathbb{Z}_{2}^{a} \times \mathbb{Z}_{2}^{b} \times \Z_{2}^{c} ,0\right)$ & $\displaystyle \mathrm{Vec}_{\mathbb{Z}_{2}^{a} \times \mathbb{Z}_{2}^{b} \times \Z_{2}^{c}}$ \\
 \hline
\end{tabular}
\caption{\label{Z230} Condensable algebras in $\mathrm{Vec}_{\mathbb{Z}_2^a \times \mathbb{Z}_2^b \times \mathbb{Z}_2^c}$ and the resulting gauged symmetries. For brevity, notation such as $(\mathbb{Z}_2, 0)$ and $(\mathbb{Z}_2 \times \mathbb{Z}_2, 0)$ denotes all distinct subgroups isomorphic to $\mathbb{Z}_2$ and $\mathbb{Z}_2 \times \mathbb{Z}_2$, respectively.}
\end{table}

\begin{table}[!h]
        \centering
        
\begin{tabular}{|c|c|}
\hline 
 \multicolumn{2}{|c|}{$\displaystyle \mathcal C = ( G,\omega) =\left(\mathbb{Z}_{2}^{a} \times \mathbb{Z}_{2}^{b} \times \Z_{2}^{c} ,aaa\right)$} \\
\hline \hline 
 $\displaystyle A = ( H,\gamma )$ & $ $$\displaystyle _{A}\mathcal{C}_{A}$ \\
\hline 
 $\displaystyle ( 1,0)$ & $\displaystyle \mathrm{Vec}_{\mathbb{Z}_{2}^{a} \times \mathbb{Z}_{2}^{b} \times \Z_{2}^{c}}^{aaa}$ \\
\hline 
 $\displaystyle \left(\mathbb{Z}_{2}^{b} ,0\right) ,\left(\mathbb{Z}_{2}^{c} ,0\right), \left(\mathbb{Z}_{2}^{bc} ,0\right)$ & $\displaystyle \mathrm{Vec}_{\mathbb{Z}_{2}^{a} \times \mathbb{Z}_{2}^{b} \times \Z_{2}^{c}}^{aaa}$ \\
\hline 
 $\displaystyle \left(\mathbb{Z}_{2}^{b} \times \mathbb{Z}_{2}^{c} ,0\right) ,\left(\mathbb{Z}_{2}^{b} \times \mathbb{Z}_{2}^{c} ,\gamma \right)$ & $\displaystyle \mathrm{Vec}_{\mathbb{Z}_{2}^{a} \times \mathbb{Z}_{2}^{b} \times \Z_{2}^{c}}^{aaa}$ \\
 \hline
\end{tabular}
\caption{\label{Z23aaa} Condensable algebras in $\mathrm{Vec}_{\mathbb{Z}_2^{a} \times \mathbb{Z}_2^b \times \mathbb{Z}_2^c}^{aaa}$ and the resulting gauged symmetries.}      
        \end{table}

\begin{table}[!h]
        \centering
        
\begin{tabular}{|c|c|}
\hline 
 \multicolumn{2}{|c|}{$\displaystyle \mathcal C = ( G,\omega) =\left(\mathbb{Z}_{2}^{a} \times \mathbb{Z}_{2}^{b} \times \Z_{2}^{c} ,abb\right)$ ($\simeq  aaa+ abb $)} \\
\hline \hline 
 $\displaystyle A = ( H,\gamma )$ & $ $$\displaystyle _{A}\mathcal{C}_{A}$ \\
\hline 
 $\displaystyle ( 1,0)$ & $\displaystyle \mathrm{Vec}_{\mathbb{Z}_{2}^{a} \times \mathbb{Z}_{2}^{b} \times \Z_{2}^{c}}^{abb}$ \\
\hline 
 $\displaystyle \left(\mathbb{Z}_{2}^{a} ,0\right) ,\left(\mathbb{Z}_{2}^{b} ,0\right)$ & $\displaystyle \mathrm{Vec}_{\mathbb{Z}_{2} \times \mathbb{Z}_{4}}$ \\
\hline 
 $\displaystyle \left(\mathbb{Z}_{2}^{c} ,0\right)$ & $\displaystyle \mathrm{Vec}_{\mathbb{Z}_{2}^{a} \times \mathbb{Z}_{2}^{b} \times \Z_{2}^{c}}^{abb}$ \\
\hline 
 $\displaystyle \left(\mathbb{Z}_{2}^{ac} ,0\right) ,\left(\mathbb{Z}_{2}^{bc} ,0\right)$ & $\displaystyle \mathrm{Vec}_{\mathbb{Z}_{2}^{a} \times \mathbb{Z}_{2}^{b} \times \Z_{2}^{c}}^{abb}$ \\
\hline 
 $\displaystyle \left(\mathbb{Z}_{2}^{b} \times \mathbb{Z}_{2}^{c} ,0\right) ,\left(\mathbb{Z}_{2}^{b} \times \mathbb{Z}_{2}^{c} ,\gamma \right) ,\left(\mathbb{Z}_{2}^{a} \times \mathbb{Z}_{2}^{c} ,0\right) ,\left(\mathbb{Z}_{2}^{a} \times \mathbb{Z}_{2}^{c} ,\gamma \right)$ & $\displaystyle \mathrm{Vec}_{\mathbb{Z}_{2} \times \mathbb{Z}_{4}}$ \\
 \hline
\end{tabular}
\caption{\label{Z23abb} Condensable algebras in $\mathrm{Vec}_{\mathbb{Z}_2^{a} \times \mathbb{Z}_2^b \times \mathbb{Z}_2^c}^{abb}$ and the resulting gauged symmetries.}     
        \end{table}

\begin{table}[!h]
        \centering
        
\begin{tabular}{|c|c|}
\hline 
 \multicolumn{2}{|c|}{$\displaystyle \mathcal C = ( G,\omega) =\left(\mathbb{Z}_{2}^{a} \times \mathbb{Z}_{2}^{b} \times \Z_{2}^{c} ,abb+bcc+acc\right)$($\displaystyle \simeq aaa+bcc$)} \\
\hline \hline 
 $\displaystyle A = ( H,\gamma )$ & $ $$\displaystyle _{A}\mathcal{C}_{A}$ \\
\hline 
 $\displaystyle ( 1,0)$ & $\displaystyle \mathrm{Vec}_{\mathbb{Z}_{2}^{a} \times \mathbb{Z}_{2}^{b} \times \Z_{2}^{c}}^{abb+bcc+acc}$ \\
\hline 
 $\displaystyle \left(\mathbb{Z}_{2}^{a} ,0\right) ,\left(\mathbb{Z}_{2}^{b} ,0\right) ,\left(\mathbb{Z}_{2}^{c} ,0\right)$ & $\displaystyle \mathrm{Vec}_{\mathbb{Z}_{2} \times \mathbb{Z}_{4}}^{\nu ^{2}}$ \\
 \hline
\end{tabular}
\caption{\label{Z23abb+bcc+acc} Condensable algebras in $\mathrm{Vec}_{\mathbb{Z}_2^{a} \times \mathbb{Z}_2^b \times \mathbb{Z}_2^c}^{abb+bcc+acc}$ and the resulting gauged symmetries.}        
        \end{table}

\begin{table}[!h]
        \centering
        
\begin{tabular}{|c|c|}
\hline 
 \multicolumn{2}{|c|}{$\displaystyle \mathcal C = ( G,\omega) =\left(\mathbb{Z}_{2}^{a} \times \mathbb{Z}_{2}^{b} \times \Z_{2}^{c} ,aaa+bbb+abb\right)$} \\
\hline \hline 
 $\displaystyle A = ( H,\gamma )$ & $ $$\displaystyle _{A}\mathcal{C}_{A}$ \\
\hline 
 $\displaystyle ( 1,0)$ & $\displaystyle \mathrm{Vec}_{\mathbb{Z}_{2}^{a} \times \mathbb{Z}_{2}^{b} \times \Z_{2}^{c}}^{aaa+bbb+abb}$ \\
\hline 
 $\displaystyle \left(\mathbb{Z}_{2}^{c} ,0\right)$ & $\displaystyle \mathrm{Vec}_{\mathbb{Z}_{2}^{a} \times \mathbb{Z}_{2}^{b} \times \Z_{2}^{c}}^{aaa+bbb+abb}$ \\
\hline 
 $\displaystyle \left(\mathbb{Z}_{2}^{ac} ,0\right) ,\left(\mathbb{Z}_{2}^{ab} ,0\right)$ & $\displaystyle \mathrm{Vec}_{\mathbb{Z}_{2}^{a} \times \mathbb{Z}_{2}^{b} \times \Z_{2}^{c}}^{aaa+bbb+abb}$ \\
 \hline
\end{tabular}
\caption{\label{Z23aaa+bbb+abb} Condensable algebras in $\mathrm{Vec}_{\mathbb{Z}_2^{a} \times \mathbb{Z}_2^b \times \mathbb{Z}_2^c}^{aaa+bbb+abb}$ and the resulting gauged symmetries.}  
        \end{table}

\begin{table}[!h]
        \centering
        
\begin{tabular}{|c|c|}
\hline 
 \multicolumn{2}{|c|}{$\displaystyle \mathcal C = ( G,\omega) =\left(\mathbb{Z}_{2}^{a} \times \mathbb{Z}_{2}^{b} \times \Z_{2}^{c} ,abb+acc+abc\right)$($\displaystyle \simeq aaa+abc$)} \\
\hline \hline 
 $\displaystyle A = ( H,\gamma )$ & $ $$\displaystyle _{A}\mathcal{C}_{A}$ \\
\hline 
 $\displaystyle ( 1,0)$ & $\displaystyle \mathrm{Vec}_{\mathbb{Z}_{2}^{a} \times \mathbb{Z}_{2}^{b} \times \Z_{2}^{c}}^{abb+acc+abc}$ \\
\hline 
 $\displaystyle \left(\mathbb{Z}_{2}^{b} ,0\right) ,\left(\mathbb{Z}_{2}^{c} ,0\right) ,\left(\mathbb{Z}_{2}^{bc} ,0\right)$ & $\displaystyle \mathrm{Vec}_{D_{8}}^{\beta }$ \\
\hline 
 $\displaystyle \left(\mathbb{Z}_{2}^{a} ,0\right)$ & $\displaystyle \mathrm{Vec}_{Q_{8}}$ \\
\hline 
 $\displaystyle \left(\mathbb{Z}_{2}^{b} \times \mathbb{Z}_{2}^{c} ,0\right)\overset{\mathrm{M}}{\sim} \left(\mathbb{Z}_{2}^{b} \times \mathbb{Z}_{2}^{c} ,\gamma \right)$ & $\displaystyle \mathrm{Rep}(Q_{8})$ \\
 \hline
\end{tabular}
\caption{\label{Z23aaa+abc} Condensable algebras in $\mathrm{Vec}_{\mathbb{Z}_2^{a} \times \mathbb{Z}_2^b \times \mathbb{Z}_2^c}^{aaa+abc}$ and the resulting gauged symmetries.} 
        \end{table}

\begin{table}[!h]
        \centering
        
\begin{tabular}{|c|c|}
\hline 
 \multicolumn{2}{|c|}{$\displaystyle \mathcal C = ( G,\omega) =\left(\mathbb{Z}_{2}^{a} \times \mathbb{Z}_{2}^{b} \times \Z_{2}^{c} ,aaa+bcc+abc\right)$} \\
\hline \hline 
 $\displaystyle A = ( H,\gamma )$ & $ $$\displaystyle _{A}\mathcal{C}_{A}$ \\
\hline 
 $\displaystyle ( 1,0)$ & $\displaystyle \mathrm{Vec}_{\mathbb{Z}_{2}^{a} \times \mathbb{Z}_{2}^{b} \times \Z_{2}^{c}}^{aaa+bcc+abc}$ \\
\hline 
 $\displaystyle \left(\mathbb{Z}_{2}^{b} ,0\right) ,\left(\mathbb{Z}_{2}^{c} ,0\right)$ & $\displaystyle \mathrm{Vec}_{D_{8}}^{\alpha }$ \\
 \hline
\end{tabular}
\caption{\label{Z23aaa+bcc+abc} Condensable algebras in $\mathrm{Vec}_{\mathbb{Z}_2^{a} \times \mathbb{Z}_2^b \times \mathbb{Z}_2^c}^{aaa+bcc+abc}$ and the resulting gauged symmetries.}   
        \end{table}

\begin{table}[!h]
        \centering
        
\begin{tabular}{|c|c|}
\hline 
 \multicolumn{2}{|c|}{$\displaystyle \mathcal C = ( G,\omega) =\left(\mathbb{Z}_{2}^{a} \times \mathbb{Z}_{2}^{b} \times \Z_{2}^{c} ,abb+bcc+acc+abc\right)$} \\
\hline \hline 
 $\displaystyle A = ( H,\gamma )$ & $ $$\displaystyle _{A}\mathcal{C}_{A}$ \\
\hline 
 $\displaystyle ( 1,0)$ & $\displaystyle \mathrm{Vec}_{\mathbb{Z}_{2}^{a} \times \mathbb{Z}_{2}^{b} \times \Z_{2}^{c}}^{abb+bcc+acc+abc}$ \\
\hline 
 $\displaystyle \left(\mathbb{Z}_{2}^{a} ,0\right) ,\left(\mathbb{Z}_{2}^{b} ,0\right) ,\left(\mathbb{Z}_{2}^{c} ,0\right)$ & $\displaystyle \mathrm{Vec}_{Q_{8}}^{4z}$ \\
 \hline
\end{tabular}
\caption{\label{Z23abb+bcc+acc+abc} Condensable algebras in $\mathrm{Vec}_{\mathbb{Z}_2^{a} \times \mathbb{Z}_2^b \times \mathbb{Z}_2^c}^{abb+bcc+acc+abc}$ and the resulting gauged symmetries.}     
        \end{table}

\begin{table}[!h]
        \centering
        
\begin{tabular}{|c|c|}
\hline 
 \multicolumn{2}{|c|}{$\displaystyle \mathcal C = ( G,\omega) =\left(\mathbb{Z}_{2}^{a} \times \mathbb{Z}_{2}^{b} \times \Z_{2}^{c} ,abc\right)$($\simeq abb+abc \simeq aaa+abb+acc+abc$)} \\
\hline \hline 
 $\displaystyle A = ( H,\gamma )$ & $ $$\displaystyle _{A}\mathcal{C}_{A}$ \\
\hline 
 $\displaystyle ( 1,0)$ & $\displaystyle \mathrm{Vec}_{\mathbb{Z}_{2}^{a} \times \mathbb{Z}_{2}^{b} \times \Z_{2}^{c}}^{abc}$ \\
\hline 
 $\displaystyle \left(\mathbb{Z}_{2}^{a} ,0\right) ,\left(\mathbb{Z}_{2}^{b} ,0\right) ,\left(\mathbb{Z}_{2}^{c} ,0\right) ,\left(\mathbb{Z}_{2}^{ac} ,0\right) ,\left(\mathbb{Z}_{2}^{ab} ,0\right) ,\left(\mathbb{Z}_{2}^{bc} ,0\right)$ & $\displaystyle \mathrm{Vec}_{D_{8}}$ \\
\hline 
 $\displaystyle  \begin{array}{{>{\displaystyle}l}}
\left(\mathbb{Z}_{2}^{a} \times \mathbb{Z}_{2}^{b} ,0\right) \overset{\mathrm{M}}{\sim}\left(\mathbb{Z}_{2}^{a} \times \mathbb{Z}_{2}^{b} ,\gamma \right) ,\left(\mathbb{Z}_{2}^{b} \times \mathbb{Z}_{2}^{c} ,0\right) \overset{\mathrm{M}}{\sim}
\left(\mathbb{Z}_{2}^{b} \times \mathbb{Z}_{2}^{c} ,\gamma \right) ,\\ \left(\mathbb{Z}_{2}^{a} \times \mathbb{Z}_{2}^{c} ,0\right) \overset{\mathrm{M}}{\sim}\left(\mathbb{Z}_{2}^{a} \times \mathbb{Z}_{2}^{c} ,\gamma \right), \left(\mathbb{Z}_{2}^{ab} \times \mathbb{Z}_{2}^{bc} ,0\right) \overset{\mathrm{M}}{\sim}\left(\mathbb{Z}_{2}^{ab} \times \mathbb{Z}_{2}^{bc} ,\gamma \right),
\end{array}$ & $\displaystyle \mathrm{Rep}(D_{8})$ \\
 \hline
\end{tabular}
\caption{\label{Z23abc} Condensable algebras in $\mathrm{Vec}_{\mathbb{Z}_2^{a} \times \mathbb{Z}_2^b \times \mathbb{Z}_2^c}^{abc}$ and the resulting gauged symmetries.}       
        \end{table}

\begin{table}[!h]
        \centering
        
\begin{tabular}{|c|c|}
\hline 
 \multicolumn{2}{|c|}{$\displaystyle \mathcal C = ( G,\omega) =\left(\mathbb{Z}_{2}^{a} \times \mathbb{Z}_{2}^{b} \times \Z_{2}^{c} ,aaa+bbb+ccc+abb+bcc+acc+abc\right)$} \\
\hline \hline 
 $\displaystyle A = ( H,\gamma )$ & $ $$\displaystyle _{A}\mathcal{C}_{A}$ \\
\hline 
 $\displaystyle ( 1,0)$ & $\displaystyle \mathrm{Vec}_{\mathbb{Z}_{2}^{a} \times \mathbb{Z}_{2}^{b} \times \Z_{2}^{c}}^{aaa+bbb+ccc+abb+bcc+acc+abc}$ \\
 \hline
\end{tabular}
\caption{\label{Z23aaa+bbb+ccc+abb+bcc+acc+abc} Condensable algebras in $\mathrm{Vec}_{\mathbb{Z}_2^{a} \times \mathbb{Z}_2^b \times \mathbb{Z}_2^c}^{aaa+bbb+ccc+abb+bcc+acc+abc}$ and the resulting gauged symmetries.} 
        \end{table}

Table~\ref{Morita} lists Morita-equivalent pointed fusion categories of dimension 8~\cite{Mu_oz_2018}. We use the following notation for the relevant 3-cocycle classes.

For $\mathbb{Z}_8$, we write
$
H^*(\mathbb{Z}_8,\mathbb{Z})=\mathbb{Z}[t]/(8t),
$
where $|t|=2$.
For $\mathbb{Z}_{4}\times \mathbb{Z}_2$, with generators $\langle v \rangle\times \langle u \rangle$, we use the following notation:
  $$
    H^k(\mathbb{Z}_{4}\times \mathbb{Z}_2,\mathbb{Z}) = 
     \begin{cases}
      \mathbb{Z}, &k=0 \\
      0 & k=1 \\
      \mathbb{Z}_{4}\times \mathbb{Z}_2= \langle v \rangle\times \langle u \rangle  & k=2 \\
      \mathbb{Z}_{4}\times \mathbb{Z}_2 \times\mathbb{Z}_2 =\langle v^2 \rangle\times \langle uv \rangle \times \langle v^2 \rangle & k=4 \\

      \cdots \\
      \end{cases}
    $$

The cohomology ring of $Q_8$ is~\cite[Theorem 3.6]{tomoda2008remarks}:
$$
H^*(Q_8, \mathbb{Z}) = \mathbb{Z}[x,y,z] / (2x, 2y, 8z, xy-4z, x^2, y^2),
$$
where the generators have degrees $|x| = |y| = 2$ and $|z| = 4$. 

Let $D_8 = \langle x,a \mid x^2 = a^4 = 1, xa^i = a^{-i}x \rangle$ be the dihedral group of order 8. We parameterize its third cohomology group as $H^3(D_8, U(1)) = \mathbb{Z}_2 \times \mathbb{Z}_2 \times \mathbb{Z}_4 \simeq \langle \alpha \rangle \times \langle \beta \rangle \times \langle \zeta \rangle$. The automorphism group $\mathrm{Aut}(D_8)$ is generated by the maps $\phi$ and $\theta$, defined by
$$
\phi: a \mapsto a, \quad x \mapsto xa,
$$
$$
\theta: a \mapsto a^{-1}, \quad x \mapsto x.
$$
It is known that $\mathrm{Aut}(D_8) \simeq D_8$, whose inner automorphism subgroup is generated by $\phi^2$ and $\theta$. We choose the 3-cocycles $\alpha, \beta, \zeta$ so that their pullbacks satisfy $\phi^*(\beta) = \alpha + \beta$, $\phi^*(\alpha) = \alpha$, and $\phi^*(\zeta) = \zeta$~\cite{Mu_oz_2018}. Their restrictions to the relevant subgroups are $\alpha|_{\langle a \rangle} = 0$, $\beta|_{\{1, x, a^2, xa^2\}} = 0$, $(\alpha + \beta)|_{\{1, ax, a^2, a^3x\}} = 0$, and $\zeta|_{\langle x \rangle}=\zeta|_{\langle ax \rangle}=0$.

\begin{table}[!h]
        \centering
        
\begin{tabular}{|c|c|}
\hline 
 $\displaystyle ( G_{1} ,\omega _{1})$ & $\displaystyle ( G_{2} ,\omega _{2})$ \\
\hline \hline 
 $\displaystyle \left(\mathbb{Z}_{2}^{3} ,acc\right)$ & $\displaystyle (\mathbb{Z}_{4} \times \mathbb{Z}_{2} ,0)$ \\
\hline 
 $\displaystyle \left(\mathbb{Z}_{2}^{3} ,abb+bcc+acc\right)$ & $\displaystyle \left(\mathbb{Z}_{4} \times \mathbb{Z}_{2} ,v^{2}\right)$ \\
\hline 
 $\displaystyle \left(\mathbb{Z}_{2}^{3} ,abc\right)$ & $\displaystyle ( D_{8} ,0)$ \\
\hline 
 $\displaystyle \left(\mathbb{Z}_{2}^{3} ,aaa+bcc+abc\right)$ & $\displaystyle ( D_{8},\alpha)$ \\
\hline 
 $\displaystyle \left(\mathbb{Z}_{2}^{3} ,abb+bcc+acc+abc\right)$ & $\displaystyle ( Q_{8} ,4z =xy )$ \\
\hline 
 $\displaystyle (\mathbb{Z}_{4} \times \mathbb{Z}_{2} ,uv)$ & $\displaystyle (\mathbb{Z}_{8} ,0)$ \\
\hline 
 $\displaystyle \left(\mathbb{Z}_{4} \times \mathbb{Z}_{2} ,uv+u^{2}\right)$ & $\displaystyle \left(\mathbb{Z}_{8} ,4t^{2}\right)$ \\
 \hline
\end{tabular}
        \caption{Morita-equivalent pointed fusion categories of dimension 8.}
        \label{Morita}
        \end{table}
        
In addition, $(\mathbb{Z}_2^3,abb+acc+abc)\simeq_M (D_8, \beta)\simeq_M (Q_8,0)$. Here, $\simeq_M$ denotes Morita equivalence of fusion categories.

\section{Lyndon-Hochschild-Serre spectral sequence for $D_8=(\Z_2^{ax}\times\Z_2^{a^3x})\rtimes \Z_2^x$}
\label{App:LHS}

Let $A=\mathbb{Z}_2^{ax}\times \mathbb{Z}_2^{a^3x}=\{1,ax,a^2,a^3x\}$ be the normal Klein subgroup of $D_8$. The quotient is $Q=D_8/A=\mathbb{Z}_2^x=\{1,\bar{x}\}$, where we use $x$ as a lift of $\bar{x}$. Conjugation by $x$ acts nontrivially on $A$ by exchanging the two reflection generators:
$$
\quad ax\mapsto a^3x,\quad a^3x\mapsto ax.
$$
Thus the LHS spectral sequence for $A\to D_8\to Q$ is
$$
E_2^{p,q}=H^p(Q,H^q(A,U(1)))\Longrightarrow H^{p+q}(D_8,U(1)),
$$
where $Q$ acts on the coefficient group $H^q(A,U(1))$ through this conjugation action. From the explicit cocycles of $\mathbb{Z}_2\times \mathbb{Z}_2$ in degrees $1,2,3$, the invariant subgroups are
$$
H^0(Q,H^1(A,U(1)))= H^1(A,U(1))^Q=\mathbb{Z}_2,
$$
$$
H^0(Q,H^2(A,U(1)))= H^2(A,U(1))^Q=\mathbb{Z}_2,
$$
$$
H^0(Q,H^3(A,U(1)))= H^3(A,U(1))^Q=\mathbb{Z}_2 \times \mathbb{Z}_2.
$$

The higher terms $H^p(Q,H^1(A,U(1)))=H^p(\mathbb{Z}_2,\mathbb{Z}_2\times \mathbb{Z}_2)$ can be computed from the normalized bar complex. A normalized $n$-cochain is determined by its value on $(\bar{x},\dots,\bar{x})$, so
$C^n(Q,\mathbb{Z}_2\times \mathbb{Z}_2)\simeq \mathbb{Z}_2\times \mathbb{Z}_2$. If $f(\bar{x},\dots,\bar{x})=m$, then
$d f(\bar{x},\dots,\bar{x})=\bar{x}\cdot m+m$. Hence the complex is
$$
0\to C^0(Q,\mathbb{Z}_2\times \mathbb{Z}_2) \stackrel{d}{\longrightarrow} C^1(Q,\mathbb{Z}_2\times \mathbb{Z}_2) \stackrel{d}{\longrightarrow} C^2(Q,\mathbb{Z}_2\times \mathbb{Z}_2)\stackrel{d}{\longrightarrow} ...
= 0\to \mathbb{Z}_2\times \mathbb{Z}_2 \stackrel{1+\bar{x}}{\longrightarrow} \mathbb{Z}_2\times \mathbb{Z}_2 \stackrel{1+\bar{x}}{\longrightarrow}\mathbb{Z}_2\times \mathbb{Z}_2 \stackrel{1+\bar{x}}{\longrightarrow}... 
$$

The map $1+\bar{x}$ has the same image and kernel, namely the diagonal copy of $\mathbb{Z}_2$. Therefore
$$
H^n(Q,H^1(A,U(1)))=0, \quad n>0.
$$
The relevant part of the $E_2$ page is therefore
   \begin{equation}\label{E2page}
    \begin{tikzpicture}
\matrix [matrix of math nodes,row sep=6mm]
{
 3 &  [5mm]  |(a)|  {\Z_2\times \Z_2} & [5mm]   & [5mm]  & [5mm] & [5mm] & [5mm] \\
2 & |(b)| \Z_2 & |(c)|  {\Z_2}  & \Z_2 & & & \\
1&  \Z_2 & |(d)|  0 & |(e)| {0} & 0 & & \\
0& U(1) &  \Z_2 & |(f)| 0 & |(g)|  {\Z_2} & 0 \\
& 0 & 1 & 2 & 3& 4&\\
};
\tikzstyle{every node}=[midway,auto,font=\scriptsize]
\draw[thick] (-2.8,-1.7) -- (-2.8,2.8) ;
\draw[thick] (-2.8,-1.7) -- (4.0,-1.7) ;
\end{tikzpicture}
\end{equation}

Recall that $H^1(D_8,U(1))=\mathbb{Z}_2\times \mathbb{Z}_2$, $H^2(D_8,U(1))=\mathbb{Z}_2$, and $H^3(D_8,U(1))=\mathbb{Z}_2^2\times \mathbb{Z}_4$. These cohomology groups are obtained from the associated graded pieces of the LHS spectral sequence by solving the corresponding extension problems. In the notation of Appendix~\ref{App:gaugingZ23}, we have $(\alpha+ \beta)|_{\{ 1,ax,a^2,a^3x\}} =0$. Thus $\alpha+\beta$ represents the $E_2^{3,0}$ contribution, while $\langle \alpha\rangle\times \langle\zeta\rangle$ realizes the extension of $E_2^{1,2}$ by $E_2^{0,3}$. In particular, the order-two class $2\zeta$ lies in the $E_2^{1,2}$ filtration piece.

For the other short exact sequence $1\to \{1,x,a^2,a^2x\}\to D_8\to \{1,\bar{a}\}\to 1$, the same computation gives the same $E_2$ page as in Eq.~\eqref{E2page}. In this filtration, $\beta$ represents the $E_2^{3,0}$ contribution, and $\langle \alpha\rangle\times \langle\zeta\rangle$ again realizes the extension of $E_2^{1,2}$ by $E_2^{0,3}$. The class $2\zeta$ is again the nontrivial class in the $E_2^{1,2}$ filtration piece.

\section{Explicit formula for the 3-cocycle $\zeta$ of $D_8$}
\label{App:nu3D8}

We use the resolution constructed in Ref.~\cite{handel1993products} to obtain an explicit representative of the 3-cocycle $\zeta$ for
$D_8 = \langle a,x\mid a^4=1,\ x^2=1,\ xa = a^{-1}x\rangle$, where $a$ is the rotation and $x$ is the reflection, as in the main text. The resolution $C$ consists of free right $\mathbb{Z}D_8$-modules $C_q$, $q\geq0$, with generators $c_q^1,c_q^2,\ldots,c_q^{q+1}$. Let $N = \sum_{i=0}^{3}a^{i}$, and let $N_j = \sum_{i=0}^{j-1}a^{i}$ for $1\leq j \leq3$, with $N_0 = 0$. We also use the convention $N_{j+4k}=N_j$.

The relevant boundary maps in this resolution are
\begin{equation}
\begin{aligned}
\partial c_3^1&=c_2^1(a-1) \\
\partial c_3^2&=c_2^1(x+1)-c_2^2 N \\
\partial c_3^3&=c_2^2(xa-1)+c_2^3(a-1) \\
\partial c_3^4&=c_2^3(x-1)
\end{aligned}
\end{equation}
\begin{equation}
\begin{aligned}
\partial c_4^1&=c_3^1 N \\
\partial c_4^2&=c_3^1(xa-1)+c_3^2(a-1) \\
\partial c_4^3&=c_3^2(x-1)+c_3^3 N \\
\partial c_4^4&=c_3^3(xa+1)+c_3^4(a-1) \\
\partial c_4^5&=c_3^4(x+1)
\end{aligned}
\end{equation}
Applying $\operatorname{Hom}_{\mathbb{Z}D_8}(-,U(1))$ to the resolution gives
\begin{equation}
\delta \mu_2\left(c_3^1, c_3^2, c_3^3, c_3^4\right)= \left(0,2 \mu_2\left(c_2^1\right)-4 \mu_2\left(c_2^2\right), 0,0\right),
\end{equation}
\begin{equation}
\delta \mu_3\left(c_4^1, c_4^2, c_4^3, c_4^4, c_4^5\right)=\left(4 \mu_3\left(c_3^1\right), 0, 4 \mu_3\left(c_3^3\right), 2 \mu_3\left(c_3^3\right), 2 \mu_3\left(c_3^4\right)\right).
\end{equation}
It follows that $H^3(D_8,U(1)) = \Z_4 \times \Z_2 \times \Z_2$. A generator of the $\Z_4$ summand is represented on this resolution by
\begin{equation}
    \zeta(c_3^{r}) = \frac{1}{4}\delta_{r,1}.
\end{equation}

We now express $\zeta$ as a cochain on the standard bar resolution. To do this, we construct a comparison map $h:F\rightarrow C$ from the standard bar resolution $F$ to the resolution $C$:
\begin{equation}
\begin{array}{ccccccccc}
F: \cdots& \rightarrow & F_2 & \stackrel{}{\longrightarrow} & F_1 & \stackrel{}{\longrightarrow} & F_0 & \stackrel{}{\longrightarrow} & \mathbb{Z} \\
 & & \downarrow h_2& & \downarrow h_1 & & \downarrow h_0 & & \downarrow \mathrm{id} \\
C: \cdots& \rightarrow & C_2 & \stackrel{}{\longrightarrow} & C_1 & \stackrel{}{\longrightarrow} & C_0 & \stackrel{}{\longrightarrow} & \mathbb{Z}
\end{array}    
\end{equation}
We use the $\mathbb{Z}$-linear contracting homotopy $T$ of $C$. Starting with $h_{-1}=\operatorname{id}$, we define $h$ inductively by
\begin{equation}
    h_q([g_1|...|g_q]) = T_{q-1}h_{q-1}(\partial [g_1|...|g_q]).
\end{equation}
For $D_8$, the relevant components of $T$ are
\begin{equation}
    T_{-1}(1) = c_0^1
\end{equation}
\begin{equation}
\begin{aligned}
    T_0(c_0^1 a^{i}) &= c_1^{1}N_i \\
    T_0(c_0^1xa^{i}) &= c_1^1 N_i + c_1^2 a^{i}
\end{aligned}
\end{equation}
\begin{equation}
\begin{aligned}
    T_1(c_1^1 a^{i}) &= \delta_{i=3}c_2^1  \\
    T_1(c_1^1xa^{i}) &= \delta_{i=0}(-c_2^1 +c_2^2a^{-1}) + \delta_{1\leq i\leq3}c_2^2 a^{i-1} \\
    T_1(c_1^2 a^{i}) &= 0  \\
    T_1(c_1^2xa^{i}) &=  c_2^3 a^{i}
\end{aligned}
\end{equation}
\begin{equation}
\begin{aligned}
    T_2(c_2^1 a^{i}) &= c_3^1 N_i  \\
    T_2(c_2^1xa^{i}) &= -c_3^1 N_i +c_3^2 a^i  \\
    T_2(c_2^2 a^{i}) &=   0\\
    T_2(c_2^2xa^{i}) &=   c_3^3a^{i-1}\\
    T_2(c_2^3 a^{i}) &=   0\\
    T_2(c_2^3xa^{i}) &=   c_3^4 a^{i}
\end{aligned}
\end{equation}
We view $F_q$ as a right $\Z D_8$-module by identifying $g m=m g^{-1}$. With this convention, the $\Z D_8$-linear comparison map $h$ is constructed by induction:
\begin{equation}
    h_0([\quad]) = T_{-1}(1) = c_0^1
\end{equation}
\begin{equation}
\begin{aligned}
    h_1([a^{i}x^{j}]) &= T_0h_0(a^ix^j[\quad]-[\quad]) = T_0(c_0^1x^ja^{-i} -c_0^1 ) \\
    &= c_1^1N_{-i} + \delta_{j=1}c_1^2 a^{-i}
\end{aligned}
\end{equation}
\begin{equation}
\begin{aligned}    h_2([a^{i_1}x^{j_1}|a^{i_2}x^{j_2}]) &= T_1h_1(a^{i_1}x^{j_1}[a^{i_2}x^{j_2}] - [a^{i_1 + (-1)^{j_1}i_2}x^{j_1+j_2}]+[a^{i_1}x^{j_1}]) \\
    &=T_1(c_1^1(N_{-i_2}x^{j_1}a^{-i_1} -N_{-i_1-(-1)^{j_1}i_2}+N_{-i_1})\\
    &\qquad+T_1(c_1^2(\delta_{j_2=1}a^{-i_2}x^{j_1}a^{-i_1}-\delta_{j_1+j_2=1}a^{-i_1-(-1)^{j_1}i_2}+\delta_{j_1=1}a^{-i_1})\\
    &=\delta_{j_1=0}\delta_{i_2\neq 0}\delta_{[-i_2]_4+[-i_1]_4\geq4}c_2^1+\delta_{j_1=1}\delta_{i_2\neq0} T_1(c_1^1 x N_{-i_2}^{-1}a^{-i_1})
    +\delta_{j_1=1}\delta_{j_2=1}c_2^3 a^{i_2-i_1}\\
    &= c_2^1a^0\delta_{i_2\neq 0}(\delta_{j_1 = 0}\delta_{[-i_2]_4+[-i_1]_4\geq4} -\delta_{j_1=1}\delta_{[i_2]_4+[-i_1]_4 < 4}) + ...
\end{aligned}
\end{equation}
\begin{equation}
\begin{aligned}
&\quad h_3([a^{i_1}x^{j_1}|a^{i_2}x^{j_2}|a^{i_3}x^{j_3}]) \\
&= T_2h_2(a^{i_1}x^{j_1}[a^{i_2}x^{j_2}|a^{i_3}x^{j_3}] - [a^{i_1+(-1)^{j_1}i_2}x^{j_1+j_2}|a^{i_3}x^{j_3}]+[a^{i_1}x^{j_1}|a^{i_2+(-1)^{j_2}i_3}x^{j_2+j_3}] - [a^{i_1}x^{j_1}|a^{i_2}x^{j_2}])\\
&=T_2(c_2^1x^{j_1}a^{-i_1} \delta_{i_3\neq0}(\delta_{j_2 = 0}\delta_{[-i_3]_4+[-i_2]_4\geq4}-\delta_{j_2=1}\delta_{[i_3]_4+[-i_2]_4<4}))+ T_2(c_2^1a^0(...))  + ... \\
&=c_3^1(-1)^{j_1}N_{-i_1}\delta_{i_3\neq0}(\delta_{j_2 = 0}\delta_{[-i_3]_4+[-i_2]_4\geq4}-\delta_{j_2=1}\delta_{[i_3]_4+[-i_2]_4<4}) + ...
\end{aligned}
\end{equation}
In the last line we have omitted all terms whose $c_3^1$ component is zero, since $\zeta$ vanishes on $c_3^2,c_3^3,c_3^4$. This gives
\begin{equation}
\begin{aligned}
h_3^{*}\zeta([a^{i_1}x^{j_1}|a^{i_2}x^{j_2}|a^{i_3}x^{j_3}]) &= \zeta(h_3([a^{i_1}x^{j_1}|a^{i_2}x^{j_2}|a^{i_3}x^{j_3}])) \\ &= \frac{1}{4} (-1)^{j_1}\delta_{i_1\neq 0}\delta_{i_3\neq0}(4-i_1)(\delta_{j_2 = 0}\delta_{[-i_3]_4+[-i_2]_4\geq4}-\delta_{j_2=1}\delta_{[i_3]_4+[-i_2]_4<4})\\
&=\frac{1}{4}(-1)^{j_1}[-i_1]_4\left(\delta_{j_2=0}\left\lfloor\frac{\left[-i_2\right]_4+\left[-i_3\right]_4}{4}\right\rfloor - \delta_{j_2=1}\delta_{i_3\neq0}\left(1-\left\lfloor\frac{\left[-i_2\right]_4+\left[i_3\right]_4}{4}\right\rfloor\right)\right)
\end{aligned}  
\end{equation}

As a consistency check, consider the subgroup $\Z_4^a \stackrel{i}{\hookrightarrow} D_8$ generated by the rotation $a$. The restriction of $\zeta$ to $\mathbb{Z}_4^a$ is a generator of $H^3(\mathbb{Z}_4,U(1))$:
\begin{equation}
i^{*}h_3^{*}\zeta([a^{i_1}|a^{i_2}|a^{i_3}]) = \frac{1}{4}[-i_1]_4 \left\lfloor\frac{[-i_2]_4+[-i_3]_4}{4}\right\rfloor
\end{equation}

Finally, after composing with the automorphism $a^ix^j\mapsto a^{-i}x^j$ of $D_8$, we obtain the following representative for the generator of the $\Z_4$ summand in $H^3(D_8,\U)=\Z_4\times\Z_2\times\Z_2$:
\begin{align}
\zeta(a^{i_1}x^{j_1},a^{i_2}x^{j_2},a^{i_3}x^{j_3})
&=
\frac{1}{4}(-1)^{j_1}i_1\left(\delta_{j_2=0}\left\lfloor\frac{i_2+i_3}{4}\right\rfloor - \delta_{j_2=1}\delta_{i_3\neq0}\left(1-\left\lfloor\frac{i_2+\left[-i_3\right]_4}{4}\right\rfloor\right)\right).
\end{align}
The 3-cocycle of interest, $2\zeta$, is therefore represented by
\begin{align}
(2\zeta)(a^{i_1}x^{j_1},a^{i_2}x^{j_2},a^{i_3}x^{j_3})
&=
\frac{1}{2}i_1\left(\delta_{j_2=0}\left\lfloor\frac{i_2+i_3}{4}\right\rfloor - \delta_{j_2=1}\delta_{i_3\neq0}\left(1-\left\lfloor\frac{i_2+\left[-i_3\right]_4}{4}\right\rfloor\right)\right).
\end{align}

\section{Generalized gauging in $\mathrm{Vec}_{D_8}^{\omega}$ for all $\omega\in H^3(D_8,\U)$}
\label{App:D8ACA}
Write $H^3(D_8,U(1)) = \langle \alpha \rangle \times \langle \beta \rangle \times \langle \zeta \rangle$. Under the outer automorphism defined in Appendix~\ref{App:gaugingZ23}, the generators transform as $\phi^*(\beta) = \alpha + \beta$, $\phi^*(\alpha) = \alpha$, and $\phi^*(\zeta) = \zeta$. Hence the outer automorphism action partitions the 3-cocycle classes into 12 equivalence classes of pointed fusion categories $\mathrm{Vec}_{D_8}^{\omega}$. Following Ref.~\cite{Mu_oz_2018}, these orbits are
    $$
    \{0\},\{\zeta\},\{2\zeta\},\{3\zeta\},\{\alpha\},\{\alpha+\beta,\beta\},\{\zeta+\alpha+\beta,\zeta+\beta\},\{2\zeta+\alpha+\beta,2\zeta+\beta\},\{3\zeta+\alpha+\beta,3\zeta+\beta\},\{\zeta+\alpha\},\{2\zeta+\alpha\},\{3\zeta+\alpha\}
    $$

Using the code in \href{https://github.com/junkicide/fusionrules}{fusionrules}, we compute the group-theoretical fusion categories obtained by gauging all condensable algebras in $\mathrm{Vec}_{D_8}^{\omega}$. The results are listed in the tables below.

\begin{center}

\begin{table}[!h]
        \centering
        
\begin{tabular}{|c|c|}
\hline 
 \multicolumn{2}{|c|}{$\displaystyle ( G,\omega) =( D_{8} ,0)$} \\
\hline \hline 
 $\displaystyle A=( H,\gamma )$ & $\displaystyle _{A}\mathcal{C}_{A}$ \\
\hline 
 $\displaystyle ( 1,0)$ & $\displaystyle \mathrm{Vec}_{D_{8}}$ \\
\hline 
 $\displaystyle ( \langle x\rangle ,0)$ $\overset{\mathrm{M}}{\sim}$ $ $$\displaystyle \left( \langle xa^{2} \rangle =\mathbb{Z}_{2} ,0\right)$ & $\displaystyle \mathrm{Rep}_{D_{8}}$ \\
\hline 
 $\displaystyle ( \langle xa\rangle ,0)$ $\overset{\mathrm{M}}{\sim}$ $ $$\displaystyle \left( \langle xa^{3} \rangle =\mathbb{Z}_{2} ,0\right)$ & $\displaystyle \mathrm{Rep}_{D_{8}}$ \\
\hline 
 $\displaystyle \left( \langle a^{2} \rangle =\mathbb{Z}_{2} ,0\right)$  & $\displaystyle \mathrm{Vec}_{\mathbb{Z}_{2}^{3}}^{w}$ \\
\hline 
 $\displaystyle ( \langle a\rangle =\mathbb{Z}_{4} ,0)$  & $\displaystyle \mathrm{Vec}_{D_{8}}$ \\
\hline 
 $\displaystyle \left( \langle x,a^{2} \rangle =\mathbb{Z}_{2} \times \mathbb{Z}_{2} ,0\right)$  & $\displaystyle \mathrm{Vec}_{D_{8}}$ \\
\hline 
 $\displaystyle \left( \langle x,a^{2} \rangle =\mathbb{Z}_{2} \times \mathbb{Z}_{2} ,\gamma \right)$  & $\displaystyle \mathrm{Vec}_{D_{8}}$ \\
\hline 
 $\displaystyle \left( \langle xa,a^{2} \rangle =\mathbb{Z}_{2} \times \mathbb{Z}_{2} ,0\right)$  & $\displaystyle \mathrm{Vec}_{D_{8}}$ \\
\hline 
 $\displaystyle \left( \langle xa,a^{2} \rangle =\mathbb{Z}_{2} \times \mathbb{Z}_{2} ,\gamma \right)$  & $\displaystyle \mathrm{Vec}_{D_{8}}$ \\
\hline 
 $\displaystyle ( D_{8} ,0)$ & $\displaystyle \mathrm{Rep}_{D_{8}}$ \\
\hline 
 $\displaystyle ( D_{8} ,\gamma )$ & $\displaystyle \mathrm{Rep}_{D_{8}}$ \\
 \hline
\end{tabular}
\caption{\label{D80} Condensable algebras in $\mathrm{Vec}_{D_8}$ and the corresponding dual categories after gauging.} 
        \end{table}
\end{center}
        
\begin{center}

\begin{table}[!h]
        \centering
        
\begin{tabular}{|c|c|}
\hline 
 \multicolumn{2}{|c|}{$\displaystyle ( G,\omega) =( D_{8} ,\zeta )$} \\
\hline \hline 
 $\displaystyle A=( H,\gamma )$ & $\displaystyle _{A}\mathcal{C}_{A}$ \\
\hline 
 $\displaystyle ( 1,0)$ & $\displaystyle \mathrm{Vec}_{D_{8}}^{\zeta }$ \\
\hline 
 $\displaystyle ( \langle x\rangle ,0)$ $\overset{\mathrm{M}}{\sim}$ $ $$\displaystyle \left( \langle xa^{2} \rangle =\mathbb{Z}_{2} ,0\right)$ & $\displaystyle \mathrm{TY}(\mathbb{Z}_{4})$ \\
\hline 
 $\displaystyle ( \langle xa\rangle ,0)$ $\overset{\mathrm{M}}{\sim}$ $ $$\displaystyle \left( \langle xa^{3} \rangle =\mathbb{Z}_{2} ,0\right)$ & $\displaystyle \mathrm{TY}(\mathbb{Z}_{4})$ \\
 \hline
\end{tabular}
\caption{\label{D8z} Condensable algebras in $\mathrm{Vec}_{D_8}^{\zeta}$ and the corresponding dual categories after gauging.}      
        \end{table}
\end{center}

\begin{center}

\begin{table}[!h]
        \centering
        
\begin{tabular}{|c|c|}
\hline 
 \multicolumn{2}{|c|}{\color{red}$\displaystyle ( G,\omega) =( D_{8} ,2\zeta )$} \\
\hline \hline 
 $\displaystyle A=( H,\gamma )$ & $\displaystyle _{A}\mathcal{C}_{A}$ \\
\hline 
 $\displaystyle ( 1,0)$ & $\displaystyle \mathrm{Vec}_{D_{8}}^{2\zeta }$ \\
\hline 
 $\displaystyle ( \langle x\rangle ,0)$ $\overset{\mathrm{M}}{\sim}$ $ $$\displaystyle \left( \langle xa^{2} \rangle =\mathbb{Z}_{2} ,0\right)$ & $\displaystyle \mathrm{Rep} H_{8}$ \\
\hline 
 $\displaystyle ( \langle xa\rangle ,0)$ $\overset{\mathrm{M}}{\sim}$ $ $$\displaystyle \left( \langle xa^{3} \rangle =\mathbb{Z}_{2} ,0\right)$ & $\displaystyle \mathrm{Rep} H_{8}$ \\
\hline 
 $\displaystyle \left( \langle a^{2} \rangle =\mathbb{Z}_{2} ,0\right)$  & $\displaystyle \mathrm{Vec}_{D_{8}}^{2\zeta }$ \\
\hline 
 $\displaystyle \left( \langle x,a^{2} \rangle =\mathbb{Z}_{2} \times \mathbb{Z}_{2} ,0\right)\overset{\mathrm{M}}{\sim} \left( \langle x,a^{2} \rangle =\mathbb{Z}_{2} \times \mathbb{Z}_{2} , \gamma \right) $ & $\displaystyle \mathrm{Rep} H_{8}$ \\
\hline 
 $\displaystyle \left( \langle xa,a^{2} \rangle =\mathbb{Z}_{2} \times \mathbb{Z}_{2} ,0\right) \overset{\mathrm{M}}{\sim} \left( \langle xa,a^{2} \rangle =\mathbb{Z}_{2} \times \mathbb{Z}_{2} ,\gamma\right)$  & $\displaystyle \mathrm{Rep} H_{8}$ \\
\hline 
\end{tabular}
\caption{\label{D82z} Condensable algebras in $\mathrm{Vec}_{D_8}^{2\zeta}$ and the corresponding dual categories after gauging. The cocycle $2\zeta$ realizes the LSM anomaly of the spin-chain DQCP candidate discussed in Sec.~\ref{sec:D8}.}      
        \end{table}
\end{center}

\begin{center}

\begin{table}[!h]
        \centering
        
\begin{tabular}{|c|c|}
\hline 
 \multicolumn{2}{|c|}{$\displaystyle ( G,\omega) =( D_{8} ,3\zeta )$} \\
\hline \hline 
 $\displaystyle A=( H,\gamma )$ & $\displaystyle _{A}\mathcal{C}_{A}$ \\
\hline 
 $\displaystyle ( 1,0)$ & $\displaystyle \mathrm{Vec}_{D_{8}}^{3\zeta }$ \\
\hline 
 $\displaystyle ( \langle x\rangle ,0)$ $\overset{\mathrm{M}}{\sim}$ $ $$\displaystyle \left( \langle xa^{2} \rangle =\mathbb{Z}_{2} ,0\right)$ & $\displaystyle \mathrm{TY}(\mathbb{Z}_{4})$, not equivalent to the one for $\zeta$ \\
\hline 
 $\displaystyle ( \langle xa\rangle ,0)$ $\overset{\mathrm{M}}{\sim}$ $ $$\displaystyle \left( \langle xa^{3} \rangle =\mathbb{Z}_{2} ,0\right)$ & $\displaystyle \mathrm{TY}(\mathbb{Z}_{4})$, not equivalent to the one for $\zeta$\\
 \hline
\end{tabular}
\caption{\label{D83z} Condensable algebras in $\mathrm{Vec}_{D_8}^{3\zeta}$ and the corresponding dual categories after gauging.}          
        \end{table}
\end{center}

\begin{center}

\begin{table}[!h]
        \centering
        
\begin{tabular}{|c|c|}
\hline 
 \multicolumn{2}{|c|}{$\displaystyle ( G,\omega) =( D_{8} ,\alpha )$} \\
\hline \hline 
 $\displaystyle A=( H,\gamma )$ & $\displaystyle _{A}\mathcal{C}_{A}$ \\
\hline 
 $\displaystyle ( 1,0)$ & $\displaystyle \mathrm{Vec}_{D_{8}}^{\alpha }$ \\
\hline 
 $\displaystyle \left( \langle a^{2} \rangle =\mathbb{Z}_{2} ,0\right)$  & $\displaystyle \mathrm{Vec}_{\mathbb{Z}_{2}^{3}}^{aaa + bcc + abc}$ \\
\hline 
 $\displaystyle ( \langle a\rangle =\mathbb{Z}_{4} ,0)$  & $\displaystyle \mathrm{Vec}_{D_{8}}^{\alpha }$ \\
 \hline
\end{tabular}
\caption{\label{D8a} Condensable algebras in $\mathrm{Vec}_{D_8}^{\alpha}$ and the corresponding dual categories after gauging.}           
        \end{table}
\end{center}

\begin{center}

\begin{table}[!h]
        \centering
        
\begin{tabular}{|c|c|}
\hline 
 \multicolumn{2}{|c|}{$\displaystyle ( G,\omega) =( D_{8} ,\beta )$} \\
\hline \hline 
 $\displaystyle A=( H,\gamma )$ & $\displaystyle _{A}\mathcal{C}_{A}$ \\
\hline 
 $\displaystyle ( 1,0)$ & $\displaystyle \mathrm{Vec}_{D_{8}}^{\beta }$ \\
\hline 
 $\displaystyle ( \langle x\rangle ,0)$ $\overset{\mathrm{M}}{\sim}$ $ $$\displaystyle \left( \langle xa^{2} \rangle =\mathbb{Z}_{2} ,0\right)$ & $\displaystyle \mathrm{Rep} Q_{8}$ \\
\hline 
 $\displaystyle \left( \langle a^{2} \rangle =\mathbb{Z}_{2} ,0\right)$  & $\displaystyle \mathrm{Vec}_{\mathbb{Z}_{2}^{3}}^{a+abc}$ \\
\hline 
 $\displaystyle \left( \langle x,a^{2} \rangle =\mathbb{Z}_{2} \times \mathbb{Z}_{2} ,0\right)$ & $\displaystyle \mathrm{Vec}_{D_{8}}^{\beta }$ \\
\hline 
 $\displaystyle \left( \langle x,a^{2} \rangle =\mathbb{Z}_{2} \times \mathbb{Z}_{2} ,\gamma \right)$  & $\displaystyle \mathrm{Vec}_{D_{8}}^{\beta }$ \\
\hline 
 $\displaystyle ( \langle a\rangle =\mathbb{Z}_{4} ,0)$  & $\displaystyle \mathrm{Vec}_{Q_{8}}$ \\
 \hline
\end{tabular}
\caption{\label{D8b} Condensable algebras in $\mathrm{Vec}_{D_8}^{\beta}$ and the corresponding dual categories after gauging.}          
        \end{table}
\end{center}

\begin{center}

\begin{table}[!h]
        \centering
        
\begin{tabular}{|c|c|}
\hline 
 \multicolumn{2}{|c|}{$\displaystyle ( G,\omega) =( D_{8} ,\beta +\zeta )$} \\
\hline \hline 
 $\displaystyle A=( H,\gamma )$ & $\displaystyle _{A}\mathcal{C}_{A}$ \\
\hline 
 $\displaystyle ( 1,0)$ & $\displaystyle \mathrm{Vec}_{D_{8}}^{\beta +\zeta }$ \\
\hline 
 $\displaystyle ( \langle x\rangle ,0)$ $\overset{\mathrm{M}}{\sim}$ $ $$\displaystyle \left( \langle xa^{2} \rangle =\mathbb{Z}_{2} ,0\right)$ & $\displaystyle \mathrm{TY}(\mathbb{Z}_{4})$, not equivalent to the ones for $\zeta$ or $3\zeta$ \\
 \hline
\end{tabular}
\caption{\label{D8bz} Condensable algebras in $\mathrm{Vec}_{D_8}^{\beta+\zeta}$ and the corresponding dual categories after gauging.}          
        \end{table}
\end{center}

\begin{center}

\begin{table}[!h]
        \centering
        
\begin{tabular}{|c|c|}
\hline 
 \multicolumn{2}{|c|}{$\displaystyle ( G,\omega) =( D_{8} ,\beta +2\zeta )$} \\
\hline \hline 
 $\displaystyle A=( H,\gamma )$ & $\displaystyle _{A}\mathcal{C}_{A}$ \\
\hline 
 $\displaystyle ( 1,0)$ & $\displaystyle \mathrm{Vec}_{D_{8}}^{\beta +2\zeta }$ \\
\hline 
 $\displaystyle ( \langle x\rangle ,0)$ $\overset{\mathrm{M}}{\sim}$ $ $$\displaystyle \left( \langle xa^{2} \rangle =\mathbb{Z}_{2} ,0\right)$ & $\displaystyle \mathrm{TY}(\mathbb{Z}_{2} \times \mathbb{Z}_{2})$, with no fiber functor \\
\hline 
 $\displaystyle \left( \langle a^{2} \rangle =\mathbb{Z}_{2} ,0\right)$  & $\displaystyle \mathrm{Vec}_{D_{8}}^{\beta +2\zeta }$ \\
\hline 
 $\displaystyle \left( \langle x,a^{2} \rangle =\mathbb{Z}_{2} \times \mathbb{Z}_{2} ,0\right)$ $\overset{\mathrm{M}}{\sim}$ $\left( \langle x,a^{2} \rangle =\mathbb{Z}_{2} \times \mathbb{Z}_{2} ,\gamma \right) $ & $\displaystyle \mathrm{TY}(\mathbb{Z}_{2} \times \mathbb{Z}_{2})$, with no fiber functor \\
\hline 
\end{tabular}
\caption{\label{D8b2z} Condensable algebras in $\mathrm{Vec}_{D_8}^{\beta+2\zeta}$ and the corresponding dual categories after gauging.}        
        \end{table}
\end{center}

\begin{center}

\begin{table}[!h]
        \centering
        
\begin{tabular}{|c|c|}
\hline 
 \multicolumn{2}{|c|}{$\displaystyle ( G,\omega) =( D_{8} ,\beta +3\zeta )$} \\
\hline \hline 
 $\displaystyle A=( H,\gamma )$ & $\displaystyle _{A}\mathcal{C}_{A}$ \\
\hline 
 $\displaystyle ( 1,0)$ & $\displaystyle \mathrm{Vec}_{D_{8}}^{\beta +3\zeta }$ \\
\hline 
 $\displaystyle ( \langle x\rangle ,0)$ or $ $$\displaystyle \left( \langle xa^{2} \rangle =\mathbb{Z}_{2} ,0\right)$ & $\displaystyle \mathrm{TY}(\mathbb{Z}_{4})$, not equivalent to the ones for $\zeta$, $3\zeta$, or $\beta+3\zeta$ \\
 \hline
\end{tabular}
\caption{\label{D8b3z} Condensable algebras in $\mathrm{Vec}_{D_8}^{\beta+3\zeta}$ and the corresponding dual categories after gauging.}         
        \end{table}
\end{center}

\begin{center}

\begin{table}[!h]
        \centering
        
\begin{tabular}{|c|c|}
\hline 
 \multicolumn{2}{|c|}{$\displaystyle ( G,\omega) =( D_{8} ,\alpha +\zeta )$} \\
\hline \hline 
 $\displaystyle A=( H,\gamma )$ & $\displaystyle _{A}\mathcal{C}_{A}$ \\
\hline 
 $\displaystyle ( 1,0)$ & $\displaystyle \mathrm{Vec}_{D_{8}}^{\alpha +\zeta }$ \\
 \hline
\end{tabular}
\caption{\label{D8az} Condensable algebras in $\mathrm{Vec}_{D_8}^{\alpha+\zeta}$ and the corresponding dual categories after gauging.}          
        \end{table}
\end{center}

\begin{center}

\begin{table}[!h]
        \centering
        
\begin{tabular}{|c|c|}
\hline 
 \multicolumn{2}{|c|}{$\displaystyle ( G,\omega) =( D_{8} ,\alpha +2\zeta )$} \\
\hline \hline 
 $\displaystyle A=( H,\gamma )$ & $\displaystyle _{A}\mathcal{C}_{A}$ \\
\hline 
 $\displaystyle ( 1,0)$ & $\displaystyle \mathrm{Vec}_{D_{8}}^{\alpha +2\zeta }$ \\
\hline 
 $\displaystyle \left( \langle a^{2} \rangle =\mathbb{Z}_{2} ,0\right)$  & $\displaystyle \mathrm{Vec}_{D_{8}}^{\alpha +2\zeta }$ \\
 \hline
\end{tabular}
\caption{\label{D8a2z} Condensable algebras in $\mathrm{Vec}_{D_8}^{\alpha+2\zeta}$ and the corresponding dual categories after gauging.} 
        \end{table}
\end{center}

\begin{center}

\begin{table}[!h]
        \centering
        
\begin{tabular}{|c|c|}
\hline 
 \multicolumn{2}{|c|}{$\displaystyle ( G,\omega) =( D_{8} ,\alpha +3\zeta )$} \\
\hline \hline 
 $\displaystyle A=( H,\gamma )$ & $\displaystyle _{A}\mathcal{C}_{A}$ \\
\hline 
 $\displaystyle ( 1,0)$ & $\displaystyle \mathrm{Vec}_{D_{8}}^{\alpha +3\zeta }$ \\
 \hline
\end{tabular}
\caption{\label{D8a3z} Condensable algebras in $\mathrm{Vec}_{D_8}^{\alpha+3\zeta}$ and the corresponding dual categories after gauging.}       
        \end{table}
\end{center}

Every $\mathrm{TY}(\mathbb{Z}_4)$ category appearing in the preceding tables has no fiber functor. The properties of the four $\text{TY}(\mathbb{Z}_2 \times \mathbb{Z}_2)$ categories are summarized in Table~\ref{TYZ2Z2}~\cite{tambara1998tensor}. Here, $H_8$ denotes the 8-dimensional Kac-Paljutkin Hopf algebra. The last category has no fiber functor and therefore cannot be realized as the representation category of any Hopf algebra.   

\begin{table}[!h]
        \centering
        
\begin{tabular}{|c|c|c|c|c|}
\hline 
 $\displaystyle \mathrm{TY}(\mathbb{Z}_{2} \times \mathbb{Z}_{2})$ & bicharacter $\chi$ & $\displaystyle \tau=\varkappa/2 $ & \begin{tabular}{c}higher Frobenius--Schur indicators\\of the non-invertible object\end{tabular} & number of fiber functors \\
\hline \hline 
 $\displaystyle \mathrm{Rep} D_{8}$ & trivial & $\displaystyle 1/2$ & $\displaystyle 0,1,0,2$ & $\displaystyle 3$ \\
\hline 
 $\displaystyle \mathrm{Rep} Q_{8}$ & trivial & $\displaystyle -1/2$ & $\displaystyle 0,-1,0,2$ & $\displaystyle 1$ \\
\hline 
 $\displaystyle \mathrm{Rep} H_{8}$ & nontrivial & $\displaystyle 1/2$ & $\displaystyle 0,1,0,0,0,1,0,2$ & $\displaystyle 1$ \\
\hline 
 no standard name & nontrivial & $\displaystyle -1/2$ & $\displaystyle 0,-1,0,0,0,-1,0,2$ & $\displaystyle 0$ \\
 \hline
\end{tabular}
\caption{\label{TYZ2Z2}Properties of the four $\mathrm{TY}(\mathbb{Z}_2 \times \mathbb{Z}_2)$ categories.} 
\end{table}



%

\end{document}